\documentclass[a4paper,12pt]{article}
\usepackage{amsmath}
\usepackage{cases}
\usepackage{mathrsfs}
\usepackage{amsfonts}
\usepackage{latexsym}
\usepackage{indentfirst}
\usepackage[dvips]{graphicx}
\usepackage{txfonts}

\makeatletter

\newcommand{\Rmnum}[1]{\expandafter\@slowromancap\romannumeral #1@}
\title{Tricyclic graphs with exactly two main eigenvalues\thanks{This work was supported by Hunan
Provincial Natural Science Foundation of China (09JJ6009) and the
Program for Science and Technology Innovative Research Team in
Higher Educational Institution of Hunan  Province.}}
\author{He Huang, Hanyuan Deng\thanks{Corresponding author :
hydeng@hunnu.edu.cn.}}
\date{}
\begin{document}
\maketitle

\begin{abstract}
An eigenvalue of a graph $G$ is called a main eigenvalue if it has
an eigenvector the sum of whose entries is not equal to zero. In
this paper, all connected tricyclic graphs with exactly two main
eigenvalues are determined.



\end{abstract}

\baselineskip=0.30in

\section{Introduction}

Let $G$ be a simple graph with vertex set $V$ and edge set $E$.
$A=A(G)$ is its adjacency matrix with eigenvalues $\lambda_1,
\lambda_2, \cdots, \lambda_n$, which is also called the
\textit{eigenvalues} of $G$. An eigenvalue of a graph $G$ is call a
\textit{main eigenvalue} if it has an eigenvector the sum of whose
entries is not equal to zero.

It is well known that a graph has exactly one main eigenvalue if and
if it is regular. A long-standing problem posed by Cvetkovi\'c
(see~\cite{cv}) is that of how to characterize graphs with exactly
$k(k\geq 2)$ main eigenvalues. Hagos~\cite{ha} gave an alternative
characterization of graphs with exactly two main eigenvalues.
Recently, Hou and Zhou~\cite{hz} characterized the tree with exactly
two main eigenvalues. Hou and Tian~\cite{ht} showed that all
connected unicyclic graphs with exactly two main eigenvalues is
$C_{r}^{k}$, where $C_{r}^{k}$ is the graph attained from $C_{r}$ by
attaching $k>0$ pendant vertices to every vertex of $C_{r}$. Zhu and
Hu~\cite{zhu} characterized all connected bicyclic graphs with
exactly two main eigenvalues. Rowlinson~\cite{pr} surveyed results
relating main eigenvalues and main angles to the structure of a
graph, and discussed graphs with just two main eigenvalues in the
context of measures of irregularity and in the context of harmonic
graphs.

We call graph $G$ a \textit{tricyclic graph}, if it is a simple and
connected graph in which the number of edges equals the number of
vertices plus two. The aim of this work is to characterize all
connected tricyclic graphs with exactly two main eigenvalues, i.e.,
we determine the 2-walk linear tricyclic graphs.

\section{2-walk linear graphs}
In this section, we recall the definition of 2-walk-linear graphs
and list some basic facts about 2-walk linear graphs. For a graph
$G$, $d(v)$ is the degree of vertex $v$ in $G$, $S(v)$ is the sum of
degree of all vertices adjacent to $v$.

A graph $G$ is called 2-walk $(a, b)$-linear (~\cite{ha},
~\cite{zhu}) if there exist unique rational numbers $a, b$ such that
\begin{equation}
S(\xi)=ad(\xi)+b
\end{equation}
holds for every vertex $\xi\in V(G)$.

In~\cite{ha}, Hagos showed that a graph $G$ has exactly two main
eigenvalues if and only if $G$ is 2-walk linear. Hence, in order to
find all graphs with exactly two main eigenvalues, it is sufficient
to find all 2-walk linear graphs. Hou and Tian obtained the
following results.

{\bf Lemma 2.1}(~\cite{ht}). Let $G$ be a 2-walk $(a, b)$-linear
graph. Then both $a$ and $b$ are integers. Furthermore, if $G$ is
connected, then $a\geq 0$.

From (1), we have directly,

{\bf Lemma 2.2}. Let $G$ be a 2-walk $(a, b)$-linear graph. If $\xi$
and $\eta$ are two vertices of a graph $G$ with degree $d(\xi)$ and
$d(\eta)$, respectively, $d(\xi)\neq d(\eta)$, then
\begin{equation}
a=\frac{S(\eta)-S(\xi)}{d(\eta)-d(\xi)},\ \ \ \ \ \
b=\frac{d(\xi)S(\eta)-d(\eta)S(\xi)}{d(\xi)-d(\eta)}.
\end{equation}

If $X$ is a cycle or a path of $G$, the length of $X$, denoted by
$l(X)$, is defined as the number of edges of $X$.

The following strengthens Lemma 3.1 in~\cite{zhu}.

{\bf Lemma 2.3}. Let $G$ be a connected and 2-walk $(a, b)$-linear
graph. $R=x_1x_2\cdots x_t$ is a path or cycle of length at least 2
in $G$ such that $d(x_1)\geq 3$, $d(x_t)\geq 3$ and
$d(x_2)=\cdots=d(x_{t-1})=2$.

(i)(\cite{zhu}) If $d(x_1)=d(x_t)$, then $l(R)\leq 3$; If $l(R)=3$,
then there exists no path $Q=y_1y_2y_3$ in $G$ such that
$d(y_1)=d(y_3)=d(x_1)$ and $d(y_2)=2$.

(ii) If $d(x_1)\neq d(x_t)$, then $l(R)\leq 2$.

{\bf Proof}. (ii) By way of contradiction, assume that $l(R)\geq 3$,
then $d(x_2)=d(x_{t-1})=2$. Applying (1) with $\xi=x_2$ and
$x_{t-1}$, respectively, we get $S(x_2)=2a+b=S(x_{t-1})$. And
$S(x_2)=d(x_1)+d(x_3)=d(x_1)+2$,
$S(x_{t-1})=d(x_{t-2})+d(x_t)=d(x_t)+2$, which implies
$d(x_1)=d(x_t)$, a contradiction. Hence, (ii) is true. \hfill $\Box$

\begin{figure}
\includegraphics{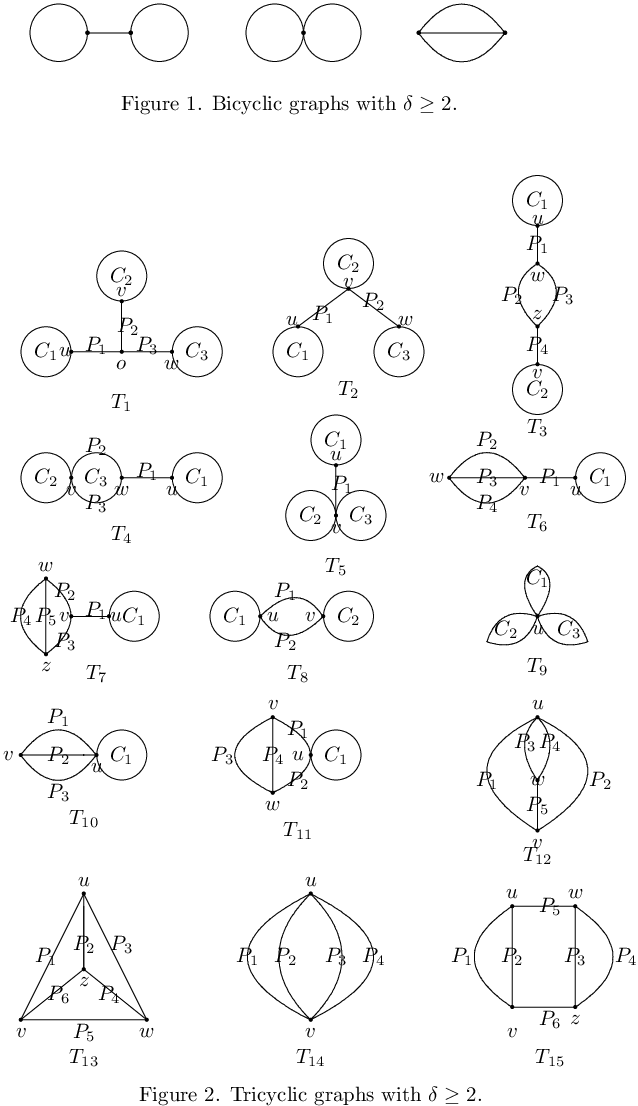}
\end{figure}
\begin{figure}
\includegraphics{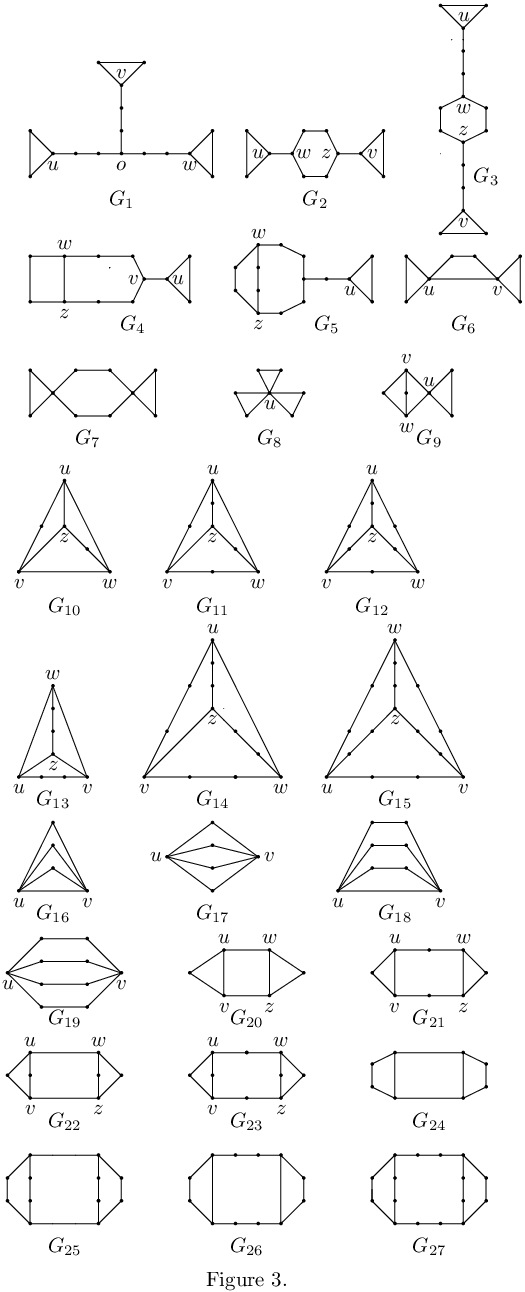}
\end{figure}

\section{2-walk linear tricyclic graphs with minimum degree
$\delta\geq 2$}

In this section, we shall determine all the 2-walk (a, b)-linear
tricyclic graphs with minimum degree $\delta\geq 2$.

Note that all the bicyclic graphs with minimum degree $\delta\geq 2$
can be partitioned into three classes by the arrangement of cycles,
see Figure 1. So, all the tricyclic graphs with minimum degree
$\delta\geq 2$ can be partitioned into fifteen classes by the
arrangement of cycles, see Figure 2.

{\bf Theorem 3.1}. The graphs $G_1, G_2, \cdots, G_{27}$ (see Figure
3) are the only 2-walk linear tricyclic graphs with minimum degree
$\delta\geq 2$.

{\bf Proof}. Let $G$ be a connected and 2-walk (a,b)-linear
tricyclic graph with minimum degree $\delta(G)\geq 2$. Then
$\left|E(G)\right|=\left|V(G)\right|+2$.

In the following, we always assume that $C_1=u_0u_1\cdots u_ru_0$,
$C_2=v_0v_1\cdots v_sv_0$, $C_3=w_0w_1\cdots w_tw_0$,
$P_1=x_0x_1\cdots x_j$, $P_2=y_0y_1\cdots y_k$, $P_3=z_0z_1\cdots
z_l$, $P_4=p_0p_1\cdots p_m$, $P_5=q_0q_1\cdots q_n$,
$P_6=f_0f_1\cdots f_e$.

We consider thirteen cases. In these cases, we assume $G$ is
2-walk-linear.

{\bf Case 1}. $G=T_1$, where $u_0=x_0=u, v_0=y_0=v, w_0=z_0=w,
x_j=y_k=z_l=o$.

Applying Lemma 2.3(i) with $R=C_1, C_2, C_3, P_1, P_2$ and $P_3$,
respectively, we get $l(C_1)=l(C_2)=l(C_3)=3$, and $2\neq l(P_i)\leq
3$, $\forall i\in \{1, 2, 3\}$.

If $l(P_1)$, $l(P_2)$ and $l(P_3)$ are not all the same, without
loss of generality, we may assume that $l(P_1)=3$ and $l(P_2)=1$,
then $j=3, k=1$. Applying (1) with $\xi=u_0$ and $v_0$,
respectively, we get $S(u_0)=3a+b=S(v_0)$. But
$S(u_0)=d(u_1)+d(u_2)+d(x_2)=2+2+2=6$,
$S(v_0)=d(v_1)+d(v_2)+d(o)=2+2+3=7$, a contradiction. Hence
$l(P_1)=l(P_2)=l(P_3)$.

If $l(P_1)=l(P_2)=l(P_3)=1$, then $j=k=l=1$. Applying (1) with
$\xi=u_0$ and $o$, respectively, we get $S(u_0)=3a+b=S(o)$. But
$S(u_0)=d(u_1)+d(u_2)+d(o)=2+2+3=7$, $S(o)=d(u)+d(v)+d(w)=9$, a
contradiction. Hence $l(P_1)=l(P_2)=l(P_3)=3$, which implies that
$G=G_1$. And it can be checked immediately that $G_1$ is 2-walk (1,
3)-linear.

{\bf Case 2}. $G=T_2, T_4$ or $T_5$.

In $T_2$, we take $u_0=x_0=u$, $v_0=x_j=y_k=v$ and $w_0=y_0=w$; In
$T_4$, we take $u_0=x_0=u$, $v_0=z_0=y_0=v$ and $x_j=y_k=z_l=w$; In
$T_5$, we take $u_0=x_0=u$ and $v_0=w_0=x_j=v$.

Applying Lemma 2.3(i) with $R=C_1$ and $C_2$, respectively, we get
$l(C_1)=l(C_2)=3$. Applying (1) with $\xi=u_1$ and $v_1$,
respectively, we get $S(u_1)=2a+b=S(v_1)$. But
$S(u_1)=d(u_0)+d(u_1)=5$, $S(v_1)=d(v_0)+d(v_2)=d(v)+2\geq 6$, a
contradiction. Hence, $G$ is not 2-walk linear.

{\bf Case 3}. $G=T_3$, where $u_0=x_0=u$, $x_j=y_0=z_0=w$,
$y_k=z_l=p_m=z$ and $v_0=p_0=v$.

Applying Lemma 2.3(i) with $R=C_1, C_2, P_1, P_2, P_3$ and $P_4$,
respectively, we have $l(C_1)=l(C_2)=3$ and
\begin{equation}
2\neq l(P_i)\leq 3, \ \ \ \ \forall i\in \{1,2,3,4\}
\end{equation}

If $l(P_1)\neq l(P_4)$, without loss of generality, we may assume
that $l(P_1)=3$ and $l(P_4)=1$, i.e., $j=3, m=1$. Applying (1) with
$\xi=u$ and $w$, respectively, we have $S(u)=3a+b=S(w)$. But
$S(u)=d(u_1)+d(u_2)+d(x_1)=2+2+2=6$,
$S(w)=d(v_1)+d(v_2)+d(z)=2+2+3=7$, a contradiction. Hence,
$l(P_1)=l(P_4)$.

If $l(P_1)=l(P_4)=1$, then $j=m=1$. Applying (1) with $\xi=u$ and
$w$, $S(u)=3a+b=S(w)$, $S(u)=d(u_1)+d(u_2)+d(x_1)=2+2+3=7$. So
$S(w)=7$, then $d(x_0)+d(y_1)+d(z_1)=7$. Since $d(x_0)=3$, we get
$d(y_1)+d(z_1)=4$, and $d(y_1)=d(z_1)=2$. So $l(P_2)>1$, $l(P_3)>1$.
Together with (3), it is easily to know that $l(P_2)=3$, $l(P_3)=3$,
which implies that $G=G_2$. It can be checked that $G_2$ is 2-walk
(2, 1)-linear.

If $l(P_1)=l(P_4)=3$, then $j=3,m=3$. Applying (1) with $\xi=u$ and
$w$, $S(u)=3a+b=S(w)$, $S(u)=d(u_1)+d(u_2)+d(x_1)=2+2+2=6$. And
$S(w)=6$, i.e., $d(x_2)+d(y_1)+d(z_1)=6$. Since $d(x_2)=2$, we get
$d(y_1)+d(z_1)=4$, and $d(y_1)=d(z_1)=2$. So $l(P_2)>1$, $l(P_3)>1$.
Together with (3), it is easily to know that $l(P_2)=3$, $l(P_3)=3$,
which implies that $G=G_3$. It can be checked that $G_3$ is 2-walk
(1, 3)-linear.

{\bf Case 4}. $G=T_6$, where $u_0=x_0=u$, $x_j=y_0=z_0=p_0=v$,
$y_k=z_l=p_m=w$.

Applying Lemma 2.3 with $R=C_1, P_1, P_2, P_3$ and $P_4$,
respectively, $l(C_1)=3$, $l(P_i)\leq 2$, $\forall i\in
\{1,2,3,4\}$. If there exist an $i\in \{1,2,3,4\}$ such that
$l(P_i)=2$, without loss of generality, we may assume that
$l(P_1)=2$, i.e., $j=1$. Applying (1) with $\xi=x_1$ and $u_1$, we
have $S(x_1)=2a+b=S(u_1)$, but $S(u_1)=d(u_2)+d(u_0)=2+3=5$,
$S(x_1)=d(x_0)+d(x_2)=3+4=7$, a contradiction. Hence
$l(P_1)=l(P_2)=l(P_3)=l(P_4)=1$. But $G$ is simple, a contradiction.
Hence, $G$ is not 2-walk linear.

{\bf Case 5}. $G=T_7$, where $u_0=x_0=u$, $x_j=y_0=z_0=v$,
$y_k=p_0=q_0=w$ and $z_l=p_m=q_n=z$.

Applying Lemma 2.3(i) with $R=C_1, P_1, P_2, P_3, P_4$ and $P_5$
respectively, we have $l(C_1)=3$ and
\begin{equation}
2\neq l(P_i)\leq 3,\ \ \ \ \forall i\in \{1,2,3,4,5\}
\end{equation}
So, $l(P_1)=1$ or $3$.

In the following, we prove $d(x_1)=d(x_{j-1})$. If $l(P_1)=1$, then
$x_{j-1}=u$, $x_1=v$ and $j=1$. So $d(x_{j-1})=d(u)=3$,
$d(x_1)=d(v)=3$. Hence, $d(x_1)=d(x_{j-1})$. If $l(P_1)=3$, then
$x_{j-1}=x_2$ and $j=3$, we get $d(x_{j-1})=d(x_2)=2$, $d(x_1)=2$.
So $d(x_1)=d(x_{j-1})$. Hence, $d(x_1)=d(x_{j-1})$ whether
$l(P_1)=1$ or $l(P_1)=3$.

Applying (1) with $\xi=u$ and $v$, we get $S(u)=3a+b=S(v)$,
$S(u)=d(x_1)+d(u_1)+d(u_2)=d(x_1)+2+2=d(x_1)+4$. Since
$S(v)=d(x_{j-1})+d(y_1)+d(z_1)=d(x_1)+d(y_1)+d(z_1)$, we have
$d(y_1)+d(z_1)=4$, and $d(y_1)=d(z_1)=2$. So $l(P_2)>1$, $l(P_3)>1$.
Together with (4), it is easily to know that $l(P_2)=3$, $l(P_3)=3$,
then $k=l=3$. Applying (1) with $\xi=w$ and $u$, we get
$S(w)=3a+b=S(u)$, $S(u)=d(u_1)+d(u_2)+d(x_1)=2+2+d(x_1)=4+d(x_1)$.
Since $S(w)=d(p_1)+d(q_1)+d(y_2)=d(p_1)+d(q_1)+2$, we get
$d(p_1)+d(q_1)=d(x_1)+2$. By the symmetry, we may assume that
$d(p_1)\geq d(q_1)$. If $l(P_1)=1$, then $d(x_1)=3$,
$d(p_1)+d(q_1)=5$. So $d(p_1)=3, d(q_1)=2$, i.e., $m=1, n>1$.
Together with (4), it is easily to know that $l(P_5)=3$, i.e.,
$n=3$, which implies that $G=G_4$; If $l(P_1)=3$, then $d(x_1)=2$,
$d(p_1)+d(q_1)=4$. So $d(p_1)=2,d(q_1)=2$. Together with (4), we get
$l(P_4)=3,l(P_5)=3$, which implies that $G=G_5$. It can be checked
that $G_4$ is 2-walk (2,1)-linear and $G_5$ is 2-walk (1, 3)-linear.

{\bf Case 6}. $G=T_8$, where $u_0=x_0=y_0=u$ and $v_0=x_j=y_k=v$.

Applying Lemma 2.3(i) with $R=C_1, C_2, P_1$ and $P_2$,
respectively, we get $l(C_1)=3$, $l(C_2)=3$, and $2\neq l(P_i)\leq
3, \forall i\in \{1, 2\}$. By the symmetry, we may assume that
$l(P_1)\geq l(P_2)$. Since $G$ is simple, $l(P_1)=3$, $l(P_2)=1$ or
$l(P_1)=3$, $l(P_2)=3$, i.e., $j=3$, $k=1$ or $j=3$, $k=3$, which
implies that $G=G_6$ or $G=G_7$. It can be checked that $G_6$ is
2-walk (2, 2)-linear and $G_7$ is 2-walk (1, 4)-linear.

{\bf Case 7}. $G=T_9$, where $u_0=v_0=w_0=u$.

Applying Lemma 2.3(i) with $R=C_1, C_2$ and $C_3$, respectively, we
get $l(C_1)=3$, $l(C_2)=3$, $l(C_3)=3$, which implies that $G=G_8$.
It can be checked that $G_8$ is 2-walk (1, 6)-linear.

{\bf Case 8}. $G=T_{10}$, where $u_0=x_0=y_0=z_0=u$ and
$x_j=y_k=z_l=v$.

Applying Lemma 2.3 with $R=C_1, P_1, P_2$ and $P_3$, respectively,
we get $l(C_1)=3$ and $l(P_i)\leq 2$, $\forall i\in \{1,2,3\}$. By
the symmetry, we may assume that $l(P_1)\geq l(P_2)\geq l(P_3)$.
Since $G$ is simple, $l(P_2)\geq 2$. So $l(P_1)=l(P_2)=2$. Applying
(1) with $\xi=u_1$ and $x_1$, we get $S(u_1)=2a+b=S(x_1)$. But
$S(u_1)=d(u_2)+d(u_0)=2+5=7$,
$S(x_1)=d(x_0)+d(x_2)=d(u)+d(v)=5+3=8$, a contradiction. Hence,
$G=T_{10}$ is not 2-walk linear.

{\bf Case 9}. $G=T_{11}$, where $u_0=x_0=y_0=u$, $x_j=z_0=p_0=v$ and
$y_k=z_l=p_m=w$.

Applying Lemma 2.3 with $R=C_1, P_1, P_2, P_3$ and $P_4$,
respectively, we get $l(C_1)=3$, $l(P_1)\leq 2$, $l(P_2)\leq 2$,
$l(P_3)\leq 3$ and $l(P_4)\leq 3$.

If $l(P_1)=2$, then $j=2$. Applying (1) with $\xi=u_1$ and $x_1$, we
get $S(u_1)=2a+b=S(x_1)$. But $S(u_1)=d(u_0)+d(u_2)=d(u)+2=4+2=6$,
$S(x_1)=d(x_0)+d(x_2)=d(u)+d(v)=4+3=7$, a contradiction. Hence,
$l(P_1)=1$. Similarly, $l(P_2)=1$. Applying (2) with $(\xi,
\eta)=(u_0, u_1)$ and $(\xi, \eta)=(v, u_1)$, respectively, we get
\begin{displaymath}
a=\frac{S(u_0)-S(u_1)}{d(u_0)-d(u_1)}=\frac{d(u_1)+d(u_2)+d(v)+d(w)-d(u_0)-d(u_2)}{4-2}=\frac{10-6}{2}=2,
\end{displaymath}
\begin{displaymath}
a=\frac{S(v)-S(u_1)}{d(v)-d(u_1)}=\frac{d(u)+d(z_1)+d(p_1)-d(u)-d(u_1)}{3-2}=d(z_1)+d(p_1)-2
\end{displaymath}
Then $d(z_1)+d(p_1)=a+2=4$, which implies that $d(z_1)=2$,
$d(p_1)=2$. Hence $l(P_3)\geq 2$, $l(P_4)\geq 2$. Applying (1) with
$\xi=z_1$ and $u_1$, we get $S(z_1)=2a+b=S(u_1)$. And
$S(u_1)=d(u_0)+d(u_2)=4+2=6$, $S(z_1)=d(z_0)+d(z_2)=3+d(z_2)$, and
$d(z_2)=3$. So $l(P_3)=2$, i.e., $l=2$. Similarly, $m=2$, which
implies that $G=G_9$. It can be checked that $G_9$ is 2-walk (2,
2)-linear.

{\bf Case 10}. $G=T_{12}$, where $x_0=y_0=z_0=p_0=u$,
$z_l=p_m=q_0=w$ and $x_j=y_k=q_n=v$.

By the symmetry, we may assume that $l(P_1)\geq l(P_2)$ and
$l(P_3)\geq l(P_4)$. Since $G$ is simple, $l(P_1)\geq 2$,
$l(P_3)\geq 2$. Applying Lemma 2.3(ii) with $R=P_1, P_2, P_3$ and
$P_4$, respectively, we get $l(P_i)\leq 2$ for $i\in\{1,2,3,4\}$.
Then $l(P_1)=2$, $l(P_3)=2$, i.e., $j=2,l=2$. If $l(P_5)\geq 2$,
then $d(q_1)=2$. Applying (1) with $\xi=x_1$ and $q_1$,
$S(x_1)=2a+b=S(q_1)$. But
$S(x_1)=d(x_0)+d(x_2)=4+3=7,S(q_1)=d(q_0)+d(q_2)=3+d(q_2)\leq6$, a
contradiction. Hence $l(P_5)=1$, i.e., $n=1$. Applying (2) with
$(\xi, \eta)=(x_0, x_2)$ and $(x_2, x_1)$, respectively,
\begin{displaymath}
a=\frac{S(x_0)-S(x_2)}{d(x_0)-d(x_2)}=\frac{d(x_1)+d(y_1)+d(z_1)+d(p_1)-d(x_1)-d(w)-d(y_{k-1})}{4-3}
\end{displaymath}
\begin{equation}
=d(y_1)+2+d(p_1)-3-d(y_{k-1})=d(y_1)+d(p_1)-d(y_{k-1})-1
\end{equation}
\begin{equation}
a=\frac{S(x_2)-S(x_1)}{d(x_2)-d(x_1)}=\frac{d(x_1)+d(w)+d(y_{k-1})-d(x_0)-d(v)}{3-2}=d(y_{k-1})-2
\end{equation}
If $l(P_2)=l(P_4)=1$, by (5), we get
$a=d(y_1)+d(p_1)-d(y_{k-1})-1=3+3-4-1=1$; by (6), we get
$a=d(y_{k-1})-2=4-2=2$, a contradiction. Hence, $l(P_2)=2$ or
$l(P_4)=2$. By the symmetry, we may assume that $l(P_2)=2$. Applying
(1) with $\xi=v$ and $w$, respectively, we get $S(v)=3a+b=S(w)$.
Since
$S(w)=S(q_1)=d(q_0)+d(x_1)+d(y_1)=3+2+2=7,S(v)=S(q_0)=d(q_1)+d(z_1)+d(p_{m-1})=3+2+d(p_{m-1})$,
$d(p_{m-1})=2$. So $l(P_4)\geq 2$. By (5), we get
$a=d(y_1)+d(p_1)-d(y_{k-1})-1=2+2-2-1=1$; by (6), we get
$a=d(y_{k-1})-2=2-2=0$, a contradiction. Hence, $G$ is not 2-walk
linear.

{\bf Case 11}. $G=T_{13}$, where $x_0=y_0=z_0=u$, $x_j=q_0=f_0=v$,
$z_l=p_0=q_n=w$ and $y_k=p_m=f_e=z$.

{\bf Claim 1}.  There exists $i\in\{1,2,3,4,5,6\}$ such that
$l(P_i)>1$.

{\bf Proof}. By way of contradiction, assume that $l(P_i)=1$ for
$i\in\{1,2,3,4,5,6\}$. Then $G$ is 3-regular, and it has exactly one
main eigenvalue, a contradiction. Hence, Claim 1 is true. \hfill
$\Box$

{\bf Claim 2}. (i) $j, k, l, m, n, e\leq 3$; (ii)
$d(x_1)=d(x_{j-1})=d(p_1)=d(p_{m-1})$,
$d(y_1)=d(y_{k-1})=d(q_1)=d(q_{n-1})$ and
$d(z_1)=d(z_{l-1})=d(f_1)=d(f_{e-1})$.

{\bf Proof}. (i) Applying Lemma 2.3(i) with $R=P_1, P_2, P_3, P_4,
P_5$ and $P_6$, respectively, we get $l(P_i)\leq 3$ for $i\in\{1, 2,
3, 4, 5, 6\}$, i.e., $j, k, l, m, n, e\leq3$.

  (ii) If $j=1$, then $x_1=v$, $x_{j-1}=u$. Since $d(u)=3=d(v)$,
$d(x_1)=d(x_{j-1})$. If $2\leq j\leq 3$, then $d(x_1)=2$,
$d(x_{j-1})=2$. So $d(x_1)=d(x_{j-1})$.

Similarly, $d(y_1)=d(y_{k-1})$, $d(z_1)=d(z_{l-1})$,
$d(p_1)=d(p_{m-1})$, $d(q_1)=d(q_{n-1})$, $d(f_1)=d(f_{e-1})$. Let
$d(x_1)=d(x_{j-1})=t_1$, $d(y_1)=d(y_{k-1})=t_2$,
$d(z_1)=d(z_{l-1})=t_3$, $d(p_1)=d(p_{m-1})=t_4$,
$d(q_1)=d(q_{n-1})=t_5$, $d(f_1)=d(f_{e-1})=t_6$. Applying (1) with
$\xi=x_0, f_0, p_0$ and $y_k$, respectively, we get $S(x_0)=3a+b$,
$S(f_0)=3a+b$, $S(p_0)=3a+b$, $S(y_k)=3a+b$. Then
$S(x_0)=S(f_0)=S(p_0)=S(y_k)$. So $S(x_0)+S(f_0)=S(p_0)+S(y_k)$.
Since $S(x_0)=d(x_1)+d(y_1)+d(z_1)=t_1+t_2+t_3$,
$S(f_0)=d(x_{j-1})+d(f_1)+d(q_{n-1})=t_1+t_6+t_5$,
$S(p_0)=d(z_{l-1})+d(p_0)+d(q_{n-1})=t_3+t_4+t_5$ and
$S(y_k)=d(y_{k-1})+d(p_{m-1})+d(f_{e-1})=t_2+t_4+t_6$, we have
$t_1+t_2+t_3+t_1+t_6+t_5=t_3+t_4+t_5+t_2+t_4+t_6$, and $t_1=t_4$.
Similarly, $t_2=t_5,t_3=t_6$. Hence,
$d(x_1)=d(x_{j-1})=d(p_1)=d(p_{m-1})$,
$d(y_1)=d(y_{k-1})=d(q_1)=d(q_{n-1})$,
$d(z_1)=d(z_{l-1})=d(f_1)=d(f_{e-1})$. \hfill $\Box$

{\bf Claim 3}. $l(P_1)=l(P_4)$, $l(P_2)=l(P_5)$, $l(P_3)=l(P_6)$.

{\bf Proof}. By Claim 2(i), we get $l(P_1)\leq3$. If $l(P_1)=1$,
then $x_1=v$ and $d(x_1)=d(v)=3$. By Claim 2(2), we get
$d(p_1)=d(x_1)=3$. So $l(P_4)=1$. Hence, $l(P_1)=l(P_4)$. If
$l(P_1)=2$, then $d(x_1)=2$. By Claim 2(ii), we get
$d(p_1)=d(x_1)=2$, then $l(p_4)\geq2$. Together with Lemma 2.3(i),
we can prove $l(P_4)=2$. So $l(P_1)=l(P_4)$. If $l(P_1)=3$, then
$d(x_1)=2$. By Claim 2(ii), we get $d(p_1)=d(x_1)=2$, then
$l(p_4)\geq2$. Together with Lemma 2.3(i), we can prove $l(P_4)=3$.
So $l(P_1)=l(P_4)$. Similarly, we can prove $l(P_2)=l(P_5)$,
$l(P_3)=l(P_6)$.

Similarly, $l(P_2)=l(P_5)$, $l(P_3)=l(P_6)$. \hfill $\Box$

By the symmetry, we may assume that $j\geq k \geq l$. It follows
from Claim 1 and Claim 3 that $j\geq 2$. From Claim 2(i), we get
$j=2$ or $3$.

If $j=2$, i.e., $l(P_1)=2$, then $(l(P_2), l(P_3))\in \{(1, 1), (2,
1), (2, 2)\}$. Together with Claim 3, we know that $(l(P_1), l(P_2),
l(P_3), l(P_4), l(P_5), l(P_6))\in\{(2, 1, 1, 2, 1, 1)$, $(2, 2, 1,
2, 2, 1), (2, 2, 2, 2, 2, 2)\}$, which implies that $G=G_{10},
G_{11}, G_{12}$, respectively. It can be checked that $G_{10}$ is
2-walk (2, 2)-linear, $G_{11}$ is 2-walk (1, 4)-linear and $G_{12}$
is 2-walk (0, 6)-linear.

If $j=3$, i.e., $l(P_1)=3$, then $(l(P_2), l(P_3))\in\{(1, 1), (3,
1), (3, 3)\}$. Together with Claim 3, we know that $(l(P_1), l(P_2),
l(P_3), l(P_4)$, $l(P_5), l(P_6))\in \{(3,1,1,3,1,1)$,
$(3,3,1,3,3,1), (3,3,3,3,3,3)\}$, which implies that $G=G_{13},
G_{14}, G_{15}$, respectively. It can be checked that $G_{13}$ is
2-walk (3, -1)-linear, $G_{14}$ is 2-walk (2, 1)-linear, $G_{15}$ is
2-walk (1, 3)-linear.

{\bf Case 12}. $G=T_{14}$, where $x_0=y_0=z_0=p_0=u$ and
$x_j=y_k=z_l=p_m=v$.

By the symmetry, we may assume that $l(P_1)\geq l(P_2)\geq
l(P_3)\geq l(P_4)$. Since $G$ is simple, $l(P_3)\geq2$. Applying
Lemma 2.3(i) with $R=P_3$, we get $l(P_3)\leq3$.

If $l(P_3)=2$, then by Lemma 2.3(i) with $R=P_3, P_1, P_2$ and
$P_4$, respectively, we get $l(P_i)\leq 3$ and $l(P_i)\neq 3$ for
$i\in\{1,2,4\}$. So $l(P_1)=2$, $l(P_2)=2$, $l(P_4)=1$ or
$l(P_4)=2$, which implies that $G=G_{16}, G_{17}$. It can be checked
that $G_{16}$ is 2-walk (1, 6)-linear, $G_{17}$ is 2-walk (0,
8)-linear.

If $l(P_3)=3$, then by Lemma 2.3(i) with $R=P_3, P_1, P_2$ and
$P_4$, we get $l(P_i)\leq 3$ and $l(P_i)\neq 2$ for $i\in
\{1,2,4\}$. So $l(P_1)=3$, $l(P_2)=3$, $l(P_4)=1$ or $l(P_4)=3$,
which implies that $G=G_{18}, G_{19}$. It can be checked that
$G_{18}$ is 2-walk (2, 2)-linear, $G_{19}$ is 2-walk (1, 4)-linear.

{\bf Case 13}. $G=T_{15}$, where $x_0=y_0=q_0=u$, $x_j=y_k=f_0=v$,
$q_n=z_0=p_0=w$ and $f_e=z_l=p_m=z$.

If $l(P_1)=1$, i.e., $j=1$, then $x_1=x_j=v$, $x_{j-1}=x_0=u$,
$d(x_1)=d(v)=3$, $d(x_{j-1})=d(u)=3$. So $d(x_1)=d(x_{j-1})$. If
$l(P_1)>1$, i.e., $j>1$, then $d(x_1)=2$, $d(x_{j-1})=2$. In a word,
$d(x_1)=d(x_{j-1})$.

Similarly, $d(y_1)=d(y_{k-1})$, $d(z_1)=d(z_{l-1})$,
$d(p_1)=d(p_{m-1})$, $d(q_1)=d(q_{n-1})$ and $d(f_1)=d(f_{e-1})$.

Hence, we have
\begin{displaymath}
d(x_1)=d(x_{j-1}),\ \ \ \ \ d(y_1)=d(y_{k-1}),\ \ \ \ \
d(z_1)=d(z_{l-1}),
\end{displaymath}
\begin{equation}
d(p_1)=d(p_{m-1}),\ \ \ \ \ d(q_1)=d(q_{n-1}),\ \ \ \ \
d(f_1)=d(f_{e-1})
\end{equation}
By the symmetry, we may assume that $l(P_1)\geq l(P_2)$, $l(P_3)\geq
l(P_4)$. Since $G$ is simple, $l(P_1)\geq 2$, $l(P_3)\geq 2$. Then
$d(x_1)=2$, $d(z_1)=2$. Applying (1) with $\xi=x_0$ and $z_0$, we
get $S(x_0)=3a+b=S(z_0)$. And
$S(x_0)=d(x_1)+d(y_1)+d(p_1)=2+d(y_1)+d(q_1)$,
$S(z_0)=d(z_1)+d(P_1)+d(q_{n-1})=2+d(p_1)+d(q_{n-1})$. Together with
(7), we get $d(y_1)=d(p_1)$.

Applying (1) with $\xi=x_0$ and $f_0$, we get $S(x_0)=3a+b=S(z_0)$.
And $S(x_0)=d(x_1)+d(y_1)+d(p_1)=2+d(y_1)+d(q_1)$,
$S(f_0)=d(x_{j-1})+d(y_{k-1})+d(f_1)=2+d(y_{k-1})+d(f_1)$. Together
with (7), we get $d(q_1)=d(f_1)$.

Hence,
\begin{equation}
d(x_1)=d(z_1)=2,\ \ \ d(y_1)=d(p_1),\ \ \ d(q_1)=d(f_1).
\end{equation}
By applying Lemma 2.3(i) with $R=P_1, P_2, P_3, P_4, P_5$ and $P_6$,
respectively, we get $l(P_i)\leq 3$ for $i\in\{1,2,3,4,5,6\}$.

If $l(P_1)=2$, by applying Lemma 2.3(i) with $R=P_1, P_2, P_3, P_4,
P_5$ and $P_6$, respectively, we get $l(P_i)\neq 3$ for
$i\in\{2,3,4,5,6\}$. Then $\forall i\in \{1,2,3,4,5,6\}$,
$l(P_i)\leq 2$, together with (8), we get $l(P_1)=l(P_3)$,
$l(P_2)=l(P_4)$, $l(P_5)=l(P_6)$. So, $(l(P_1), (P_2), l(P_3),
l(P_4), l(P_5), l(P_6))\in \{(2,1,1,2,1,1), (2,2,1,2,2,1), (2,1,2$,
$2,1,2), (2,2,2,2,2,2)\}$, which implies $G=G_{20}, G_{21}, G_{22},
G_{23}$, respectively. It can be checked that $G_{20}$ is 2-walk (2,
2)-linear, $G_{21}$ is 2-walk (1, 4)-linear, $G_{22}$ is 2-walk (1,
4)-linear, $G_{23}$ is 2-walk (0, 6)-linear.

If $l(p_1)=3$, by applying Lemma 2.3(i) with $R=P_1, P_2, P_3, P_4,
P_5$ and $P_6$, respectively, we get $l(P_i)\neq 2$ for
$i\in\{2,3,4,5,6\}$. Then $l(P_i)=1$ or $3$, together with (8), we
get $l(P_1)=l(P_3)$, $l(P_2)=l(P_4)$, $l(P_5)=l(P_6)$. So, $(l(P_1),
l(P_2), l(P_3), l(P_4)$, $l(P_5), l(P_6))\in \{(3, 1, 1, 3, 1, 1),
(3, 3, 1, 3, 3, 1), (3, 1, 3, 3, 1, 3), (3, 3, 3, 3, 3, 3)\}$, which
implies $G=G_{24}, G_{25}, G_{26}, G_{27}$, respectively. It can be
checked that $G_{24}$ is 2-walk (3, -1)-linear, $G_{25}$ is 2-walk
(2, 1)-linear, $G_{26}$ is 2-walk (2, 1)-linear, $G_{27}$ is 2-walk
(1, 3)-linear.

\section{2-walk linear tricyclic graphs with minimum degree
$\delta=1$}

In the following, we shall determine all 2-walk $(a, b)$-linear
tricyclic graph with minimum degree $\delta=1$. For convenience, we
define
\begin{center}
$\mathscr{G}_{a, b}$=\{$G$ is a connected and 2-walk (a, b)-linear
tricyclic graph with $\delta(G)=1\}$
\end{center}
and for each $G\in\mathscr{G}_{a, b}$, let $G_0$ be the graph
obtained from $G$ by deleting all pendant vertices. If $v\in
V(G_0)$, we use $d_{G_0}(v)$ to denote the degree of the vertex $v$
in $G_0$. Given a graph $G\in\mathscr{G}_{a,b}$, we suppose,
throughout the rest of this paper, that {\bf x} is a pendant vertex of
$G$ and {\bf xy} is the unique edge incident with {\bf x}.

The proofs of the following Lemmas 4.1-4.6 are similar to Lemmas
3.3-3.8 in~\cite{zhu}, we omit them here.

{\bf Lemma 4.1}. Let $G\in\mathscr{G}_{a, b}$ and $v \in V(G_0)$. If
$d(v)\neq d_{G_0}(v)$, then $d(v)=a+b$.

{\bf Lemma 4.2}. Let $G\in\mathscr{G}_{a, b}$. Then (i)
$\delta(G_0)\geq 2$; (ii) $|E(G_0)|=|V(G_0)|+2$.

{\bf Lemma 4.3}. Let $G\in\mathscr{G}_{a, b}$. Then $S({\bf
x})=d({\bf y})=a+b\geq3$ and $a\geq2$.

{\bf Lemma 4.4}. Let $G\in\mathscr{G}_{a,b}$ and $R=u_0u_1\cdots
u_k$ a path or cycle of $G_0$ with length at least 3 such that
$d_{G_0}(u_0)=d_{G_0}(u_k)=3$ and $d_{G_0}(u_i)=2$ for $1\leq i \leq
{k-1}$. Then

(i) $d(u_1)=d(u_2)=\cdots=d(u_{k-1})\in \{2,a+b\}$;

(ii) If $d(u_1)=2$, then $l(R)=3$ and $d(u_1)=d(u_2)=2$.

{\bf Corollary 1}.  Let $G\in\mathscr{G}_{a, b}$ and
$R=u_0u_1u_2\cdots u_k$ a path of $G_0$ with length at least 3 such
that $d_{G_0}(u_0)=d_{G_0}(u_k)=3$ and $d_{G_0}(u_i)=2$ for $1\leq
i\leq k-1$.

(i) If $d(u_0)=d(u_k)$, then $d(u_1)=d(u_{k-1})$;

(ii) If $d(u_0)=d(u_1)=d(u_k)=a+b$, then $d(u_i)=a+b$ for $i\in
\{0,1,2,\cdots,k\}$.

{\bf Proof}. (i) If $k=1$, then $u_{k-1}=u_0$, $u_1=u_k$. So
$d(u_{k-1})=d(u_0)$, $d(u_1)=d(u_k)$. Since $d(u_0)=d(u_k)$,
$d(u_1)=d(u_{k-1})$. If $k=2$, then $u_{k-1}=u_1$, and
$d(u_1)=d(u_{k-1})$. If $k>2$, then, applying Lemma 4.4(i), we get
$d(u_1)=d(u_{k-1})$.

(ii) If $k=1$, then $d(u_0)=d(u_1)=a+b$. If $k=2$, then
$d(u_1)=a+b$. So $d(u_0)=d(u_1)=d(u_2)=a+b$. If $k>2$, then,
applying Lemma 4.4, we get $d(u_1)=d(u_2)=\cdots=d(u_{k-1})=a+b$. So
$d(u_0)=d(u_1)=\cdots=d(u_k)=a+b$. Hence, $d(u_i)=a+b$ for $i\in
\{0, 1, 2, \cdots, k\}$.

{\bf Lemma 4.5}. Let $G\in\mathscr{G}_{a, b}$ and $C=u_0u_1\cdots
u_ku_0$ a cycle of $G_0$ such that $d_{G_0}(u_0)\geq 3$,
$d_{G_0}(u_1)=2$. Then there exists an integer $i\in \{0, 1, 2,
\cdots, k\}$ such that $d(u_i)\neq a+b$.

{\bf Lemma 4.6}. Let $G\in\mathscr{G}_{a, b}$ and $C=u_0u_1\cdots
u_ku_0$ a cycle of $G_0$ such that $3\leq d_{G_0}(u_0)\leq 4$ and
$d_{G_0}(u_1)=d_{G_0}(u_2)=\cdots=d_{G_0}(u_k)=2$. Then there exists
no integer $j\in \{0, 1, 2, \cdots, k\}$ such that
$d(u_j)=d(u_{j+1})=a+b$, where $u_{k+1}=u_0$.

{\bf Corollary 2}. Let $G\in\mathscr{G}_{a, b}$ and $C=u_0u_1\cdots
u_ku_0$ a cycle of $G_0$ such that $d_{G_0}(u_0)=3$ and
$d_{G_0}(u_1)=d_{G_0}(u_2)=\cdots=d_{G_0}(u_k)=2$. Then $l(C)=3$,
$d(u_1)=d(u_2)=2$, $d(u_0)=a+b$, $a=2$.

{\bf Proof}. Applying Lemma 4.4(i) with $R=C$, we get
$d(u_1)=d(u_2)=\cdots=d(u_k)\in \{2, a+b\}$. By Lemma 4.6, we get
$d(u_1)=d(u_2)=\cdots=d(u_k)\neq a+b$,then
$d(u_1)=d(u_2)=\cdots=d(u_k)=2$. Applying Lemma 4.4(ii) with $R=C$,
we get $l(C)=3$, $d(u_1)=d(u_2)=2$. Applying (2) with $(\xi,
\eta)=(u_1,{\bf x})$, we get
\begin{equation}
a=\frac{S(u_1)-S({\bf x})}{d(u_1)-d({\bf
x})}=\frac{d(u_0)+d(u_2)-(a+b)}{2-1}=d(u_0)+2-(a+b)
\end{equation}
By Lemma 4.3, we get $d(u_0)\geq a+b$, then $d(u_0)\geq (a+b) \geq
3=d_{G_0}(u_0)$. Since $d(u_0)\in \{d_{G_0}(u_0), a+b)\}$,
$d(u_0)=a+b$. Together with (9), we get $a=2$.

{\bf Lemma 4.7}. Let $G\in\mathscr{G}_{a, b}$ and $R=u_0u_1\cdots
u_k$ a path or cycle of $G_0$ with length at least 3 such that
$d_{G_0}(u_0)=d_{G_0}(u_k)\geq 4$ and $d_{G_0}(u_i)=2$ for $1\leq i
\leq {k-1}$. If there exists an integer $i\in \{1,2,\cdots,k-2\}$
such that $d(u_i)\neq d(u_{i+1})$, then $d(u_0)=d(u_k)=d_{G_0}(u_0)$
is even, and $a+b=\frac{d_{G_0}(u_0)}{2}+1$ or
$\frac{1+\sqrt{1+4d_{G_0}(u_0)}}{2}$. Moreover,

(i) If $a+b=\frac{d_{G_0}(u_0)}{2}+1$, then $a=2$;

(ii) If $a+b=\frac{1+\sqrt{1+4d_{G_0}(u_0)}}{2}$, then
$a=\frac{1+\sqrt{1+4d_{G_0}(u_0)}}{2}$.

{\bf Proof}. Applying Lemma 4.1 with $v=u_i$ for
$i\in\{1,2,\cdots,k\}$, we get $d(u_i)\in \{d_{G_0}(u_i),
a+b\}=\{2,a+b\}$.

Since $d_{G_0}(u_0)=d_{G_0}(u_k), d(u_i)\neq d(u_{i+1})$, by the
symmetry, we may assume that $d(u_i)=2$, $d(u_{i+1})=a+b$. Applying
(2) with $(\xi, \eta)=(u_{i+1},{\bf x})$,
\begin{displaymath}
a=\frac{S(u_{i+1})-S({\bf x})}{d(u_{i+1})-d({\bf
x})}=\frac{d(u_i)+d(u_{i+2})+d(u_{i+1})-2-(a+b)}{a+b-1}
\end{displaymath}
\begin{equation}
 =\frac{2+d(u_{i+2})-2}{a+b-1}=\frac{d(u_{i+2})}{a+b-1}
\end{equation}
we get $d(u_{i+2})=a(a+b-1)\geq2(a+b-1)=a+b+(a+b-2)>a+b$. Together
with Lemma 4.1, $d(u_{i+2})=d_{G_0}(u_{i+2})$. And
$d_{G_0}(u_{i+2})>a+b>2=d_{G_0}(u_j)$ for $2\leq j\leq k-1$, then
$u_{i+2}=u_k$, $u_{i+1}=u_{k-1}$, $u_i=u_{k-2}$, $d(u_k)=a(a+b-1)$,
$d(u_{k-1})=a+b$, $d_{u_{k-2}}=2$. So, $d(u_k)>a+b$. By Lemma 4.1,
$d(u_k)=d_{G_0}(u_k)$. Since $d_{G_0}(u_0)=d_{G_0}(u_k)>a+b$,
$d(u_0)=d_{G_0}(u_0)$,so $d(u_k)=d(u_0)=d_{G_0}(u_0)$

Applying (2) with $(\xi, \eta)=(u_{k-2}, {\bf x})$,
\begin{displaymath}
a=\frac{S(u_{k-2})-S({\bf x})}{d(u_{k-2})-d({\bf
x})}=\frac{d(u_{k-1})+d(u_{k-3})-(a+b)}{2-1}=d(u_{k-3})
\end{displaymath}
then $d(u_{k-3})=a<a(a+b-1)=d(u_0)$.So $u_{k-3}\neq u_0$,
$d(u_{k-3})\in \{d_{G_0}(u_{k-3}),a+b\}=\{2,a+b\}$.

If $d(u_{k-3})=2$, then $a=2$. By (10), we get
$a+b=\frac{d_{G_0}(u_0)}{2}+1$.

If $d(u_{k-3})=a+b$, then $a=a+b$. By (10), we get
$a+b=\frac{1+\sqrt{1+4d_{G_0}(u_0)} }{2}$.

{\bf Corollary 3}. Let $G\in\mathscr{G}_{a,b}$ and $R=u_0u_1\cdots
u_k$ a path or cycle of $G_0$ with length at least 3 such that
$d_{G_0}(u_1)=d_{G_0}(u_k)=4$ and $d_{G_0}(u_i)=2$ for $2\leq i \leq
{k-1}$. If there exists an integer $i\in \{1,2,\cdots,k-2\}$ such
that $d(u_i)\neq d(u_{i+1})$, then $l(R)=5$, $a+b=3$, $a=2$,
$d(u_0)=d(u_5)=4$, $d(u_1)=d(u_4)=3$, $d(u_2)=d(u_3)=2$.

{\bf Proof}. From Lemma 4.7, we get $d(u_0)=d(u_k)=d_{G_0}(u_0)=4$,
$a=2$,$a+b=\frac{d_{G_0}(u_0)}{2}+1=3$, or
$a+b=\frac{1+\sqrt{1+4d_{G_0}(u_0)} }{2}=\frac{1+\sqrt{17}}{2}$.
$a+b=\frac{1+\sqrt{17}}{2}$ is not an integer. So $a+b=3$.

By the symmetry, we may assume that $d(u_{i+1})=a+b$, $d(u_i)=2$,
$i\in\{1,2,\cdots,{k-2}\}$. Applying (2) with $(\xi,
\eta)=(u_{i+1},{\bf x})$, we get
\begin{displaymath}
a=\frac{S(u_{i+1})-S({\bf x})}{d(u_{i+1})-d({\bf
x})}=\frac{d(u_{i+2})+d(u_{i})+d(u_{i+1})-2-(a+b)}{d(u_{i+1})-d({\bf
x})} =\frac{d(u_{i+2})}{a+b-1}
\end{displaymath}
we get $d(u_{i+2})=a(a+b-1)=4$, then $i+2=k$, and
$d(u_{k-1})=a+b=3$, $d(u_{k-2})=2$. By applying (2) with $(\xi,
\eta)=(u_{k-2},{\bf x}),(u_{k-3},{\bf x}),(u_{k-4},{\bf x})$, we get
$d(u_{k-3})=2$,$d(u_{k-4})=3$,$d(u_{k-5})=4$,respectively. Then
$u_{k-5}=u_0$, i.e., $k=5$. So $l(R)=5$.

Hence $l(R)=5$, $a+b=3$, $a=2$, $d(u_0)=d(u_5)=4$,
$d(u_1)=d(u_4)=3$, $d(u_2)=d(u_3)=2$.

\begin{figure}
\includegraphics{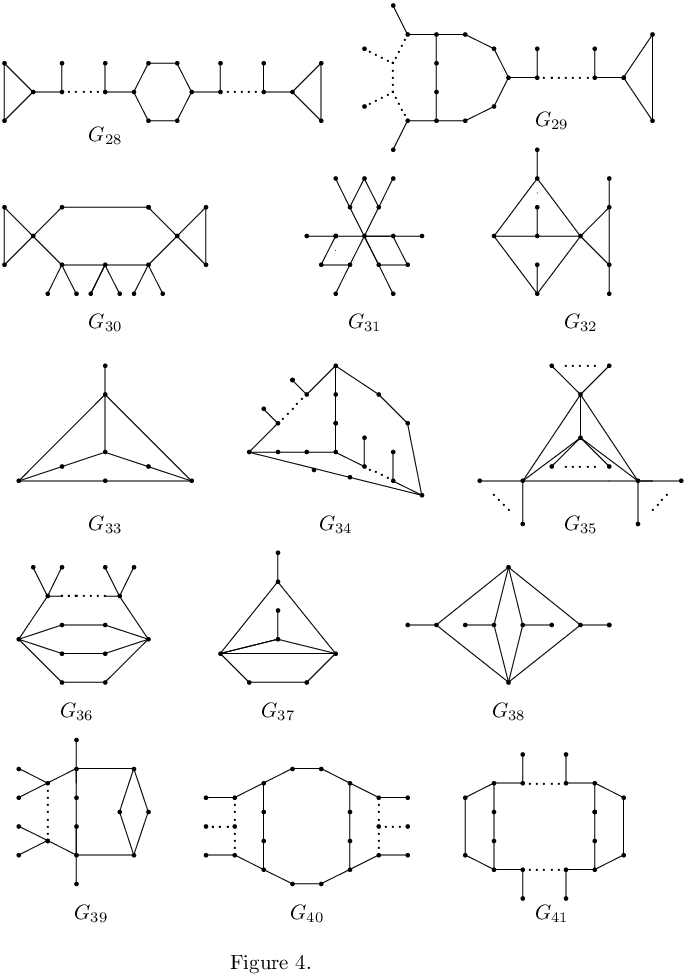}
\end{figure}

{\bf Theorem 4.1}. Let $G$ be a 2-walk tricyclic graph with minimum
degree $\delta=1$. Then $G$ is one of $G_{28}, G_{29},\cdots,
G_{41}$ (see Figure 4).

{\bf Proof}. Let $G$ be a 2-walk tricyclic graph with minimum degree
$\delta(G)=1$ and $G_0$ the graph obtained from $G$ by deleting all
pendent vertices. Then $G_0$ is a tricyclic graph with minimum
degree $\delta(G)=2$ from Lemma 4.2.

{\bf Case 1}. $G_0=T_1$, where $u_0=x_0=u$, $v_0=y_0=v$, $w_0=z_0=w$
and $x_j=y_k=z_l=o$. By applying Corollary 2 with $R=C_1, C_2$ and
$C_3$, respectively, we get $a=2$; $l(C_1)=3$, $l(C_2)=3$ and
$l(C_3)=3$; $d(u_1)=d(u_2)=2$, $d(v_1)=d(v_2)=2$ and
$d(w_1)=d(w_2)=2$; $d(u_0)=a+b$, $d(v_0)=a+b$, and $d(w_0)=a+b$.

By applying (2) with $(\xi, \eta)=(u_0, {\bf x})$,
\begin{displaymath}
a=\frac{S(u_0)-S({\bf x})}{d(u_0)-d({\bf
x})}=\frac{d(u_1)+d(u_2)+d(x_1)+d(u_0)-3-(a+b)}{a+b-1}=\frac{d(x_1)+1}{a+b-1}
\end{displaymath}
Together with Lemma 4.3, we get
\begin{equation}
d(x_1)=2(a+b-1)-1=a+b+(a+b-3)\geq a+b\geq3\geq d_{G_0}(x_1)
\end{equation}
By Lemma 4.1, we get $d(x_1)\in \{d_{G_0}(x_1),a+b\}$. Together with
(11), we get $d(x_1)=a+b$, then $a+b=3$.  By applying Corollary 1,
we get $d(x_{j-1})=a+b$. In the same way, $d(y_{k-1})=a+b$ and
$d(z_{l-1})=a+b$.By Lemma 4.1, we get $d(o)\in
\{a+b,d_{G_0}(o)\}=\{3\}$.  Applying (2) with $(\xi, \eta)=(o, {\bf x})$,
we get
\begin{displaymath}
a=\frac{S(o)-S({\bf x})}{d(o)-d({\bf
x})}=\frac{d(x_{j-1})+d(y_{k-1})+d(z_{l-1})+d(o)-3-(a+b)}{d(o)-1}=\frac{6}{2}=3,
\end{displaymath}
a contradiction.

{\bf Case 2}. $G_0=T_2, T_5, T_6$.

In $T_2$, let $u_0=x_0=u$, $v_0=x_j=y_k=v$ and $w_0=y_0=w$; In
$T_5$, let $u_0=x_0=u$ and $v_0=w_0=x_j=v$; In $T_6$, let
$u_0=x_0=u$, $x_j=y_0=z_0=p_0=v$ and $y_k=z_l=p_m=w$.

By applying Corollary 2 with $C=C_1$, we get $a=2$, $l(C_1)=3$,
$d(u_1)=d(u_2)=2$, $d(u_0)=a+b$. By applying (2) with $(\xi,
\eta)=(u,{\bf x})$, we get
\begin{displaymath}
a=\frac{S(u_0)-S({\bf x})}{d(u_0)-d({\bf
x})}=\frac{d(u_1)+d(u_2)+d(x_1)+d(u_0)-3-(a+b)}{a+b-1}=\frac{d(x_1)+1}{a+b-1}
\end{displaymath}
then $d(x_1)=a(a+b-1)-1=2(a+b-1)-1=a+b+(a+b-3)\geq a+b$.

{\bf Subcase 1}. $j=1$. Then $x_1=v$
\begin{equation}
d(v)=d(x_1)=2(a+b)-3,\ \ \ i.e.,\ \ \ a+b=\frac{d(v)+3}{2}
\end{equation}
By Lemma 4.1, we get $d(v)\in\{d_{G_0}(v), a+b\}$. If $d(v)=a+b$, by
(12), we get $a+b=3<d_{G_0}(v)$, a contradiction. So
$d(v)=d_{G_0}(v)$. Since $d_{G_0}(v)=4$ or $5$, by (12), we get
$a+b=\frac{7}{2}$ or $4$. $a+b=\frac{7}{2}$ is not an integer, so
$a+b=4$. Since $d(v)=5\neq a+b$, there exists no pendent vertex
adjacent to $v$. So we may assume $v$ is adjacent to
$x_1^{(0)},x_1^{(1)},x_1^{(2)},x_1^{(3)},x_1^{(4)}$ in $G_0$, where
$x_1^{(0)}=x_0$. By applying (2) with $(\xi, \eta)=(v, {\bf x})$
\begin{displaymath}
a=\frac{S(v)-S({\bf x})}{d(v)-d({\bf
x})}=\frac{d(x_0)+\sum_{i=1}^{i=4}d(x_1^{(i)})-(a+b)}{5-1}
=\frac{\sum_{i=1}^{i=4}d(x_1^{(i)})}{4}
\end{displaymath}
then $d(x_1^{(1)})=d(x_1^{(2)})=d(x_1^{(3)})=d(x_1^{(4)})=2$. Let
$\overline{x_1}$ be another neighbor of $x_1^{(1)}$ different from
$x_1$. By applying (2), we get
\begin{displaymath}
a=\frac{S(x_1^{(1)})-S({\bf x})}{d(x_1^{(1)})-d({\bf
x})}=\frac{d(x_1)+d(\overline{x_1})-(a+b)}{2-1}=d(\overline{x_1})+1
\end{displaymath}
then $d(\overline{x_1})=1$, implying that $x_1^{(1)}$ has degree 1
in $G_0$, which contradicts with Lemma 4.2.

{\bf Subcase 2}. $j>1$. $d(x_1)=2(a+b)-3\geq (a+b)\neq
3>d_{G_0}(x_1)$, then $d(x_1)=a+b$, and $a+b=3$. By applying (2)
with $(\xi, \eta)=(x_1, {\bf x})$,
\begin{displaymath}
a=\frac{S(x_1)-S({\bf x})}{d(x_1)-d({\bf
x})}=\frac{d(x_0)+d(x_2)+d(x_1)-2-(a+b)}{3-1}=\frac{d(x_2)+1}{2}
\end{displaymath}
then $d(x_2)=2a-1=3$. Similarly, applying (2) with $(\xi,
\eta)=(x_2,{\bf x}),(x_3,{\bf x}),\cdots,(x_{j-1},{\bf x})$, we get $d(x_3)=3$,
$d(x_4)=3$, $d(x_5)=3,\cdots, d(x_j)=3$. But $d_{G_0}(x_j)>3$ for
$G=T_2, T_5, T_6$ (see Figure 2), a contradiction.

{\bf Case 3}. $G_0=T_3$, where $u_0=x_0=u$, $x_j=y_0=z_0=w$,
$y_k=z_l=p_m=z$ and $p_0=v_0=v$.

By applying Corollary 2 with $C=C_1$ and $C_2$, we get $a=2$,
$l(C_1)=3$, $d(u_1)=d(u_2)=2$, $d(u_0)=a+b$ and $l(C_2)=3$,
$d(v_1)=d(v_2)=2$, $d(v_0)=a+b$.

By applying (2) with $(\xi, \eta)=(u, {\bf x})$,
\begin{displaymath}
a=\frac{S(u)-S({\bf x})}{d(u)-d({\bf
x})}=\frac{d(u_1)+d(u_2)+d(x_1)+d(u)-3-(a+b)}{a+b-1}=\frac{d(x_1)+1}{a+b-1}
\end{displaymath}
then
\begin{equation}
 d(x_1)=a(a+b-1)-1=2(a+b-1)-1=a+b+a+b-3\geq a+b\geq3\geq d_{G_0}(x_1)
\end{equation}
By Lemma 4.1, we get $d(x_1)\in{\{d_{G_0}(x_1),a+b\}}$. This
together with (13) implies that $d(x_1)=a+b$, and $a+b=3$, then
$d(x_1)=a+b=3$, $d(u_0)=3$ and $d(v_0)=3$. By Lemma 4.1, $\forall
 \alpha \in \{x_0,x_j\} $, $d(\alpha)\in \{d_{G_0}(\alpha),a+b\}={3}$.
So $d(x_0)=3=a+b$, $d(x_j)=3=a+b$. Since $d(x_1)=a+b$ , we get
$d(x_1)=d(x_2)=\cdots=d(x_{j-1})=d(x_j)=a+b$ by applying Corollary
1(ii) $R=P_1$. By applying (1) with $\xi=x_0, x_j$, we get
$S(x_0)=3a+b=S(x_j)$. And $S(x_0)=d(x_1)+d(u_1)+d(u_2)$,
$S(x_j)=d(x_{j-1})+d(y_1)+d(z_1)$, then
$d(x_{j-1})+d(y_1)+d(z_1)=d(x_1)+d(u_1)+d(u_2)$, $d(y_1)+d(z_1)=4$,
which implies $d(y_1)=d(z_1)=2$. By applying(2) with $(\xi,
\eta)=(y_1,{\bf x})$
\begin{displaymath}
a=\frac{S(y_1)-S({\bf x})}{d(y_1)-d({\bf
x})}=\frac{d(y_0)+d(y_2)-(a+b)}{2-1}=d(y_2)
\end{displaymath}
then $d(y_2)=2$. In the same way, $d(z_2)=2$.
 So $l(P_2)>2$, $l(P_3)>2$. By
Lemma 4.4(ii), we get $l(P_2)=3$, $d(y_1)=d(y_2)=2$ and $l(P_3)=3$,
$d(z_1)=d(z_2)=2$. By applying (1) with $\xi=u_0$ and $v_0$, we get
$S(v_0)=3a+b=S(u_0)$. And $S(v_0)=d(p_1)+d(v_1)+d(v_2)=4+d(p_1)$,
$S(u_0)=d(u_1)+d(u_2)+d(x_1)=7$, then $d(p_1)=3$. Together with
Corollary 1(ii), it implies that $d(p_i)=3$ for $1\leq i\leq m$.

By the symmetry, we may assume that $j\geq m\geq 1$. From
$\delta(G)=1$, we get $j\neq 2$. Hence, $G=G_{28}$. It is easy to
check that $G=G_{28}$ is 2-walk (2, 1)-linear.

{\bf Case 4}. $G_0=T_4$, where $u_0=x_0=u$, $v_0=y_0=z_0=v$ and
$y_k=z_l=x_j=w$.

By applying Corollary 2 with $C=C_1$, we get $a=2$, $l(C_1)=3$,
$d(u_1)=d(u_2)=2$ and $d(u_0)=a+b$.

If $d(v_1)=d(v_2)=\cdots=d(v_s)$, by Lemma 4.1, we get
$d(v_i)\in\{d_{G_0}(v_i),a+b\}=\{2,a+b\}$ for $1\leq i\leq s$.
Together with Lemma 4.6, we get $d(v_1)=d(v_2)=\cdots=d(v_s)\neq
a+b$, then $d(v_1)=d(v_2)=\cdots=d(v_s)=2$. By applying Lemma 2.3(i)
with $R=C_2$, we get $l(C_2)=3$, $d(v_1)=d(v_2)=2$. By applying (1)
with $\xi=u_1$ and $v_1$, we get $S(u_1)=2a+b=S(v_1)$. And
$S(u_1)=d(u_0)+d(u_2)=a+b+2$, $S(v_1)=d(v_0)+d(v_2)=d(v_0)+2$, then
$d(v_0)=a+b$. Since $d_{G_0}(v_0)=4$, $a+b\geq 4$.

By applying (2) with $(\xi, \eta)=(u_0, {\bf x})$,
\begin{displaymath}
a=\frac{S(u_0)-S({\bf x})}{d(u_0)-d({\bf
x})}=\frac{d(u_1)+d(u_2)+d(x_1)+d(u_0)-3-(a+b)}{a+b-1}=\frac{d(x_1)+1}{a+b-1}
\end{displaymath}
then $d(x_1)=a(a+b-1)-1=2(a+b-1)-1=a+b+(a+b-3)>max
\{a+b,d_{G_0}(x_1)\}$, it is impossible. So there exists $i\in
\{1,2,\dots,s-1\}$ such that $d(v_i)\neq d(v_{i+1})$. By applying
Corollary 3 with $R=C_2$, we get $a+b=3$, $a=2$, $l(C_2)=5$,
$d(v_1)=d(v_4)=a+b=3$, $d(v_0)=d_{G_0}(v_0)=4$. By applying (2) with
$(\xi, \eta)=(v_0, {\bf x})$,
\begin{displaymath}
a=\frac{S(v_0)-S({\bf x})}{d(v_0)-d({\bf
x})}=\frac{d(v_1)+d(v_2)+d(y_1)+d(z_1)-(a+b)}{4-1}=\frac{3+d(y_1)+d(z_1)}{3}
\end{displaymath}
then $d(y_1)+d(z_1)=3$, it is impossible.

{\bf Case 5}. $G_0=T_7$, where $u_0=x_0=u$, $x_j=y_0=z_0=v$,
$y_k=p_0=q_0=w$ and $z_l=p_m=q_n=z$.

By applying Corollary 2, we get $a=2$, $l(C_1)=3$,
$d(u_1)=d(u_2)=2$, $d(u_0)=a+b$. By applying (2) with $(\xi,
\eta)=(u_0,{\bf x})$,
\begin{displaymath}
a=\frac{S(u_0)-S({\bf x})}{d(u_0)-d({\bf
x})}=\frac{d(u_1)+d(u_2)+d(x_1)+d(u_0)-3-(a+b)}{a+b-1}=\frac{d(x_1)+1}{a+b-1}
\end{displaymath}
then
\begin{equation}
d(x_1)=a(a+b-1)-1=2(a+b-1)-1\geq a+b\geq 3\geq d_{G_0}(x_1)
\end{equation}
By Lemma 4.1, we get $d(x_1)\in\{d_{G_0}(x_1),a+b\}$, this together
with (14) implies that $d(x_1)=a+b,a+b=3$, and $d(x_1)=a+b=3$. By
Lemma 4.1 again, we get $\forall \alpha\in \{y_0,p_0,p_m\}$,
$d(\alpha)\in \{d_{G_0}(\alpha),a+b\}=\{3\}$. So
$d(y_0)=d(p_0)=d(p_m)=3$.

By Corollary 2 with $R=P_1$, we get $\forall i\in \{1,2,\cdots,j\}$,
$d(x_i)=a+b=3$. By applying (1) with $\xi=u_0,y_0$, we get
$S(u_0)=3a+b=S(y_0)$. And $S(u_0)=d(u_1)+d(u_2)+d(x_1)=7$,
$S(y_0)=d(x_{j-1})+d(y_1)+d(z_1)=3+d(y_1)+d(z_1)$, then
$d(y_1)=d(z_1)=2$. By applying Lemma 2.3(i) with $R=C_1, Q=P_2$, we
get $l(P_2)\neq 2$, then $l(P_2)>3$. By applying Lemma 4.4(ii) with
$R=P_2$, we get $l(P_2)=3$, $d(y_1)=d(y_2)=2$. In the same way,
$l(P_3)=3$, $d(z_1)=d(z_2)=2$. By applying (2) with $(\xi,
\eta)=(p_0,{\bf x})$,
\begin{displaymath}
a=\frac{S(p_0)-S({\bf x})}{d(p_0)-d({\bf
x})}=\frac{d(p_1)+d(q_1)+d(y_2)-(a+b)}{3-1}=\frac{d(p_1)+d(q_1)-1}{2}
\end{displaymath}
then $d(p_1)+d(q_1)=5$. By the symmetry, we may assume that
$d(p_1)\geq d(q_1)$. So $d(p_1)=3$, $d(q_1)=2$. Since $d(p_1)=3$, by
applying Corollary 1(ii) with $R=P_4$, we get
$d(p_1)=d(p_2)=\cdots=d(p_m)=3$. Since $d(q_1)=2$, by applying Lemma
2.3(i) with $R=C_1,Q=P_5$, we get $l(P_5)\neq2$, then $l(P_5)\geq
3$. By applying Lemma 4.4(ii) with $R=P_5$, we get $l(P_5)=3$,
$d(q_1)=d(q_2)=2$. Since $\delta(G)=1$, $j\geq 1,$ or $m\geq 1$.
Hence $G=G_{29}$, and it is easy to check that $G_{29}$ is 2-walk
(2, 1)-linear.

{\bf Case 6}. $G_0=T_8$, where $u_0=x_0=y_0=u$ and $x_j=y_k=v_0=v$.

If there exists $i\in\{1, 2, \cdots, r-1\}$ such that $d(u_i)\neq
d(u_{i+1})$. By applying Corollary 3 with $C=C_1$, we get
$d(u_0)=d_{G_0}(u_0)=4$, $a+b=3$, $a=2$, $l(C_1)=5$ and
$d(u_1)=d(u_4)=3$.

By applying (2) with $(\xi, \eta)=(u_0, {\bf x})$,
\begin{displaymath}
a=\frac{S(u_0)-S({\bf x})}{d(u_0)-d({\bf
x})}=\frac{d(u_1)+d(u_4)+d(x_1)+d(y_1)-(a+b)}{4-1}=\frac{3+d(x_1)+d(y_1)}{3}
\end{displaymath}
then $d(x_1)+d(y_1)=3$, it is impossible. So
$d(u_1)=d(u_2)=\cdots=d(u_r)$. In the same way,
$d(v_1)=d(v_2)=\cdots=d(v_s)$.

By applying Lemma 4.1, we get $\forall i\in\{1, 2, \cdots, r\}$,
$d(u_i)\in \{d_{G_0}(u_i), a+b\}=\{2, a+b\}$, then
$d(u_1)=d(u_2)=\cdots=d(u_r)=2$ or
$d(u_1)=d(u_2)=\cdots=d(u_r)=a+b$. By applying Lemma 4.6, we get
$d(u_1)=d(u_2)=\cdots=d(u_r)\neq a+b$, then
$d(u_1)=d(u_2)=\cdots=d(u_r)=2$. By applying 2.3(i) with $R=C_1$, we
get $l(C_1)=3$. By applying (2) with $(\xi, \eta)=(u_1, {\bf x})$,
\begin{displaymath}
a=\frac{S(u_1)-S({\bf x})}{d(u_1)-d({\bf
x})}=\frac{d(u_0)+d(u_2)-(a+b)}{2-1}=d(u_0)+2-(a+b)
\end{displaymath}
By Lemma 4.3, we get $d(u_0)\geq (a+b)$. In the same way, we get
$l(C_2)=3$, $d(v_1)=d(v_2)=2$ and $d(v_0)\geq a+b$.By applying (1)
with $\xi=u_1$ and $v_1$, we get $S(u_1)=2a+b=S(v_1)$. Since
$S(u_1)=d(u_2)+d(u_0)=2+d(u_0),S(v_1)=d(v_2)+d(v_1)=2+d(v_0)$,
$d(u_0)=d(v_0)$.

If $d(u_0)>a+b$, then $d(u_0)=d_{G_0}(u_0)=4$, $a+b=3$. So $a=3$. By
applying (2) with $(\xi, \eta)=(u_0, {\bf x})$
\begin{displaymath}
a=\frac{S(u_0)-S({\bf x})}{d(u_0)-d({\bf
x})}=\frac{d(u_1)+d(u_2)+d(x_1)+d(y_1)+d(u_0)-4-(a+b)}{4-1}=\frac{d(x_1)+d(y_1)+1}{3}
\end{displaymath}
then $d(x_1)+d(y_1)=8$, which implies that $d(x_1)=4,d(y_1)=4$,then
$x_1=v,y_1=v$,and $G$ is not simple,a contradiction.

If $d(u_0)=a+b$, then $a=2$, $a+b\geq4$. By applying (2) with $(\xi,
\eta)=(u_0, {\bf x})$
\begin{displaymath}
a=\frac{S(u_0)-S({\bf x})}{d(u_0)-d({\bf
x})}=\frac{d(u_1)+d(u_2)+d(x_1)+d(y_1)+d(u_0)-4-(a+b)}{a+b-1}=\frac{d(x_1)+d(y_1)}{a+b-1}
\end{displaymath}
then
\begin{equation}
d(x_1)+d(y_1)=2(a+b-1)
\end{equation}

By the symmetry, we may assume that $d(y_1)\geq d(x_1)$. By Lemma
4.1, we get $d(x_1)\in \{d_{G_0}(x_1),a+b\}$. Since
$d_{G_0}(x_1)\leq 3$ and $a+b\geq 4$, $d(x_1)\leq a+b$. In the same
way, $d(y_1)\leq a+b$. It together with (15) implies that
$d(x_1)=d(y_1)=a+b-1$ or $d(x_1)=a+b-2$, $d(y_1)=a+b$. If
$d(x_1)=d(y_1)=a+b-1$, then $d(x_1)=d_{G_0}(x_1)$ and
$d(y_1)=d_{G_0}(y_1)$. So $d_{G_0}(x_1)=d_{G_0}(y_1)$. Since $G$ is
simple, $d_{G_0}(x_1)=d_{G_0}(y_1)=2$. Then $d(x_1)=d(y_1)=2$. So
$a+b=3$, it contradicts with $a+b\geq 4$. Hence, $d(x_1)=a+b-2$,
$d(y_1)=a+b$, then $x_1\neq x_j$. So $d(x_1)=2$, $a+b=4$ and
$d(y_1)=a+b=4$. By applying (2) with $(\xi, \eta)=(x_1, {\bf x}),(x_2,
{\bf x})$, we get $d(x_2)=2,d(x_3)=4$, respectively. If $x_j\neq x_3$,
i.e., $j>3$. By applying (2) with $(x_3, {\bf x})$
\begin{displaymath}
a=\frac{S(x_3)-S({\bf x})}{d(x_3)-d({\bf
x})}=\frac{d(x_2)+d(x_4)-(a+b)}{d(x_3)-d({\bf
x})}=\frac{2+d(x_4)-(a+b)}{4-1}
\end{displaymath}
then $d(x_4)=8$, it is impossible. So $x_3=x_j$, i.e., $j=3$. Since
$\delta({G})=1$, $l(P_4)>1$.
\begin{displaymath}
a=\frac{S(y_1)-S({\bf x})}{d(y_1)-d({\bf
x})}=\frac{d(u_0)+d(y_1)+d(y_2)-2-(a+b)}{d(y_1)-1}=\frac{2+d(y_2)}{3}
\end{displaymath}
then $d(y_2)=4$. Similarly, applying (2) with $(\xi,
\eta)=(y_2,{\bf x}),(y_3,{\bf x}),\cdots,(y_{k-1},{\bf x})$, we get
$d(y_3)=d(y_4)=\cdots=d(y_k)=4$. So $G=G_{30}$. It is easy to check
that $G_{30}$ is 2-walk (2, 2)-linear.

{\bf Case 7}. $G_0=T_9$, where $u_0=v_0=w_0=u$.

{\bf Subcase 7.1}. $d(u_1)=d(u_2)=\cdots=d(u_r)$,
$d(v_1)=d(v_2)=\cdots=d(v_s)$ and $d(w_1)=d(w_2)=\cdots=d(w_t)$.

By Lemma 4.1, we get $\forall i\in\{1,2,\cdots,r\},d(u_i) \in
\{d_{G_0}(u_i),a+b\}=\{2,a+b\}$. If
$d(u_1)=d(u_2)=\cdots=d(u_r)=a+b$, by applying Lemma 4.5 with
$C=C_1$, we get $d(u_0)\neq a+b$, then $d(u_0)=d_{G_0}(u_0)=6$ and
$a+b\neq 6$. By applying (2) with $(\xi, \eta)=(u_1, {\bf x})$
\begin{displaymath}
a=\frac{S(u_1)-S({\bf x})}{d(u_1)-d({\bf
x})}=\frac{d(u_2)+d(u_0)+d(u_1)-2-(a+b)}{2-1}
=\frac{a+b+4}{a+b-1}=1+\frac{5}{a+b-1}
\end{displaymath}
Since $a+b$ is integer, and $a+b\geq 3, a+b\neq 6$,
$\frac{5}{a+b-1}$ is not an integer, we get $a$ is not an integer, a
contradiction. If $d(u_1)=d(u_2)=\dots=d(u_r)=2$, by applying Lemma
2.3 with $R=C_1$, we get $l(C_1)=3$, $d(u_1)=2$ and $d(u_2)=2$. In
the same way, we get $l(C_2)=3$, $d(v_1)=d(v_2)=2$ and $l(C_3)=3$,
$d(w_1)=d(w_2)=2$. Then $u_1, u_2, v_1, v_2, w_1, w_2$ are not
adjacent to any pendent vertex, $u_0$ is adjacent to pendent
vertices. So $d(u_0)=a+b$ and $a+b>6$. By applying (2) with $(\xi,
\eta)=(u_0,u_1)$, we get
\begin{displaymath}
a=\frac{S(u_0)-S(u_1)}{d(u_0)-d(u_1)}
\end{displaymath}
\begin{displaymath}
=\frac{d(u_1)+d(u_2)+d(v_1)+d(v_2)+d(w_1)+d(w_2)+d(u_0)-6-(a+b+2)}{a+b-2}
\end{displaymath}
\begin{displaymath}
=\frac{4}{a+b-2}<1
\end{displaymath}
it is contradicts with Lemma 4.3.

{\bf Subcase 7.2}. There exist $i_1\in\{1, 2, \cdots, r-1\}$ such
that $d(u_{i_1})\neq d(u_{{i_1}+1})$; or $i_2\in\{1, 2, \cdots,
s-1\}$ such that $d(v_{i_2})\neq d(v_{{i_2}+1})$; or $i_3\in\{1, 2,
\cdots, t-1\}$ such that $d(w_{i_3})\neq d(w_{{i_3}+1})$. By the
symmetry, we may assume that there exists $i_1\in\{1, 2, \cdots,
r-1\}$ such that $d(u_{i_1})\neq d(u_{{i_1}+1})$. By applying Lemma
4.7 with $C=C_1$,we get

(i) $d(u_0)=d_{G_0}(u_0)=6$, $a=2$; and
$a+b=\frac{d_{G_0}(u_0)}{2}+1=4$.

(ii) $d(u_0)=d_{G_0}(u_0)=6$, $a+b=\frac{1+\sqrt{1+4d(u_0)}}{2}=3$,
$a=3$.

 If $d(u_0)=d_{G_0}(u_0)=6$, $a=2$,
 $a+b=\frac{d_{G_0}(u_0)}{2}+1=4$;
by the symmetry, we may assume that $d(u_{i_1})=2$,
$d(u_{{i_1}+1})=a+b$. Then by applying (2) with $(\xi,
\eta)=(u_{{i_1}+1}, {\bf x})$,
\begin{displaymath}
a=\frac{S(u_{i_1+1})-S({\bf x})}{d(u_{i_1+1})-d({\bf x})}
=\frac{d(u_{i_1})+d(u_{i_1+2})+d(u_{i_1+1})-2-(a+b)}{a+b-1}
\end{displaymath}
\begin{displaymath}
=\frac{d(u_{i_1+2})}{a+b-1}=\frac{d(u_{i_1+2})}{3}
\end{displaymath}
and $d(u_{i_1+2})=6$, it implies that $u_{i_1+2}=u_0$, then
$u_{i_1+1}=u_r$ and $d(u_r)=a+b=4,d(u_{r-1})=2$. By applying (2)
with $(\xi, \eta)=(u_{r-1}, {\bf x}),(u_{r-2}, {\bf x}),(u_{r-3}, {\bf x})$, we get
$d(u_{r-2})=2$, $d(u_{r-3})=4$, $d(u_{r-4})=6$, respectively. It
implies that $u_{r-4}=u_0$, i.e., $r=4$, then $d(u_1)=d(u_{r-3})=4$.
By applying (2) with $(\xi, \eta)=(u_0, {\bf x})$,
\begin{displaymath}
a=\frac{S(u_0)-S({\bf x})}{d(u_0)-d({\bf
x})}=\frac{d(u_1)+d(u_{r})+d(v_1)+d(v_{s})+d(w_1)+d(w_{t})-(a+b)}{6-1}
\end{displaymath}
\begin{displaymath}
=\frac{d(v_1)+d(v_s)+d(w_1)+d(w_t)+4}{5}
\end{displaymath}
then $d(v_1)+d(v_s)+d(w_1)+d(w_t)=6$, it is impossible.

 If $d(u_0)=d_{G_0}(u_0)=6$, $a+b=3$, $a=3$; by the symmetry, we may
assume that $d(u_{i_1})=2$, $d(u_{i_1+1})=a+b$. By applying (2) with
$(\xi, \eta)=(u_{i_1+1}, {\bf x})$
\begin{displaymath}
a=\frac{S(u_{i_1+1})-S({\bf x})}{d(u_{i_1+1})-d({\bf x})}
=\frac{d(u_i)+d(u_{i_1+2})+d(u_{i_1+1})-2-(a+b)}{a+b-1}=\frac{d(u_{i_1+2})}{a+b-1}
\end{displaymath}
then $d(u_{i_1+2})=6$, it implies that $u_{i_1+2}=u_0$, i.e.,
${i_1}+1=r$. Then $d(u_r)=a+b=3$, $d(u_{r-1})=2$. By applying (2)
with $(\xi, \eta)=(u_{r-1}, {\bf x}),(u_{r-2}, {\bf x})$, we get $d(u_{r-2})=3$,
$d(u_{r-3})=6$. It implies that $u_{r-3}=u_0$, i.e., $r=3$. Then
$l(C_1)=4$, $d(u_1)=d(u_{r-2})=3$, $d(u_2)=d(u_{r-1})=2$ and
$d(u_3)=d(u_r)=3$. By applying (2) with $(\xi, \eta)=(u_0, {\bf x})$,
\begin{displaymath}
a=\frac{S(u_0)-S({\bf x})}{d(u_0)-d({\bf
x})}=\frac{d(u_1)+d(u_r)+d(v_1)+d(v_s)+d(w_1)+d(w_t)-(a+b)}{6-1}
\end{displaymath}
\begin{displaymath}
=\frac{3+d(v_1)+d(v_s)+d(w_1)+d(w_t)}{5}
\end{displaymath}
then
\begin{equation}
d(v_1)+d(v_s)+d(w_1)+d(w_t)=12
\end{equation}
Since $\forall \alpha\in \{v_1,v_s,w_1,w_t\}$, $d(\alpha)\in
\{d_{G_0}(\alpha),a+b\}=\{2,3\}$, $d(\alpha)\leq 3$, it together
with (16) implies that $d(v_1)=3$, $d(v_s)=3$, $d(w_1)=3$,
$d(w_t)=3$. By applying (2) with $(\xi, \eta)=(v_1, {\bf x}),(v_2,
{\bf x}),(v_3, {\bf x})$, we get $d(v_2)=2$, $d(v_3)=3$, $d(v_4)=6$. It implies
that $v_4=v_0$, i.e., $l(C_2)=4$. In the same way, $l(C_3)=4$,
$d(w_1)=d(w_3)=3$, $d(w_2)=2$, which implies that $G=G_{31}$. It is
easy to check that $G_{31}$ is 2-walk (3, 0)-linear.

{\bf Case 8}. $G_0=T_{10}$, where $u_0=x_0=y_0=z_0=u$,
$x_j=y_k=z_l=v$.

If there exist $i\in\{1,2,\cdots,r-1\}$ such that $d(u_i)\neq
d(u_{i+1})$, then it is contradicts with Lemma 4.7 since
$d_{G_0}(u_0)$ is odd. So $d(u_1)=d(u_2)=\cdots=d(u_r)$. By Lemma
4.1, $d(u_i)\in \{d_{G_0}(u_i),a+b\}=\{2,a+b\}$ for $1\leq i\leq r$.
Hence, (i) $d(u_1)=d(u_2)=\cdots=d(u_r)=2$; (ii)
$d(u_1)=d(u_2)=\cdots=d(u_r)=a+b$.

{\bf Subcase 8.1}. $d(u_1)=d(u_2)=\cdots=d(u_r)=2$, by applying
Lemma 2.3 with $R=C_1$, we get $l(C_1)=3$, $d(u_1)=d(u_2)=2$. By
applying (2) with $(\xi, \eta)=(u_1, {\bf x})$
\begin{equation}
a=\frac{S(u_1)-S({\bf x})}{d(u_1)-d({\bf
x})}=\frac{d(u_0)+d(u_2)-(a+b)}{2-1}=2+d(u_0)-(a+b)
\end{equation}
it together with Lemma 4.3 implies that $d(u_0)\geq a+b$. Since
$a+b\geq 3$, $a\geq 2$, $a+b\in N^{+}$, $a\in N^{+}$, we discuss
three cases again.

(I) $d(u_0)>a+b$.

Then $d(u_0)=d_{G_0}(u_0)=5$ by Lemma 4.1, and $a+b<5$.

 If $a+b=3$, then $a=4$ from (17). By applying (2) with
$(\xi, \eta)=(u_0, {\bf x})$,
\begin{displaymath}
a=\frac{S(u_0)-S({\bf x})}{d(u_0)-d({\bf
x})}=\frac{d(u_1)+d(u_2)+d(x_1)+d(y_1)+d(z_1)-(a+b)}{5-1}
\end{displaymath}
\begin{displaymath}
=\frac{d(x_1)+d(y_1)+d(z_1)+1}{4}
\end{displaymath}
then $d(x_1)+d(y_1)+d(z_1)=15$. So there exists $\alpha \in
\{x_1,y_1,z_1\}$, such that $d(\alpha)>4$, it is impossible.

 If $a+b=4$, then $a=3$ from (17). By applying (2) with $(\xi, \eta)=(u_0, {\bf x})$

\begin{displaymath}
a=\frac{S(u_0)-S({\bf x})}{d(u_0)-d({\bf
x})}=\frac{d(u_1)+d(u_2)+d(x_1)+d(y_1)+d(z_1)-(a+b)}{5-1}
\end{displaymath}
\begin{displaymath}
=\frac{d(x_1)+d(y_1)+d(z_1)}{4}
\end{displaymath}
then
\begin{displaymath}
d(x_1)+d(y_1)+d(z_1)=12
\end{displaymath}
By Lemma 4.1, we get $\forall \alpha \in \{x_1,y_1,z_1\}$,
$d(\alpha)\in \{d_{G_0}(\alpha),a+b\}\subseteq\{2, 3, a+b\}$,
$d(\alpha)\leq a+b=4$, it together with $d(x_1)+d(y_1)+d(z_1)=12$
implies that $d(x_1)=d(y_1)=d(z_1)=4$. $G$ is simple, by the
symmetry, we may assume that $l(P_1)>1$. By applying (2) with $(\xi,
\eta)=(x_1,{\bf x})$,
\begin{displaymath}
a=\frac{S(x_1)-S({\bf x})}{d(x_1)-d({\bf
x})}=\frac{d(x_0)+d(x_2)+d(x_1)-2-(a+b)}{4-1}=\frac{3+d(x_2)}{3}
\end{displaymath}
then $d(x_2)=6$, it is impossible.

(II) $d(u_0)=a+b$. Then $a+b=d(u_0)\geq d_{G_0}(u_0)=5$, and $a=2$
from (17). By applying (2) with $(\xi, \eta)=(u_0, {\bf x})$,
\begin{displaymath}
a=\frac{S(u_0)-S({\bf x})}{d(u_0)-d({\bf
x})}=\frac{d(u_1)+d(u_2)+d(x_1)+d(y_1)+d(z_1)+d(u_0)-5-(a+b)}{a+b-1}
\end{displaymath}
\begin{displaymath}
=\frac{d(x_1)+d(y_1)+d(z_1)-1}{a+b-1}
\end{displaymath}
then $d(x_1)+d(y_1)+d(z_1)=2(a+b-1)+1=2(a+b)-1$. By the symmetry, we
may assume that $d(x_1)\geq d(y_1)\geq d(z_1)$, then $d(y_1)<a+b$,
$d(z_1)<a+b$, Otherwise, $d(y_1)\geq a+b$, we get $d(z_1)<-1$, it is
impossible. By Lemma 4.1, we get $d(y_1)=d_{G_0}(y_1)$,
$d(z_1)=d_{G_0}(z_1)$, then $d(y_1)=2$ or $3$, $d(z_1)=2$ or $3$. If
$d(z_1)=3$, then $d(y_1)=3$, $k=1$, $l=1$, it contradicts with $G$
is simple. So $d(z_1)=2$. Since $d(x_1)+d(y_1)+d(z_1)=2(a+b)-1$,
$d(y_1)+d(x_1)=2(a+b)-1-d(z_1)=a+b+(a+b-3)\geq a+b+2\geq7
>d_{G_0}(y_1)+d_{G_0}(x_1)$, it together with $d(y_1)=d_{G_0}(y_1)$ implies that
$d(x_1)=a+b$, then $d(y_1)=a+b-3$. By applying (2) with $(\xi,
\eta)=(x_1, {\bf x})$
\begin{displaymath}
a=\frac{S(x_1)-S({\bf x})}{d(x_1)-d({\bf
x})}=\frac{d(x_0)+d(x_2)+d(x_1)-2-(a+b)}{a+b-1}=\frac{d(x_2)+a+b-2}{a+b-1}
\end{displaymath}
then $d(x_2)=a+b$. Similarly, applying(2) with $(\xi,
\eta)=(x_2,{\bf x}),(x_3,{\bf x}),\cdots,(x_{j-1},{\bf x})$, we get
$d(x_3)=\cdots=d(x_j)=a+b$, then $d(y_k)=d(v)=d(x_j)=a+b$. By Lemma
4.1, we get $d(y_1)\in \{a+b,d_{G_0}(y_1)\}\subseteq \{2,a+b\}$.
Since $d(y_1)=a+b-3<a+b$, $y_1\neq y_k$, it together with Lemma 4.1
implies that $d(y_1)=2$ and $a+b=5$. By applying (2) with $(\xi,
\eta)=(v,{\bf x})$,
\begin{displaymath}
a=\frac{S(v)-S({\bf x})}{d(v)-d({\bf
x})}=\frac{d(x_{j-1})+d(y_{k-1})+d(z_{l-1})+d(v)-3-(a+b)}{5-1}
\end{displaymath}
\begin{displaymath}
=\frac{5+d(y_{k-1})+d(z_{l-1})+5-3-5}{4}=\frac{d(y_{k-1})+d(z_{l-1})+2}{4}
\end{displaymath}
then $d(y_{k-1})+d(z_{l-1})=6$, $d(y_{k-1})<a+b$, $d(z_{l-1})<a+b$,
it together with Lemma 4.1 implies that $d(y_{k-1})=2$,
$d(z_{l-1})=2$, which contradicts with $d(y_{k-1})+d(z_{l-1})=6$.

{\bf Subcase 8.2}. $d(u_1)=d(u_2)=\cdots=d(u_r)=a+b$.

By applying Lemma 4.5 with $C=C_1$, we get $d(u_0)\neq a+b$, which
together with Lemma 4.1 implies that $d(u_0)=d_{G_0}(u_0)=5$ and
$a+b\neq 5$. Now, we prove $l(C_1)=3$. By way of contradiction, we
may assume $l(C_1)>3$. By applying (1) with $\xi=u_1,u_{r-1}$, we
get $S(u_1)=a(a+b)=S(u_{r-1})$. Since
$S(u_1)=d(u_0)+d(u_2)+d(u_1)-2=2(a+b)+3$ and
$S(u_{r-1})=d(u_r)+d(u_{r-2})+d(u_{r-1})-2=3(a+b)-2$, $a+b=5$, a
contradiction. Hence, $l(C_1)=3$, $d(u_1)=d(u_2)=a+b$.

By applying (2) with $(\xi, \eta)=(u_1, {\bf x})$
\begin{displaymath}
a=\frac{S(u_1)-S({\bf x})}{d(u_1)-d({\bf
x})}=\frac{d(u_0)+d(u_2)+d(u_1)-2-(a+b)}{a+b-1}
\end{displaymath}
\begin{displaymath}
=\frac{5+2(a+b)-2-(a+b)}{a+b-1} =1+\frac{4}{a+b-1}
\end{displaymath}
which together with Lemma 4.3 and $a, a+b\in N^+$(where $N^+$ is the
set of positive integers), $a+b\neq 5$, implies that $a+b=3$, $a=3$.
By applying (2) with $(\xi, \eta)=(u_0, {\bf x})$,
\begin{displaymath}
a=\frac{S(u_0)-S({\bf x})}{d(u_0)-d({\bf
x})}=\frac{d(u_1)+d(u_2)+d(x_1)+d(y_1)+d(z_1)-(a+b)}{5-1}
\end{displaymath}
\begin{displaymath}
=\frac{a+b+a+b+d(x_1)+d(y_1)+d(z_1)-(a+b)}{4}=\frac{3+d(x_1)+d(y_1)+d(z_1)}{4}
\end{displaymath}
then $d(x_1)+d(y_1)+d(z_1)=9$. So $d(x_1)=3,d(y_1)=3,d(z_1)=3$. By
applying (2) with $(\xi, \eta)=(x_1, {\bf x})$
\begin{displaymath}
a=\frac{S(x_1)-S({\bf x})}{d(x_1)-d({\bf
x})}=\frac{d(x_0)+d(x_2)+d(x_1)-2-(a+b)}{3-1}=\frac{d(x_2)+3}{2}
\end{displaymath}
then $d(x_2)=3$. If $x_2\neq x_j$, i.e., $j>2$, by applying (2) with
$(\xi, \eta)=(x_2, {\bf x})$
\begin{displaymath}
a=\frac{S(x_1)-S({\bf x})}{d(x_1)-d({\bf
x})}=\frac{d(x_1)+d(x_3)+d(x_2)-2-(a+b)}{3-1}=\frac{1+d(x_3)}{2}
\end{displaymath}
then $d(x_3)=5$, it is impossible. So $j=2$. In the same way, we get
$k=2$, $l=2$, $d(y_1)=3$, $d(z_1)=3$, which implies $G=G_{32}$. It
is easy to check that $G_{32}$ is 2-walk (3, 0)-linear.

{\bf Case 9}. $G_0=T_{11}$, where $u_0=x_0=y_0=u$, $x_j=z_0=p_0=v$
and $y_k=z_l=p_m=w$.

{\bf Subcase 9.1}. There exists $i\in \{1,2,\cdots,r-1\}$ such that
$d(u_i)\neq d(u_{i+1})$. By applying Corollary 3 with $C=C_1$, we
get $d(u_0)=d_{G_0}(u_0)=4$, $a+b=\frac{d_{G_0}(u_0)}{2}+1=3$,
$a=2$, $l(C_1)=5$, $d(u_1)=d(u_4)=a+b=3$, $d(u_2)=d(u_3)=2$. By
applying (2) with $(\xi, \eta)=(u_0, {\bf x})$,
\begin{displaymath}
a=\frac{S(u_0)-S({\bf x})}{d(u_0)-d({\bf
x})}=\frac{d(u_1)+d(u_4)+d(x_1)+d(y_1)-(a+b)}{4-1}
\end{displaymath}
\begin{displaymath}
=\frac{3+3+d(x_1)+d(y_1)-3}{3}=\frac{d(x_1)+d(y_1)+3}{3}
\end{displaymath}
then $d(x_1)+d(y_1)=3$, it is impossible. So,
$d(u_1)=d(u_2)=\cdots=d(u_r)$. By Lemma 4.1, $d(u_i)\in \{2,a+b\}$.
Hence,(i) $d(u_1)=d(u_2)=\cdots=d(u_r)=a+b$; (ii)
$d(u_1)=d(u_2)=\cdots=d(u_r)=2$.

{\bf Subcase 9.2}.  $d(u_1)=d(u_2)=\cdots=d(u_r)=a+b$, by applying
Lemma 4.5 with $C=C_1$, we get $d(u_0)\neq a+b$, it together with
Lemma 4.1 implies that $d(u_0)=d_{G_0}(u_0)=4$, and $a+b\neq 4$. By
applying (2) with $(\xi, \eta)=(u_1, {\bf x})$,
\begin{displaymath}
a=\frac{S(u_1)-S({\bf x})}{d(u_1)-d({\bf
x})}=\frac{d(u_0)+d(u_2)+d(u_1)-2-(a+b)}{d(u_1)-1}
\end{displaymath}
\begin{displaymath}
=\frac{4+a+b+a+b-2-(a+b)}{a+b-1}
=\frac{a+b+2}{a+b-1}=1+\frac{3}{a+b-1}
\end{displaymath}
Since $a+b\geq 3$, $a+b\neq 4$ and $a+b \in N^{+}$,
$\frac{3}{a+b-1}$ is not an integer. So $a$ is not an integer, a
contradiction.

{\bf Subcase 9.3}. $d(u_1)=d(u_2)=\cdots=d(u_{r-1})=2$, by applying
Lemma 2.3 with $R=C_1$, we get $l(C_1)=3,d(u_1)=d(u_2)=2$. By
applying (2) with $(\xi, \eta)=(u_1, {\bf x})$,
\begin{equation}
a=\frac{S(u_1)-S({\bf x})}{d(u_1)-d({\bf
x})}=\frac{d(u_0)+d(u_2)-(a+b)}{2-1}=2+d(u_0)-(a+b),
\end{equation}
it together with Lemma 4.3 implies that $d(u_0)\geq a+b$.

(I) $d(u_0)>a+b$. Then $d(u_0)=d_{G_0}(u_0)$ from Lemma 4.1. Since
$d_{G_0}(u_0)=d_{T_{11}}(u_0)=4$ (see Figure2), implying $a+b=3$
from Lemma 4.3, by (18), we get $a=3$.

By applying (2) with $(\xi, \eta)=(u_0, {\bf x})$,
\begin{displaymath}
a=\frac{S(u_0)-S({\bf x})}{d(u_0)-d({\bf
x})}=\frac{d(u_1)+d(u_2)+d(x_1)+d(y_1)-(a+b)}{4-1}
\end{displaymath}
\begin{displaymath}
=\frac{2+2+d(x_1)+d(y_1)-3}{3}
=\frac{d(x_1)+d(y_1)+1}{3}
\end{displaymath}
then
\begin{displaymath}
 d(x_1)+d(y_1)=8
\end{displaymath}
By the symmetry, we may assume that $d(x_1)\geq d(y_1)$, it together
with $d(x_1)+d(y_1)=8$ implies that $d(y_1)\geq 4$, it is
impossible.

(II) $d(u_0)=a+b$. Then $a+b=d(u_0)\geq d_{G_0}(u_0)=5$, and $a=2$
from (18).

By applying (2) with $(\xi, \eta)=(u_0, {\bf x})$,
\begin{displaymath}
a=\frac{S(u_0)-S({\bf x})}{d(u_0)-d({\bf
x})}=\frac{d(u_1)+d(u_2)+d(x_1)+d(y_1)+d(u_0)-4-(a+b)}{d(u_0)-1}
\end{displaymath}
\begin{displaymath}
=\frac{2+2+d(x_1)+d(y_1)+a+b-4-(a+b)}{a+b-1}=\frac{d(x_1)+d(y_1)}{a+b-1}
\end{displaymath}
then $d(x_1)+d(y_1)=2(a+b-1)$, by the symmetry, we may assume that
$d(x_1)\geq d(y_1)$.

(i) $d(x_1)=d(y_1)$. Then $d(x_1)=d(y_1)=a+b-1$, it together with
Lemma 4.1, implies that $d(x_1)=d_{G_0}(x_1),d(y_1)=d_{G_0}(y_1)$,
then $d_{G_0}(x_1)=a+b-1\geq 3$, $d_{G_0}(y_1)=a+b-1\geq 3$, which
implies that $x_1=v$, $y_1=w$, then
$d(v)=d_{G_0}(v)=3,d(w)=d_{G_0}(w)=3$ and $a+b=4$. By applying (2)
with $(\xi, \eta)=(z_0, {\bf x})$,
\begin{displaymath}
a=\frac{S(z_0)-S({\bf x})}{d(z_0)-d({\bf
x})}=\frac{d(u_0)+d(z_1)+d(p_1)-(a+b)}{3-1}
\end{displaymath}
\begin{displaymath}
=\frac{a+b+d(z_1)+d(p_1)-(a+b)}{2}=\frac{d(z_1)+d(p_1)}{2}
\end{displaymath}
then $d(z_1)+d(p_1)=4$. So $d(z_1)=d(p_1)=2$. By applying (2) with
$(\xi, \eta)=(z_1,{\bf x})$
\begin{displaymath}
a=\frac{S(z_1)-S({\bf x})}{d(z_1)-d({\bf
x})}=\frac{d(z_0)+d(z_2)-(a+b)}{2-1}=d(z_2)-1
\end{displaymath}
then $d(z_2)=3$, which implies that $z_2=z_l$, i.e., $l=2$. In the
same way, we get $m=2$. we get $G=G_9$, but $\delta(G_9)=2$, it
contradicts with $\delta(G)=1$.

 (ii) $d(x_1)>d(y_1)$. By Lemma 4.1, we get $d(x_1)\in
\{d_{G_0}(x_1),a+b\}$, then $d(x_1)=a+b$ (Otherwise,
$d(x_1)=d_{G_0}(x_1)\leq 3$, then $d(y_1)\leq 2$ and
$d(x_1)+d(y_1)\leq 3+2=5<2(a+b-1)$, which contradicts with
$d(x_1)+d(y_1)=2(a+b-1)$). We can see
 $d(y_1)=d_{G_0}(y_1)\in \{2,3\}$, so we discuss the following two cases.

(a)  $d(y_1)=2$. Then $a+b=4$. Since $d(x_1)+d(y_1)=2(a+b-1)$ and
$d(x_1)=a+b$. So $d(x_1)=a+b=4$. By applying (2) with $(\xi,
\eta)=(y_1,{\bf x})$
\begin{displaymath}
a=\frac{S(y_1)-S({\bf x})}{d(y_1)-d({\bf
x})}=\frac{d(y_0)+d(y_2)-(a+b)}{2-1}=a+b+d(y_2)-(a+b)=d(y_2)
\end{displaymath}
then $d(y_2)=2$. By applying (2) with $(\xi, \eta)=(y_2, {\bf x})$
\begin{displaymath}
a=\frac{S(y_2)-S({\bf x})}{d(y_2)-d({\bf
x})}=\frac{d(y_1)+d(y_3)-(a+b)}{2-1}=2+d(y_3)-4=d(y_3)-2
\end{displaymath}
then $d(y_3)=4$.

If $y_3\neq y_k$, i.e., $k>3$, by applying (2) with $(\xi,
\eta)=(y_3,{\bf x})$
\begin{displaymath}
a=\frac{S(y_3)-S({\bf x})}{d(y_3)-d({\bf
x})}=\frac{d(y_2)+d(y_4)+d(y_3)-2-(a+b)}{4-1}
\end{displaymath}
\begin{displaymath}
=\frac{2+d(y_4)+4-2-4}{3}=\frac{d(y_4)}{3}
\end{displaymath}
then $d(y_4)=6$, it is impossible.

If $d(y_3)=d(y_k)$, i.e., $k=3$, by applying(2) with $(\xi,
\eta)=(y_3,{\bf x})$,
\begin{displaymath}
a=\frac{S(y_3)-S({\bf x})}{d(y_3)-d({\bf
x})}=\frac{d(y_2)+d(p_{m-1})+d(z_{l-1})+d(y_3)-3-(a+b)}{4-1}
\end{displaymath}
\begin{displaymath}
=\frac{2+d(p_{m-1})+d(z_{l-1})+4-3-4}{3}=\frac{d(p_{m-1})+d(z_{l_1})-1}{3}
\end{displaymath}
then $d(p_{m-1})+d(z_{l-1})=7$. By the symmetry, we may assume that
$d(p_{m-1})\geq d(z_{l-1})$. By Lemma 4.1, we get $d(p_{m-1})\in
\{d_{G_0}(p_{m-1}), a+b\}\subseteq \{2,3,4\}$, then $d(p_{m-1})\leq
4$, it together with $d(p_{m-1})+d(z_{l-1})=7$ and $d(p_{m-1})\geq
d(z_{l-1})$, implies that $d(p_{m-1})=4$, $d(z_{l-1})=3$. So $l=1$
and
\begin{equation}
d(z_0)=3
\end{equation}
By applying (2) with $(\xi, \eta)=(p_{m-1}, {\bf x})$,
\begin{displaymath}
a=\frac{S(p_{m-1})-S({\bf x})}{d(p_{m-1})-d({\bf
x})}=\frac{d(p_{m-2})+d(y_k)+d(p_{m-1})-2-(a+b)}{4-1}
\end{displaymath}
\begin{displaymath}
=\frac{d(p_{m-2})+4+4-2-4}{3}=\frac{d(p_{m-2})+2}{3}
\end{displaymath}
then $d(p_{m-2})=4$. Similarly, by applying(2) with $(\xi,
\eta)=(p_{m-2}, {\bf x}),(p_{m-3}, {\bf x}),\dots$, $(p_2, {\bf x}),(p_1, {\bf x})$, we get
$d(p_{m-3})=\cdots=d(p_1)=d(p_0)=4$, it together with (19) implies
that $d(p_0)\neq d(z_0)$, it contradicts with $p_0=z_0$.

(b) $d(y_1)=3$. Then $y_1=w$. So $d(w)=3$. Since
$d(x_1)+d(y_1)=2(a+b-1)$ and $d(x_1)=a+b$, $a+b=5$, and $d(x_1)=5$.
By applying (2) with $(\xi, \eta)=(w, {\bf x})$,
\begin{displaymath}
a=\frac{S(w)-S({\bf x})}{d(w)-d({\bf
x})}=\frac{d(y_0)+d(p_{m-1})+d(z_{l-1})-(a+b)}{3-1}
=\frac{d(p_{m-1})+d(z_{l-1})}{2}
\end{displaymath}
then $d(p_{m-1})+d(z_{l-1})=4$, so $d(p_{m-1})=d(z_{l-1})=2$. By
applying (2) with $(\xi, \eta)=(p_{m-1}, {\bf x})$,
\begin{displaymath}
a=\frac{S(p_{m-1})-S({\bf x})}{d(p_{m-1})-d({\bf
x})}=\frac{d(w)+d(p_{m-2})-(a+b)}{2-1}=d(p_{m-2})-2
\end{displaymath}
then $d(p_{m-2})=4$, but by Lemma 4.1, we get $d((p_{m-2}))\in
\{d_{G_0}(p_{m-2}),a+b\}\subseteq \{2,3,5\}$, a contradiction.

{\bf Case 10}. $G_0=T_{12}$, where $x_0=y_0=z_0=p_0=u$,
$x_j=y_k=q_n=v$ and $p_m=z_l=q_0=w$.

{\bf Subcase 10.1}. $l(P_i)\leq 2$ for $i\in \{1,2,3,4\}$.

By the symmetry, we may assume that $l(P_1)\geq l(P_2)$ and
$l(P_3)\geq l(P_4)$. We discuss three cases.

(I)  $l(P_1)=2$, $l(P_2)=1$, $l(P_3)=2$ and $l(P_4)=1$.

 By Lemma 4.1, we get $d(x_1)\in\{d_{G_0}(x_1),a+b\}=\{2,a+b\}$,
$d(z_1)\in\{d_{G_0}(z_1),a+b\}=\{2,a+b\}$. If $d(x_1)\neq d(z_1)$,
without loss of generality we may assume that $d(x_1)=a+b$,
$d(z_1)=2$. By applying (2) with $(\xi, \eta)=(x_1, z_1)$,
\begin{displaymath}
a=\frac{S(x_1)-S(z_1)}{d(x_1)-d(z_1)}=\frac{d(x_0)+d(x_2)+d(x_1)-2-d(z_0)-d(z_2)-d(z_1)+2}{d(x_1)-d(z_1)}
\end{displaymath}
\begin{displaymath}
=\frac{d(v)+a+b-d(w)-2}{a+b-2} =1+\frac{d(v)-d(w)}{a+b-2}\leq
1+\frac{a+b-3}{a+b-2}<2
\end{displaymath}
it contradicts with Lemma 4.3. So $d(x_1)=d(z_1)$, then
$S(x_1)=S(z_1)$ and $d(x_2)=d(z_2)$, i.e., $d(v)=d(w)$. By (1) with
$\xi=x_0$, we get $S(x_0)=d(x_1)+d(z_1)+d(v)+d(w)+d(x_0)-4$ and
$S(x_1)=d(x_0)+d(v)+d(x_1)-2$, then $S(x_0)-S(x_1)=d(z_1)+d(w)-2>0$,
so
\begin{equation}
d(x_0)>d(x_1)
\end{equation}
Note that $d(x_0)\geq d_{G_0}(x_0)=4$, We discuss the following two
cases.

 If $d(x_0)=a+b>4$, it together with Lemma 4.1 and (20) implies that $d(x_1)=2$, then
$d(z_1)=d(x_1)=2$. By applying (2) with $(\xi, \eta)=(x_0,{\bf x})$,
\begin{displaymath}
a=\frac{d(x_1)+d(y_1)+d(z_1)+d(p_1)+d(x_0)-4-(a+b)}{a+b-1}
\end{displaymath}
\begin{displaymath}
=\frac{2+d(w)+2+d(w)+a+b-4-(a+b)}{a+b-1}=\frac{2d(w)}{a+b-1}
\end{displaymath}
then $d(w)=\frac{a(a+b-1)}{2}\geq a+b-1>3=d_{G_0}(w)$, so
$d(w)=a+b$, then $a=\frac{2(a+b)}{a+b-1}=2+\frac{2}{a+b-1}$. Since
$a+b\in N^{+}$, and $a+b>4$, $\frac{2}{a+b-1}$ is not an integer, a
contradiction.

If $d(x_0)=4$, by Lemma 4.1, we get $d(x_1)\in
\{d_{G_0}(x_1),a+b\}=\{2,a+b\}$.

(i) $d(x_1)=2$. By applying (2) with $(\xi, \eta)=(x_0, x_1)$,
\begin{displaymath}
a=\frac{S(x_0)-S(x_1)}{d(x_0)-d(x_1)}=\frac{d(x_1)+d(y_1)+d(p_1)+d(z_1)-d(x_0)-d(x_2)}{4-2}
\end{displaymath}
\begin{displaymath}
=\frac{2+2+2d(w)-4-d(w)}{2}=\frac{d(w)}{2}
\end{displaymath}
then $d(w)=2a$, from Lemma 4.1, $d(w)=3$ or $a+b$ implies that
$d(w)=a+b$. We get $a+b=2a$. By applying (2) with $(\xi, \eta)=(w,
{\bf x})$
\begin{displaymath}
a=\frac{S(w)-S({\bf x})}{d(w)-d({\bf
x})}=\frac{d(z_1)+d(p_0)+d(q_1)+d(w)-3-(a+b)}{a+b-1}
\end{displaymath}
\begin{displaymath}
=\frac{2+4+d(q_1)+a+b-3-(a+b)}{a+b-1}=\frac{3+d(q_1)}{2a-1}
\end{displaymath}
then $d(q_1)=a(2a-1)-3=2a^2-a-3$. If $a=2$, then $d(q_1)=3$,
$a+b=2a=4$. By applying (2) with $(\xi, \eta)=(x_1, {\bf x})$,
$$a=\frac{S(x_1)-S({\bf x})}{d(x_1)-d({\bf x})}=\frac{d(x_0)+d(v)-(a+b)}{2-1}=d(v),$$,
then $d(v)=2<3=d_{G_0}(v)$, a contradiction. If $a\geq3$,
$d(q_1)=2a^2-a-3\geq12>d_{G_0}(q_1)$, then $d(q_1)=a+b$, so
$a+b=2a^2-a-3$, i.e., $2a=2a^2-a-3$, we get $a$ is not an integer, a
contradiction.

(ii) $d(x_1)=a+b$. Then $d(z_1)=a+b$. By (20), Lemma 4.3 and $a+b$
is an integer, implies that $d(x_0)=4$, $a+b=3$, then $d(x_1)=3$. By
Lemma 4.1, we get $d(q_0)\in \{d_{G_0}(q_0),a+b\}=\{3\}$. By
applying (1) with $\xi=x_1$ and $q_0$, we get $S(x_1)=3a+b=S(q_0)$,
$S(x_1)=d(x_0)+d(v)+d(x_1)-2=4+3+1=8$,
$S(q_0)=d(z_1)+d(u)+d(q_1)=3+4+d(q_1)\geq7+d_{G_0}(q_1)>8$, a
contradiction.

(II) $l(P_1)=2,l(P_2)=2,l(P_3)=2,l(P_4)=1$ or
$l(P_1)=2,l(P_2)=1,l(P_3)=2,l(P_4)=2$. By the symmetry, we only need
to discuss $l(P_1)=2,l(P_2)=2,l(P_3)=2,l(P_4)=1$.

First, we prove $d(x_1)=d(y_1)=d(z_1)$. If $d(y_1)\neq d(z_1)$, by
Lemma 4.1, we get $d(y_1)\in \{d_{G_0}(y_1),a+b\}=\{2,a+b\}$,
$d(z_1)\in \{d_{G_0}(z_1),a+b\}=\{2,a+b\}$. We may assume without
loss of generality, $d(y_1)=a+b,d(z_1)=2$, by applying (2) with
$(\xi, \eta)=(y_1, z_1)$,
\begin{displaymath}
a=\frac{S(y_1)-S(z_1)}{d(y_1)-d(z_1)}=\frac{d(y_0)+d(y_2)+d(y_1)-2-(d(z_0)+d(z_2))}{d(y_1)-2}
\end{displaymath}
\begin{displaymath}
=\frac{d(u)+d(w)+a+b-2-(d(u)+d(w))}{a+b-2}=\frac{a+b-2}{a+b-2}=1<2
\end{displaymath}
which contradicts with Lemma 4.3, then $d(y_1)=d(z_1)$. If
$d(x_1)\neq d(y_1)$, by applying Lemma 4.1, we know $d(x_1)\in
\{d_{G_0}(x_1),a+b\}=\{2,a+b\}$,
$d(y_1)\in\{d_{G_0}(y_1),a+b\}=\{2,a+b\}$, so we may assume without
loss of generality, $d(x_1)=a+b$, $d(y_1)=2$. By applying (2) with
$(\xi, \eta)=(x_1, y_1)$
\begin{displaymath}
a=\frac{S(x_1)-S(y_1)}{d(x_1)-d(y_1)}=\frac{d(x_0)+d(x_2)+d(x_1)-2-(d(y_0)+d(y_2))}{a+b-2}
\end{displaymath}
\begin{displaymath}
=\frac{d(u)+d(v)+a+b-2-(d(u)+d(w))}{a+b-2}=1+\frac{d(v)-d(w)}{a+b-2}\leq
1+\frac{a+b-3}{a+b-2}<2
\end{displaymath}
which contradicts with Lemma 4.3, so $d(x_1)=d(y_1)$. Hence,
$d(x_1)=d(y_1)=d(z_1)$.

Next, we prove $d(x_0)>d(x_1)$. By applying (1) with $\xi=x_1$ and
$z_1$, we get $S(x_1)=ad(x_1)+b=ad(z_1)+b=S(z_1)$, it together with
$S(x_1)=d(x_0)+d(x_2)+d(x_1)-2=d(u)+d(v)+d(x_1)-2$ and
$S(z_1)=d(z_0)+d(z_2)+d(z_1)-2=d(u)+d(w)+d(x_1)-2$, implies that
$d(v)=d(w)$. Since
$S(x_0)-S(x_1)=d(x_1)+d(y_1)+d(z_1)+d(p_1)+d(x_0)-4-(d(x_0)+d(x_2)+d(x_1)-2)
=3d(x_1)+d(v)+d(u)-4-(d(u)+d(v)+d(x_1)-2)=2d(x_1)-2>0$, then
$d(x_0)>d(x_1)$. By Lemma 4.1, we get
$d(x_1)\in\{d_{G_0}(x_1),a+b\}=\{2,a+b\}$.

 (i) $d(x_1)=2$. Then $d(y_1)=d(z_1)=d(x_1)=2$. By applying (2) with
$(\xi, \eta)=(x_0, {\bf x})$
\begin{displaymath}
a=\frac{S(x_0)-S({\bf x})}{d(x_0)-d({\bf
x})}=\frac{d(x_1)+d(y_1)+d(z_1)+d(p_1)+d(x_0)-4-(a+b)}{d(x_0)-1}
\end{displaymath}
\begin{equation}
=\frac{2+2+2+d(w)+d(u)-4-(a+b)}{d(u)-1}=1+\frac{3+d(v)-(a+b)}{d(u)-1}
\end{equation}
By applying Lemma 4.1, we get $d(v)\in \{d_{G_0}(v),
a+b\}=\{3,a+b\}$ and $d(u)\in \{d_{G_0}(u),a+b\}=\{4,a+b\}$, we
discuss the following cases.

(a) $d(u)=d_{G_0}(u)=4$, $d(v)=3$.

By applying (21) with $(d(u),d(v))=(4,3)$, we get
\begin{displaymath}
a=1+\frac{3+d(v)-(a+b)}{d(u)-1}=1+\frac{6-(a+b)}{4-1}=3-\frac{a+b}{3},
\end{displaymath}
it together with Lemma 4.3 and $a, a+b \in N^+$, implies that
$a+b=3$, $a=2$. By (1), we get $S(v)=3a+b=S(w)$, since
$S(v)=d(x_1)+d(y_1)+d(q_{n-1})=2+2+d(q_{n-1})=4+d(q_{n-1})$ and
$S(w)=d(z_1)+d(u)+d(q_1)=2+4+d(q_1)=6+d(q_1)$,
$d(q_{n-1})=d(q_1)-2$, which contradicts with Corollary 1.

(b) $d(u)=d_{G_0}(u)=4$, $d(v)=a+b$.

By applying (21) with $(d(u),d(v))=(4,a+b)$, we get
\begin{displaymath}
a=1+\frac{3+d(v)-(a+b)}{d(u)-1}=1+\frac{3+a+b-(a+b)}{3}=2.
\end{displaymath}
By applying (2) with $(\xi, \eta)=(x_1, {\bf x})$,
\begin{displaymath}
a=\frac{S(x_1)-S({\bf x})}{d(x_1)-d({\bf
x})}=\frac{d(x_0)+d(v)-(a+b)}{2-1}=4+a+b-(a+b)=4
\end{displaymath}
which contradicts with $a=2$.

(c) $d(u)=a+b$, $d(v)=3$.

By applying (21) with $(d(u), d(v))=(a+b,3)$, we get
\begin{displaymath}
a=1+\frac{3+d(v)-(a+b)}{d(u)-1}=1+\frac{6-(a+b)}{a+b-1}
=\frac{5}{a+b-1},
\end{displaymath}
it together with Lemma 4.3 and $a+b \in N^+$, implies that $a$ is
not an integer, a contradiction.

(d) $d(u)=a+b$, $d(v)=a+b$.

Since $d(u)\geq d_{G_0}(u_0)=4$, $a+b\geq 4$. If $a+b>4$,by applying
(21) with $(d(u),d(v))=(a+b,a+b)$, we get
\begin{displaymath}
a=1+\frac{3+d(v)-(a+b)}{d(u)-1}=1+\frac{3}{a+b-1}<2,
\end{displaymath}
a contradiction. If $a+b=4$, see (ii) previously.

 (ii) $d(x_1)=a+b$. We get $d(x_0)>d(x_1)=a+b$, this together with
Lemma 4.1 and Lemma 4.3 implies that $d(x_0)=4$, $a+b=3$. By Lemma
4.1, we get $d(v)=\{d_{G_0}(v),a+b\}=\{3\}$,
$d(w)=\{d_{G_0}(w),a+b\}=\{3\}$, so $d(v)=d(w)=3$. By applying (2)
with $(\xi, \eta)=(x_1, {\bf x})$,
\begin{displaymath}
a=\frac{S(x_1)-S({\bf x})}{d(x_1)-d({\bf
x})}=\frac{d(u)+d(v)+d(x_1)-2-(a+b)}{a+b-1}
\end{displaymath}
\begin{displaymath}
=\frac{4+3+a+b-2-(a+b)}{3-1}=\frac{5}{2}
\end{displaymath}
a contradiction.

(III) Let $l(P_1)=l(P_2)=l(P_3)=l(P_4)=2$.

Now, we prove $d(x_1)=d(y_1)=d(z_1)=d(p_1)$. If $d(x_1)\neq d(y_1)$,
by the symmetry, we may assume that $d(x_1)=a+b$, $d(y_1)=2$, by
applying (2) with $(\xi, \eta)=(x_1, y_1)$,
\begin{displaymath}
a=\frac{S(x_1)-S(y_1)}{d(x_1)-d(y_1)}=\frac{d(x_0)+d(x_2)+d(x_1)-2-(d(y_0)+d(y_2))}{a+b-2}
\end{displaymath}
\begin{displaymath}
=\frac{d(u)+d(v)+a+b-2-(d(u)+d(v))}{a+b-2}=1
\end{displaymath}
which contradicts with Lemma 4.3, so $d(x_1)=d(y_1)$. In the same
way, $d(z_1)=d(p_1)$.

By Lemma 4.1, we get
$d(x_1)\in\{d_{G_0}(x_1),a+b\}=\{2,a+b\},d(z_1)\in\{d_{G_0}(z_1),a+b\}=\{2,a+b\}$.
If $d(x_1)\neq d(z_1)$, by the symmetry, we may assume that
$d(x_1)=a+b$, $d(z_1)=2$, by applying (2) with $(\xi,
\eta)=(x_1,z_1)$,
\begin{displaymath}
a=\frac{S(x_1)-S(z_1)}{d(x_1)-d(z_1)}=\frac{d(x_0)+d(x_2)+d(x_1)-2-(d(z_0)+d(z_2))}{d(x_1)-d(z_1)}
\end{displaymath}
\begin{displaymath}
=\frac{d(u)+d(v)+a+b-2-(d(u)+d(w))}{a+b-2}=1+\frac{d(v)-d(w)}{a+b-2}\leq
1+\frac{a+b-3}{a+b-2}<2
\end{displaymath}
which contradicts with Lemma 4.3, then $d(x_1)=d(z_1)$.

Hence, $d(x_1)=d(y_1)=d(z_1)=d(p_1)$. By Lemma 4.3, we get
$d(x_1)\in \{d_{G_0}(x_1),a+b\}=\{2,a+b\}$. If $d(x_1)=2$, then
$d(y_1)=d(z_1)=d(p_1)=d(x_1)=2$, by applying (2) with $(\xi,
\eta)=(x_0,{\bf x})$
\begin{displaymath}
a=\frac{S(x_0)-S({\bf x})}{d(x_0)-d({\bf
x})}=\frac{d(x_1)+d(y_1)+d(z_1)+d(p_1)+d(x_0)-4-(a+b)}{d(x_0)-1}
\end{displaymath}
\begin{displaymath}
=\frac{2+2+2+2+d(x_0)-4-(a+b)}{d(x_0)-1}=1+\frac{5-(a+b)}{d(x_0)-1}\leq 1+\frac{5-3}{4-1}=1+\frac{2}{3}<2
\end{displaymath}
which contradicts with Lemma 4.3. If $d(x_1)=a+b$, then
$d(y_1)=d(z_1)=d(p_1)=d(x_1)=a+b$. Since
$S(x_0)=d(x_1)+d(y_1)+d(z_1)+d(p_1)+d(x_0)-4=4(a+b)+d(x_0)-4,S(x_1)=d(x_0)+d(x_2)+d(x_1)-2=a+b+d(v)+d(x_0)-2\leq
a+b+a+b+d(x_0)-2$, then $S(x_0)-S(x_1)\geq2(a+b)-2>0$, so
$d(x_0)>d(x_1)=a+b$, which together with Lemma 4.1 and Lemma 4.3
implies that $d(x_0)=d_{G_0}(x_0)=4$, and $a+b=3$. By applying (2)
with $(\xi, \eta)=(x_1, {\bf x})$
\begin{displaymath}
a=\frac{S(x_1)-S({\bf x})}{d(x_1)-d({\bf
x})}=\frac{d(x_0)+d(x_2)+d(x_1)-2-(a+b)}{a+b-1}=\frac{4+3+3-2-3}{3-1}=\frac{5}{2}
\end{displaymath}
which contradicts with $a \in N^+$.

{\bf Subcase 10.2}. There exists $i\in \{1,2,3,4\}$ such that
$l(P_i)>2$.

First, we prove the following Claim:

{\bf Claim 1}.

(i) If $l(P_1)\geq 2$, then $d(x_1)=d(x_2)=\cdots=d(x_{j-1})$;

(ii) If $l(P_2)\geq 2$, then $d(y_1)=d(y_2)=\cdots=d(y_{k-1})$;

(iii) If $l(P_3)\geq 2$, then $d(z_1)=d(z_2)=\cdots=d(z_{l-1})$;

(iv) If $l(P_4)\geq 2$, then $d(p_1)=d(p_2)=\cdots=d(p_{m-1})$.

{\bf Proof}. (i)
 By way of contradiction, if there
exists $i\in \{1,2,\cdots,j-2\}$ such that $d(x_{i})\neq
d(x_{i+1})$. By Lemma 4.1, we get $d(x_i)\in
\{d_{G_0}(x_i),a+b\}=\{2,a+b\},d(x_{i+1})\in
\{d_{G_0}(x_{i_1}),a+b\}=\{2,a+b\}$. If $d(x_i)=2$, then
$d(x_{i+1})=a+b$, by applying (2) with $(\xi, \eta)=(x_{i+1}, {\bf x})$
\begin{displaymath}
a=\frac{S(x_{i+1})-S({\bf x})}{d(x_{i+1})-d({\bf
x})}=\frac{d(x_i)+d(x_{i+2})+d(x_{i+1})-2-(a+b)}{a+b-1}
\end{displaymath}
\begin{displaymath}
=\frac{2+d(x_{i+2})+a+b-2-(a+b)}{a+b-1}=\frac{d(x_{i+2})}{a+b-1}
\end{displaymath}
then $d(x_{i+2})=a(a+b-1)\geq 2(a+b-1)=a+b+a+b-2>a+b\geq
d_{G_0}(x_{i+2})$, it contradicts with Lemma 4.1. If $d(x_i)=a+b$,
then $d(x_{i+1})=2$, by applying (2) with $(\xi, \eta)=(x_i, {\bf x})$
\begin{displaymath}
a=\frac{S(x_i)-S({\bf x})}{d(x_i)-d({\bf
x})}=\frac{d(x_{i-1})+d(x_{i+1})+d(x_i)-2-(a+b)}{a+b-1}
\end{displaymath}
\begin{displaymath}
=\frac{d(x_{i-1})+2+a+b-2-(a+b)}{a+b-1}=\frac{d(x_{i-1})}{a+b-1}
\end{displaymath}
then $d(x_{i-1})=a(a+b-1)\geq 2(a+b-1)=a+b+a+b-2>a+b$, it together
with Lemma 4.1 implies that $d(x_{i-1})=d_{G_0}(x_{i-1})$ and
$x_{i-1}=x_0$, i.e., $i=1$, then $d(x_0)=d_{G_0}(x_0)=4$, $a+b=3$,
$a=2$ and $d(x_1)=d(x_i)=a+b=3$. By applying (2) with $(\xi,
\eta)=(x_0,{\bf x})$,
\begin{displaymath}
a=\frac{S(x_0)-S({\bf x})}{d(x_0)-d({\bf
x})}=\frac{d(x_1)+d(y_1)+d(z_1)+d(p_1)+d(x_0)-4-(a+b)}{d(x_0)-1}
\end{displaymath}
\begin{displaymath}
=\frac{3+d(y_1)+d(z_1)+d(p_1)+4-4-3}{4-1}=\frac{d(y_1)+d(z_1)+d(p_1)}{3}
\end{displaymath}
then $d(y_1)+d(z_1)+d(p_1)=6$, so $d(y_1)=2$, $d(z_1)=2$,
$d(p_1)=2$. By applying (2) with $(\xi, \eta)=(y_1, {\bf x})$,
\begin{displaymath}
a=\frac{S(y_1)-S({\bf x})}{d(y_1)-d({\bf
x})}=\frac{d(y_0)+d(y_2)-(a+b)}{2-1}=4+d(y_2)-3=d(y_2)+1
\end{displaymath}
then $d(y_2)=1$, it is impossible. So
$d(x_1)=d(x_2)=\cdots=d(x_{j-1})$. In the same way, we get
(ii),(iii),(iv). \hfill $\Box$

Next, we prove Claim 2:

 {\bf Claim 2}. $d(u)=d(v)=a+b$, and $a+b\geq 4$.
 By (1) with $v=x_1,x_{j-1}$, we get $S(x_1)=ad(x_1)+b=ad(x_{j-1})+b=S(x_{j-1})$.
Since $S(x_1)=d(x_0)+d(x_2)$ and
$S(x_{j-1})=d(x_{j-2})+d(x_j)=d(x_2)+d(x_j)$, $d(x_0)=d(x_j)$, i.e.,
$d(u)=d(v)$, it together with Lemma 4.1 implies that
\begin{displaymath}
d(u)=d(v)=a+b
\end{displaymath}
Since $d_{G_0}(x_0)=4$,$a+b\geq 4$.  \hfill $\Box$

By Lemma 4.1, we get $d(x_i)\in \{d_{G_0}(x_i),a+b\}$=\{2,a+b\},
then $d(x_1)=d(x_2)=\cdots=d(x_{j-1})=a+b$,or
$d(x_1)=d(x_2)=\cdots=d(x_{j-1})=2$ by Claim 2.

(I) $d(x_1)=d(x_2)=\cdots=d(x_{j-1})=a+b$, we first prove
$l(P_2)\geq 2$. Otherwise, $l(P_2)=1$, then all the vertices on the
cycle $C=P_1\cup P_2$ have the same degree $a+b$, which contradicts
with Lemma 4.5. So $l(P_2)\geq 2$. Applying Lemma 4.5 to $C=P_1\cup
P_2$,there is $y_i$ such that $d(y_i)\neq a+b$. By Claim 1(ii), we
have $d(y_1)=d(y_2)=\cdots=d(y_{k-1})=2$.

By applying (2) with $(\xi, \eta)=(x_1, y_1)$
\begin{displaymath}
a=\frac{S(x_1)-S(y_1)}{d(x_1)-d(y_1)}=\frac{d(x_0)+d(x_2)+d(x_1)-2-(d(y_0)+d(y_2))}{a+b-2}
\end{displaymath}
\begin{displaymath}
=\frac{a+b+a+b+a+b-2-(a+b+d(y_2))}{a+b-2}
\end{displaymath}
\begin{displaymath}
=\frac{2(a+b)-2-d(y_2)}{a+b-2}=1+\frac{a+b-d(y_2)}{a+b-2}
\end{displaymath}
then $(a+b)-d(y_2)=(a-1)(a+b-2)\geq (a+b-2)$, we get $d(y_2)\leq 2$.
Together with Lemma 4.2(1) and Lemma 4.3, it implies that
$d(y_2)=2$, then $l(P_2)>2$ and $a=2$. by applying Lemma 2.3 with
$R=P_2$, we get $l(P_2)\leq 3$. So $l(P_2)=3$, $d(y_1)=d(y_2)=2$. By
applying (2) with $(\xi, \eta)=(x_j, {\bf x})$
\begin{displaymath}
a=\frac{S(x_j)-S({\bf x})}{d(x_j)-d({\bf
x})}=\frac{d(x_{j-1})+d(y_{k-1})+d(q_{n-1})+d(x_j)-3-(a+b)}{a+b-1}
\end{displaymath}
\begin{displaymath}
=\frac{a+b+2+d(q_{n-1})+a+b-3-(a+b)}{a+b-1}=1+\frac{d(q_{n-1})}{a+b-1}
\end{displaymath}
then $d(q_{n-1})=a+b-1$, it together with Lemma 4.1 implies that
$d(q_{n-1})=d_{G_0}(q_{n-1})$. We have $q_{n-1}=q_0$, i.e.,$n=1$;
otherwise, $d_{G_0}(q_0)=2$, then $a+b-1=d_{G_0}(q_0)=2$, so
$a+b=3$, it contradicts with $a+b\geq 4$. So we get
$d_(q_{n-1})=d_{G_0}(q_{n-1})=d_{G_0}(q_0)=3$ and $a+b=4$. By
applying (2) with $(\xi, \eta)=(q_0, {\bf x})$,
\begin{displaymath}
a=\frac{S(q_0)-S({\bf x})}{d(q_0)-d({\bf
x})}=\frac{d(z_{l-1})+d(p_{m-1})+d(q_1)-(a+b)}{3-1}
\end{displaymath}
\begin{displaymath}
=\frac{d(z_{l-1})+d(p_{m-1})+a+b-(a+b)}{2}=\frac{d(z_{l-1})+d(p_{m-1})}{2}
\end{displaymath}
then $d(z_{l-1})=2,d(p_{m-1})=2$. By applying (2) with $(\xi,
\eta)=(z_{l-1},{\bf x})$
\begin{displaymath}
a=\frac{S(z_{l-1})-S({\bf x})}{d(z_{l-1})-d({\bf
x})}=\frac{d(z_l)+d(z_{l-2})-(a+b)}{2-1}=3+d(z_{l-2})-4=d(z_{l-2})-1
\end{displaymath}
then $d(z_{l-2})=3$, but by Lemma 4.1, $d(z_{l-2})\in
\{d_{G_0}(z_{l-2}),a+b\}=\{2,4\}$, a contradiction.

(II) $d(x_1)=d(x_2)=\cdots=d(x_{j-1})=2$. By Lemma 2.3 with $R=P_1$,
we get $l(P_1)=3$, i.e., $j=3$. By applying (2) with $(\xi,
\eta)=(x_1,{\bf x})$
\begin{displaymath}
a=\frac{S(x_1)-S({\bf x})}{d(x_1)-d({\bf
x})}=\frac{d(x_0)+d(x_2)-(a+b)}{2-1}=a+b+2-(a+b)=2
\end{displaymath}
then $a=2$. By applying (2) with $(\xi, \eta)=(v, {\bf x})$
\begin{displaymath}
a=\frac{S(v)-S({\bf x})}{d(v)-d({\bf
x})}=\frac{d(x_2)+d(q_{n-1})+d(y_{k-1})+a+b-3-(a+b)}{a+b-1}
\end{displaymath}
\begin{displaymath}
=\frac{d(q_{n-1})+d(y_{k-1})-1}{a+b-1}
\end{displaymath}
then $d(q_{n-1})+d(y_{k-1})=2(a+b-1)+1=2(a+b)-1$. So,
$d(q_{n-1})=a+b-1,d(y_{k-1})=a+b$ or
$d(q_{n-1})=a+b,d(y_{k-1})=a+b-1$. If $d(y_{k-1})=a+b-1$, by Lemma
4.1, we get $d(y_{k-1})\in\{d_{G_0}(y_{k-1}),a+b\}=\{2,a+b\}$, then
$d_{G_0}(y_{k-1})=d(y_{k-1})=a+b-1$, $d_{G_0}(y_{k-1})=2$, so
$a+b=3$, it contradicts with $a+b\geq 4$. If $d(q_{n-1})=a+b-1$,
$d(y_{k-1})=a+b$. By Lemma 4.1, we get $d(q_{n-1})\in
\{d_{G_0}(q_{n-1}),a+b\}$, it together with $a+b\geq 4$, we get
$q_{n-1}=q_0$, i.e., $n=1$,
$d(q_{n-1})=d_{G_0}(q_{n-1})=d_{G_0}(q_0)=3$, then $a+b=4$, by
applying (2) with $(\xi, \eta)=(q_0, {\bf x})$,
\begin{displaymath}
a=\frac{S(q_0)-S({\bf x})}{d(q_0)-d({\bf
x})}=\frac{d(v)+d(z_{l-1})+d(p_{m-1})-(a+b)}{3-1}
\end{displaymath}
\begin{displaymath}
=\frac{a+b+d(z_{l-1})+d(p_{m-1})-(a+b)}{2}=\frac{d(z_{l-1})+d(p_{m-1})}{2}
\end{displaymath}
then $d(z_{l-1})+d(p_{m-1})=4$, so $d(z_{l-1})=2$, $d(p_{m-1})=2$.
By applying (2) with $(\xi, \eta)=(z_{l-1}, {\bf x})$,
\begin{displaymath}
a=\frac{S(z_{l-1})-S({\bf x})}{d(z_{l-1})-d({\bf
x})}=\frac{d(z_{l-2})+d(z_l)-(a+b)}{2-1}=d(z_{l-2})+3-4=d(z_{l-2})-1
\end{displaymath}
then $d(z_{l-2})=3$, but by Lemma 4.1, $d(z_{l-2})\in
\{d_{G_0}(z_{l-2}),a+b\}=\{2,4\}$, a contradiction.

{\bf Case 11}. $G_0=T_{13}$, where $x_0=y_0=z_0=u$, $x_j=q_0=f_0=v$,
 $z_l=p_0=q_n=w$ and $y_k=p_m=f_e=z$.

{\bf Subcase 11.1} $d(u),d(v),d(w),d(z)$ are not all the same.

By the symmetry, we may assume that $d(u)\neq d(v)$, by Lemma 4.1,
we get $d(u)\in \{d_{G_0}(u),a+b\}=\{3,a+b\}$ and $d(v)\in
\{d_{G_0}(v),a+b\}=\{3,a+b\}$, then we may assume $d(u)=a+b>3$ and
$d(v)=3$. By Lemma 4.1, we get $\forall \overline{v}\in
\{d_{G_0}(\overline{v}),a+b\}$, it together with
$d_{G_0}(\overline{v})\leq 3$ implies that $d(\overline{v})\leq
a+b$.

(I)  $l(P_1)=1$, by applying (2) with $(\xi, \eta)=(u, {\bf x})$
\begin{displaymath}
a=\frac{S(u)-S({\bf x})}{d(u)-d({\bf
x})}=\frac{d(v)+d(y_1)+d(z_1)+d(u)-3-(a+b)}{a+b-1}
\end{displaymath}
\begin{equation}
=\frac{3+d(y_1)+d(z_1)+a+b-3-(a+b)}{a+b-1}=\frac{d(y_1)+d(z_1)}{a+b-1}
\end{equation}
then $d(y_1)+d(z_1)=a(a+b-1)\geq 2(a+b-1)\geq 6$.

By Lemma 4.1, we get $d(y_1)\in \{d_{G_0}(y_1),a+b\}\subseteq
\{2,3,a+b\}$ and $d(z_1)\in \{d_{G_0}(z_1),a+b\}\subseteq
\{2,3,a+b\}$. By the symmetry, we may assume that $d(y_1)\geq
d(z_1)$, We discuss the following four cases by $d(y_1)\geq d(z_1)$
and $d(y_1)+d(z_1)\geq 6$.

(i) $d(y_1)=d(z_1)=a+b$.

By applying (22) with $(y_1,z_1)=(a+b,a+b)$, we get
\begin{displaymath}
a=\frac{d(y_1)+d(z_1)}{a+b-1}=\frac{d(y_1)+d(z_1)}{a+b-1}=\frac{2(a+b)}{a+b-1}=2+\frac{2}{a+b-1}
\end{displaymath}
Since $a+b>3$ and $a+b\in N^+$, $\frac{2}{a+b-1}$ is not an integer,
a contradiction.

(ii) $d(y_1)=a+b,d(z_1)=3$.

By applying (22) with $(y_1,z_1)=(a+b,3)$, we get
\begin{displaymath}
a=\frac{d(y_1)+d(z_1)}{a+b-1}=\frac{a+b+3}{a+b-1}=1+\frac{4}{a+b-1}
\end{displaymath}
it together with Lemma 4.3, $a+b>3$ and $a,a+b\in N^+$ implies that
$a+b=5$, $a=2$.

 In the following, we prove
$d(y_0)=d(y_1)=\cdots=d(y_k)=a+b=5$.

 If $l(P_2)=1$, i.e., $k=1$,
then $y_0=u,y_1=z,d(y_0)=d(u)=a+b=5$. So $d(y_0)=d(y_1)=a+b=5$. If
$l(P_2)>1$, by applying (2) with $(\xi, \eta)=(y_1, {\bf x})$
\begin{displaymath}
a=\frac{S(y_1)-S({\bf x})}{d(y_1)-d({\bf
x})}=\frac{d(y_0)+d(y_2)+d(y_1)-2-(a+b)}{a+b-1}
\end{displaymath}
\begin{displaymath}
=\frac{a+b+d(y_2)+a+b-2-(a+b)}{a+b-1} =1+\frac{d(y_2)-1}{a+b-1}
\end{displaymath}
then $d(y_2)=a+b$. Similarly, by applying (2) with $(\xi,
\eta)=(y_2,{\bf x}),(y_3,{\bf x}),\cdots,(y_{k-1}$, ${\bf x})$, we get
$d(y_3)=d(y_4)=\cdots=d(y_k)=a+b=5$. Then
$d(y_0)=d(y_1)=\cdots=d(y_k)=a+b=5$.

 Hence, $d(y_0)=d(y_1)=\cdots=d(y_k)=a+b=5$.\hfill $\Box$

 Since $y_k=z$, $d(z)=d(y_k)=a+b=5$. By applying (2) with $(\xi,
\eta)=(z,{\bf x})$
\begin{displaymath}
a=\frac{S(z)-S({\bf x})}{d(z)-d({\bf
x})}=\frac{d(y_{k-1})+d(p_{m-1})+d(f_{e-1})+d(z)-3-(a+b)}{a+b-1}
\end{displaymath}
\begin{displaymath}
=\frac{5+d(p_{m-1})+d(f_{e-1})+5-3-5}{5-1}=\frac{d(p_{m-1})+d(f_{e-1})+2}{4}
\end{displaymath}
then $d(p_{m-1})+d(f_{e-1})=6$, by Lemma 4.1, we get
$d(p_{m-1})\in\{d_{G_0}(p_{m-1}),a+b\}\subseteq \{2,3,5\}$,
$d(f_{e-1})\in\{d_{G_0}(_{e-1}),a+b\}\subseteq\{2,3,5\}$, which
implies that $d(p_{m-1})=3$, $d(f_{e-1})=3$, then $p_{m-1}=z$,
$f_{e-1}=z$, i.e.,$m=1$, $e=1$,then $p_0=w$, $f_0=v$. By applying
(2) with $(\xi, \eta)=(v, {\bf x})$
\begin{displaymath}
a=\frac{S(v)-S({\bf x})}{d(v)-d({\bf
x})}=\frac{d(u)+d(z)+d(p_1)-(a+b)}{3-1}=\frac{5+5+d(p_1)-5}{2}
\end{displaymath}
then $d(p_1)=-1$, it is impossible.

(iii) $d(y_1)=a+b$, $d(z_1)=2$.

By applying (22) with $(y_1,z_1)=(a+b,2)$, we get
\begin{displaymath}
a=\frac{d(y_1)+d(z_1)}{a+b-1}=\frac{a+b+2}{a+b-1}=1+\frac{3}{a+b-1}
\end{displaymath}
it together with Lemma 4.3 and $a,a+b\in N^+$ implies that $a+b=4$,
$a=2$. If $l(P_2)=1$, i.e., $k=1$, then $y_1=z$, $d(z)=d(y_1)=a+b$.
If $l(p_2)>2$, by applying (2) with $(\xi, \eta)=(y_1, {\bf x})$
\begin{displaymath}
a=\frac{S(y_1)-S({\bf x})}{d(y_1)-d({\bf
x})}=\frac{d(y_0)+d(y_2)+d(y_1)-2-(a+b)}{a+b-1}=\frac{a+b+d(y_2)-2}{a+b-1}
\end{displaymath}
then $d(y_2)=a+b$. Similarly, applying (2) with $(\xi, \eta)=(y_2,
{\bf x}), (y_3, {\bf x}), \cdots, (y_{k-1},{\bf x})$, we can get
$d(y_3)=d(y_4)=\cdots=d(y_k)=a+b$. So $\forall k\in N^+$,
$d(y_0)=d(y_1)=d(y_2)=\cdots=d(y_k)=a+b=4$, i.e.,
\begin{equation}
d(z)=a+b=4
\end{equation}
By applying (2) with $(\xi, \eta)=(v, {\bf x})$
\begin{displaymath}
a=\frac{S(v)-S({\bf x})}{d(v)-d({\bf
x})}=\frac{d(f_1)+d(q_1)+d(u)-(a+b)}{d(v)-d(x)}
\end{displaymath}
\begin{displaymath}
=\frac{d(f_1)+d(q_1)+a+b-(a+b)}{3-1}=\frac{d(f_1)+d(q_1)}{2}
\end{displaymath}
then $d(f_1)=2$, $d(q_1)=2$. By applying (2) with $(\xi,
\eta)=(f_1,{\bf x})$
\begin{displaymath}
a=\frac{S(f_1)-S({\bf x})}{d(f_1)-d({\bf
x})}=\frac{d(f_0)+d(f_2)-(a+b)}{2-1}=3+d(f_2)-4=d(f_2)-1
\end{displaymath}
then $d(f_2)=3$, which implies that $f_2=z$, then $d(z)=d(f_2)=3$,
it contradicts with (23).

(iv) $d(y_1)=d(z_1)=3$.

We get $y_1=z$, $z_1=w$, by applying (22) with
$(d(y_1),d(z_1))=(3,3)$,
\begin{displaymath}
a=\frac{d(y_1)+d(z_1)}{a+b-1}=\frac{6}{a+b-1}
\end{displaymath}
it together with Lemma 4.3 and $a,a+b\in N^+$, we get $a+b=4$,
$a=2$. By applying (2) with $(\xi, \eta)=(v, {\bf x})$
\begin{displaymath}
a=\frac{S(v)-S({\bf x})}{d(v)-d({\bf
x})}=\frac{d(f_1)+d(u)+d(q_1)-(a+b)}{3-1}
\end{displaymath}
\begin{displaymath}
=\frac{d(f_1)+4+d(q_1)-4}{2}=\frac{d(f_1)+d(q_1)}{2}
\end{displaymath}
then $d(f_1)=d(q_1)=2$. By applying (2) with $(\xi, \eta)=(f_1, {\bf x})$,
\begin{displaymath}
a=\frac{S(f_1)-S({\bf x})}{d(f_1)-d({\bf
x})}=\frac{d(v)+d(f_2)-(a+b)}{2-1}=3+d(f_2)-4=d(f_2)-1
\end{displaymath}
then $d(f_2)=3\neq a+b$, so $d(f_2)=d_{G_0}(f_2)=3$ and $f_2=z$. In
the same way, $d(q_2)=d_{G_0}(q_2)=3$, and $q_2=w$. By applying (2)
with $(\xi, \eta)=(z, {\bf x})$
\begin{displaymath}
a=\frac{S(z)-S({\bf x})}{d(z)-d({\bf
x})}=\frac{d(u)+d(f_2)+d(p_{m-1})}{3-1}=\frac{4+2+d(p_{m-1})-4}{2}=\frac{2+d(p_{m-1})}{2}
\end{displaymath}
then $d(p_{m-1})=2$. By applying (2) with $(\xi, \eta)=(p_{m-1}, {\bf x})$
\begin{displaymath}
a=\frac{S(p_{m-1})-S({\bf x})}{d(p_{m-1})-d({\bf
x})}=\frac{d(p_{m-2})+d(z)-(a+b)}{2-1}=d(p_{m-2})+3-4=d(p_{m-2})-1
\end{displaymath}
then $d(p_{m-2})=3$, which implies that $p_{m-2}=p_0=w$, i.e.,
$m=2$. Hence, we get $G=G_{33}$. It is easy to check that $G_{33}$
is 2-walk (2, 2)-linear.

(II) $l(P_1)>1$.

Without loss of generality, we may assume that $d(u)=a+b>3$,
$d(v)=3$, by Lemma 2.3(ii) with $R=P_1$, we get $l(P_1)\leq 2$, then
$l(P_1)=2$, i.e., $j=2$. By Lemma 4.1, we get
$d(x_1)\in\{d_{G_0}(x_1),a+b\}=\{2,a+b\}$.

If $d(x_1)=a+b$, by applying (2) with $(\xi, \eta)=(x_1, {\bf x})$
\begin{displaymath}
a=\frac{S(x_1)-S({\bf x})}{d(x_1)-d({\bf
x})}=\frac{d(x_0)+d(x_2)+d(x_1)-2-(a+b)}{a+b-1}
\end{displaymath}
\begin{displaymath}
=\frac{a+b+3+a+b-2-(a+b)}{a+b-1}=\frac{a+b+1}{a+b-1}=1+\frac{2}{a+b-1}
\end{displaymath}
since $a+b\in N^+$, $a+b\geq 4$, then $\frac{2}{a+b-1}$ is not an
integer, it contradicts with $a$ is an integer.

If $d(x_1)=2$, by applying (2) with $(\xi, \eta)=(x_1, {\bf x})$,
\begin{displaymath}
a=\frac{S(x_1)-S({\bf x})}{d(x_1)-d({\bf
x})}=\frac{d(u)+d(v)-(a+b)}{2-1}=a+b+3-(a+b)=3
\end{displaymath}
By applying (2) with $(\xi, \eta)=(u, {\bf x})$,
\begin{displaymath}
a=\frac{S(u)-S({\bf x})}{d(u)-d({\bf
x})}=\frac{d(x_1)+d(y_1)+d(z_1)+d(u)-3-(a+b)}{a+b-1}
\end{displaymath}
\begin{displaymath}
=\frac{2+d(y_1)+d(z_1)+a+b-3-(a+b)}{a+b-1}=\frac{d(y_1)+d(z_1)-1}{a+b-1}
\end{displaymath}
then $d(y_1)+d(z_1)=3(a+b-1)+1=2(a+b)+a+b-2>2(a+b)$, it contradicts
with $d(y_1)\leq a+b,d(z_1)\leq a+b$.

{\bf Subcase 11.2}. $d(u)=d(v)=d(w)=d(z)$. Let
$d(u)=d(v)=d(w)=d(z)=t$.

By applying Corollary 1 with $R=P_1,P_2,\cdots,P_6$, respectively,
we get $d(x_1)=d(x_{j-1})$, $d(y_1)=d(y_{k-1})$,
$d(z_1)=d(z_{l-1})$, $d(p_1)=d(p_{m-1})$, $d(q_1)=d(q_{n-1})$ and
$d(f_1)=d(f_{e-1})$. Let $d(x_1)=d(x_{j-1})=t_1$,
$d(y_1)=d(y_{k-1})=t_2$, $d(z_1)=d(z_{l-1})=t_3$,
$d(p_1)=d(p_{m-1})=t_4$, $d(q_1)=d(q_{n-1})=t_5$,
$d(f_1)=d(f_{e-1})=t_6$. Then $S(u)=at+b=S(v)=S(w)=S(z)$, implying
$S(u)+S(z)=S(v)+S(w)$. Since
\begin{displaymath}
S(u)=d(x_1)+d(y_1)+d(z_1)+d(u)-3=t_1+t_2+t_3+t-3
\end{displaymath}
\begin{displaymath}
S(v)=d(x_{j-1})+d(q_1)+d(f_1)+d(v)-3=t_1+t_5+t_6+t-3
\end{displaymath}
\begin{displaymath}
S(w)=d(z_{l_1})+d(p_1)+d(q_{n-1})+d(w)-3=t_3+t_4+t_5+t-3
\end{displaymath}
\begin{displaymath}
S(z)=d(y_{k-1})+d(p_{m-1})+d(f_{e-1})+d(z)-3=t_2+t_4+t_6+t-3
\end{displaymath}
then $S(u)+S(z)=t_1+t_2+t_3+t-3+t_2+t_4+t_6+t-3$,
$S(v)+S(w)=t_1+2t_2+t_3+t_4+t_6+2t-6$, so
$t_1+2t_2+t_3+t_4+t_6+2t-6=t_1+t_3+t_4+2t_5+t_6+2t-6$, we get
$t_2=t_5$.
In the same way, we get $t_1=t_4$ and $t_3=t_6$.

(I) $t=3$. i.e., $d(u)=d(v)=d(w)=d(z)=3$. Note that $\deltaup(G)=1$,
there is $y\in V(P_1)\cup V(P_2)\cup V(P_3)\cup V(P_4)\cup
V(P_5)\cup V(P_6)\backslash \{u,v,w,z\}$ (where {\bf y} is a
neighbor of a pendant ${\bf x}$ by the preceding assumption). By the
symmetry, we may assume that $y\in V(P_1)\backslash \{u,v\}$. By
applying Lemma 4.4 with $R=P_1$, we get
$d(x_1)=d(x_2)=\cdots=d(x_{j-1})=d(y)=a+b$. By applying (2) with
$(\xi, \eta)=(x_1, {\bf x})$,
\begin{displaymath}
a=\frac{S(x_1)-S({\bf x})}{d(x_1)-d({\bf
x})}=\frac{d(u)+d(x_2)+d(x_1)-2-(a+b)}{a+b-1}
\end{displaymath}
\begin{displaymath}
=\frac{3+d(x_2)+a+b-2-(a+b)}{a+b-1}=\frac{1+d(x_2)}{a+b-1}
\end{displaymath}
then $d(x_2)=a(a+b-1)\geq 2(a+b-1)-1=a+b+(a+b-3)\geq a+b\geq 3\geq
d_{G_0}(x_2)$, which together with Lemma 4.3 implies that
$d(x_2)=a+b$, $a+b=3$, $a=2$. By applying (2) with $(\xi, \eta)=(u,
{\bf x})$
\begin{displaymath}
a=\frac{S(u)-S({\bf x})}{d(u)-d({\bf
x})}=\frac{d(x_1)+d(y_1)+d(z_1)-(a+b)}{3-1}=\frac{d(y_1)+d(z_1)}{2}
\end{displaymath}
then $d(y_1)+d(z_1)=4$, so $d(y_1)=2,d(z_1)=2$. By applying(2) with
$(\xi, \eta)=(y_1, {\bf x})$
\begin{displaymath}
a=\frac{S(y_1)-S({\bf x})}{d(y_1)-d({\bf
x})}=\frac{d(y_0)+d(y_2)-(a+b)}{2-1}=3+d(y_2)-3=d(y_2)
\end{displaymath}
then $d(y_2)=2$, so $l(P_2)\geq 3$, by Lemma 4.4, we get $l(P_2)=3$,
$d(y_1)=d(y_2)=2$. In the same way, $l(P_3)=3$, $d(z_1)=d(z_2)=2$,
so $t_1=3$, $t_2=2$, $t_3=2$ and $t_4=t_1=3$, $t_5=t_2=2$,
$t_6=t_3=2$. By applying Lemma 4.4(i) with $R=P_4$, we get
$d(p_1)=d(p_2)=\cdots=d(p_{m-1})=t_4=3$. By applying Lemma 4.4(ii)
with $R=P_5,P_6$, we get $l(P_5)=3$, $l(P_6)=3$, $d(q_1)=d(q_2)=2$,
$d(f_1)=d(f_2)=2$. Since $\delta(G)=1$, $j>1$ or $n>1$. Hence, we
get $G=G_{34}$. It is easy to check that $G_{34}$ is 2-walk (2,
1)-linear.

(II) $t>3$, then $d(u)>3$. By Lemma 4.1, we get $d(u)\in
\{d_{G_0}(u),a+b\}=\{3,a+b\}$, then $d(u)=a+b$, so
$t=a+b,d(v)=d(w)=d(z)=t=a+b$.

(i) $l(P_i)=1$ for $1\leq i \leq6$.

By applying (2) with $(\xi, \eta)=(u, {\bf x})$
\begin{displaymath}
a=\frac{S(u)-S({\bf x})}{d(u)-d({\bf
x})}=\frac{d(v)+d(w)+d(z)+d(u)-3-(a+b)}{a+b-1}
\end{displaymath}
\begin{displaymath}
=\frac{4(a+b)-3-(a+b)}{a+b-1}=\frac{3(a+b)-3}{a+b-1}=3
\end{displaymath}
then we get $G=G_{35}$, it is easy to check that $G_{35}$ is 2-walk
(3, t-3)-linear.

(ii) There exist $i\in \{1,2,\cdots,6\}$ such that $l(P_i)>1$.

By the symmetry, we may assume that $l(P_1)>1$. By applying Lemma
4.4(i) with $R=P_1$, we get $d(x_1)=d(x_2)=\cdots=d(x_{j-1})\in
\{d_{G_0}(x_0),a+b\}=\{2,a+b\}$.

(a) $d(x_1)=d(x_2)=\cdots=d(x_{j-1})=a+b$. By applying (2) with
$(\xi, \eta)=(x_1, {\bf x})$,
$$a=\frac{S(x_1)-S({\bf x})}{d(x_1)-d({\bf x})}=\frac{d(x_0)+d(x_2)+d(x_1)-2-(a+b)}{a+b-1}=\frac{2(a+b)-2}{a+b-1}=2.$$
By applying (2) with $(\xi, \eta)=(u, {\bf x})$,
\begin{displaymath}
a=\frac{S(u)-S({\bf x})}{d(u)-d({\bf
x})}=\frac{d(x_1)+d(y_1)+d(z_1)+d(u)-3-(a+b)}{a+b-1}
\end{displaymath}
\begin{displaymath}
=\frac{a+b+d(y_1)+d(z_1)+a+b-3-(a+b)}{a+b-1}=1+\frac{d(y_1)+d(z_1)-2}{a+b-1}
\end{displaymath}
then $d(y_1)+d(z_1)-2=a+b-1$, so $d(y_1)+d(z_1)=a+b+1$ and
$d(y_1)<a+b,d(z_1)<a+b$. By Lemma 4.1, we get
$d(y_1)=d_{G_0}(y_1)=2,d(z_1)=d_{G_0}(z_1)=2$, then $a+b=3$. So
$t=3$, a contradiction.

(b) $d(x_1)=d(x_2)=\cdots=d(x_{j-1})=2$. By applying (2) with $(\xi,
\eta)=(x_1, {\bf x})$
\begin{displaymath}
a=\frac{S(x_1)-S({\bf x})}{d(x_1)-d({\bf
x})}=\frac{d(u)+d(x_2)-(a+b)}{2-1}=a+b+d(x_2)-(a+b)=d(x_2)
\end{displaymath}

If $j>2$, then $d(x_2)=2$, i.e., $a=2$, by applying (2) with $(\xi,
\eta)=(u, {\bf x})$
\begin{displaymath}
a=\frac{S(u)-S({\bf x})}{d(u)-d({\bf
x})}=\frac{d(x_1)+d(y_1)+d(z_1)+d(u)-3-(a+b)}{a+b-1}
\end{displaymath}
\begin{displaymath}
=\frac{2+d(y_1)+d(z_1)+a+b-3-(a+b)}{a+b-1}=\frac{d(y_1)+d(z_1)-1}{a+b-1}
\end{displaymath}
then $d(y_1)+d(z_1)=2(a+b-1)+1=a+b+a+b-1$, which together with Lemma
4.1 implies that $a+b-1=2$, then $a+b=3$, so $t=3$, a contradiction.

If $j=2$, i.e., $x_2=v$, then $d(x_2)=d(v)=a+b$. So $a=a+b$. By
applying (2) with $(\xi, \eta)=(u, {\bf x})$
\begin{displaymath}
a=\frac{S(u)-S({\bf x})}{d(u)-d({\bf
x})}=\frac{d(x_1)+d(y_1)+d(z_1)+d(u)-3-(a+b)}{a+b-1}
\end{displaymath}
\begin{displaymath}
=\frac{2+d(y_1)+d(z_1)+a+b-3-(a+b)}{a+b-1}=\frac{d(y_1)+d(z_1)-1}{a+b-1}
\end{displaymath}
then
\begin{equation}
d(y_1)+d(z_1)=a(a+b-1)+1=(a+b)^2-(a+b)+1
\end{equation}
By Lemma 4.1, we get $d(y_1)\in \{d_{G_0}(y_1),a+b\}=\{2,a+b\}$,
$d(z_1)\in \{d_{G_0}(z_1),a+b\}=\{2,a+b\}$, by the symmetry, we may
assume that $d(y_1)\geq d(z_1)$. We discuss the following three
cases.

 If $d(y_1)=a+b$, $d(z_1)=a+b$, by applying (24) with
$(d(x_1),d(y_1))=(a+b,a+b)$, directly calculating yields $a+b$ is
not integer, a contradiction.

 If $d(y_1)=a+b$, $d(z_1)=2$, by applying (24) with
$(d(x_1),d(y_1))=(a+b,2)$, directly calculating yields $a+b$ is not
integer, a contradiction.

 If $d(y_1)=2$, $d(z_1)=2$, by applying (24) with
$(d(x_1),d(y_1))=(2,2)$, directly calculating yields $a+b\leq 3$, it
contradicts with Lemma 4.3.

{\bf Case 12}. $G=T_{14}$, where $x_0=y_0=z_0=p_0=u$ and
$x_j=y_k=z_l=p_m=v$.

We first show some claims.

{\bf Claim 1}. $d(u)=d(v)$, $d(w)=d(z)$.

{\bf Proof}. First, we prove

(i) if $j\geq 2$, then
\begin{displaymath}
 d(x_1)=d(x_2)=\cdots=d(x_{j-1});
\end{displaymath}

(ii) if $k\geq 2$, then
\begin{displaymath}
d(y_1)=d(y_2)=\cdots=d(y_{k-1});
\end{displaymath}

(iii) if $l\geq 2$, then
\begin{displaymath}
d(z_1)=d(z_2)=\cdots=d(z_{l-1});
\end{displaymath}

(iv) if $m\geq 2$, then
\begin{equation}
 d(p_1)=d(p_2)=\cdots=d(p_{m-1}).
\end{equation}

By the symmetry, we only need to prove (i). By way of contradiction,
there exist $i\in\{1,2,j-2\}$, such that $d(u_i)\neq d(u_{i+1})$.
By applying Corollary 3 with $R=P_1$, we get $a+b=3$,
$a=2$ and $l(P_1)=5$, $d(u)=d(v)=4$, $d(x_1)=d(x_4)=3$,
$d(x_2)=d(x_3)=2$. By applying (2) with $(\xi, \eta)=(u, {\bf x})$,
\begin{displaymath}
a=\frac{S(u)-S({\bf x})}{d(u)-d({\bf
x})}=\frac{d(x_1)+d(y_1)+d(z_1)+d(p_1)-(a+b)}{4-1}
\end{displaymath}
\begin{displaymath}
=\frac{3+d(y_1)+d(z_1)+d(p_1)-3}{3}=\frac{d(y_1)+d(z_1)+d(p_1)}{3}
\end{displaymath}

Since $G$ is simple, at most one of $l(P_2)$, $l(P_3)$ and $l(P_4)$ is 1.

If one of $l(P_2)$, $l(P_3)$ and $l(P_4)$ is 1, without loss of generality, $l(P_4)=1$,
then $a=\frac{d(y_1)+d(z_1)+d(p_1)}{3}=\frac{d(y_1)+d(z_1)+4}{3}$,
which implies that $d(y_1)+d(z_1)=2$, it is impossible.

If $l(P_2)$, $l(P_3)$, $l(P_4)>1$, then $a=\frac{d(y_1)+d(z_1)+d(p_1)}{3}$,
which implies that $d(y_1)=d(z_1)=d(p_1)=2$, by applying (2) with $(\xi, \eta)=(y_1,{\bf x})$
\begin{displaymath}
a=\frac{S(y_1)-S({\bf x})}{d(y_1)-d({\bf
x})}=\frac{d(y_0)+d(y_2)-(a+b)}{2-1}=4+d(y_2)-3=d(y_2)+1
\end{displaymath}
then $d(y_2)=1$, it is impossible. So
\begin{displaymath}
d(x_1)=d(x_2)=\cdots=d(x_{j-1})
\end{displaymath}

In the same way, we can prove (ii)-(iv).

Next, we prove $d(u)=d(v)$. Otherwise, $d(u)\neq d(v)$, then
$S(u)-S(v)=d(x_1)+d(y_1)+d(z_1)+d(p_1)+d(u)-4-d(x_{j-1})-d(y_{k-1})-d(z_{l-1})-d(p_{m-1})-d(v)+4$,
it together with (25) implies that
$S(u)-S(v)=d(p_1)+d(u)-d(p_{m-1})-d(v)$. If $m=1$, then
$S(u)-S(v)=0$; If $m>1$, then $S(u)-S(v)=d(u)-d(v)$. So
\begin{displaymath}
a = \left\{ \begin{array}{ll}
0 & \mbox{if $m=1$}; \\
1 & \mbox{if $m>1$}
\end{array} \right.
\end{displaymath}
we get $a\leq 1$, it contradicts with Lemma 4.3. Hence, $d(u)=d(v)$.

{\bf Claim 2}. If $d(u)=d(v)=a+b$, let $X=\{\overline{v}|
d(\overline{v})=a+b,\overline{v}\in \{x_1,y_1,z_1,p_1\}\}$, then
$|X|=1$.

{\bf Proof}. By way of contradiction, assume that $|X|\neq 1$.

If $|X|>1$, we may assume without loss of generality,
$d(x_1)=d(y_1)=a+b$. By (25) and $d(u)=d(v)=a+b$, we get
$d(x_0)=d(x_1)=\cdots=d(x_j)=a+b$,
$d(y_0)=d(y_1)=\cdots=d(y_k)=a+b$. Noting that $P_1\neq P_2$, and
$G$ is simple. We may assume without loss of generality, that
$l(P_1)\geq 2$. Then, $C=ux_1\ldots vy_1\ldots y_{k-1}u$ is a cycle
that contradicts with Lemma 4.5.

If $|X|=0$, then $d(x_1),d(y_1),d(z_1),d(p_1)\neq a+b$, it together
with Lemma 4.1 implies that $d(x_1)=d(y_1)=d(z_1)=d(p_1)=2$. Since
$d(u)=a+b$ and $d_{G_0}(u)=4$, $a+b\geq4$. By applying (2) with
$(\xi, \eta)=(u, {\bf x})$
\begin{displaymath}
a=\frac{S(u)-S({\bf x})}{d(u)-d({\bf
x})}=\frac{d(x_1)+d(y_1)+d(z_1)+d(p_1)+d(u)-4-(a+b)}{d(u)-1}
\end{displaymath}
\begin{displaymath}
=\frac{8+a+b-4-(a+b)}{a+b-1}=\frac{4}{a+b-1}\leq\frac{4}{3}<2
\end{displaymath}
which contradicts with Lemma 4.3.

Hence, $|X|=1$.

{\bf Claim 3}. $d(u)=d(v)=4$.

 {\bf Proof}. By way of contradiction,
assume that $d(u),d(v)\neq 4$. Since $d(u)\geq d_{G_0}(u)=4$ and
$d(v)\geq d_{G_0}(v)=4$, we only need to consider $d(u)>4$ and
$d(v)>4$. By Lemma 4.1 and $d_{G_0}(u)=4$, we get $d(u)=a+b>4$. By
Claim 2 we get $|X|=1$, by the symmetry, we may assume that
$d(x_1)=a+b$, $d(y_1)=d_{G_0}(y_1)=2$, $d(z_1)=d_{G_0}(z_1)=2$,
$d(p_1)=d_{G_0}(p_1)=2$. By applying (2) with $(\xi, \eta)=(u, {\bf x})$,
\begin{displaymath}
a=\frac{S(u)-S({\bf x})}{d(u)-d({\bf
x})}=\frac{d(x_1)+d(y_1)+d(z_1)+d(p_1)+d(u)-4-(a+b)}{a+b-1}
\end{displaymath}
\begin{displaymath}
=\frac{4+a+b+a+b-4-(a+b)}{a+b-1}=\frac{2+a+b}{a+b-1}=1+\frac{3}{a+b-1}\leq1+\frac{3}{4}<2
\end{displaymath}
which contradicts with Lemma 4.3. Hence, $d(v)=d(u)=4$.

By Claim 3, $d(u)=d(v)=4=d_{G_0}(u)=d_{G_0}(v)$. Note that
$\delta(G)=1$, there is $y\in V(P_1)\cup V(P_2)\cup V(P_3)\cup
V(P_4)\backslash \{u,v\}$, (where {\bf y} is a neighbor of a pendant
${\bf x}$ by the preceding assumption). By the symmetry, we may
assume that $y\in V(P_1)\backslash \{u,v\}$, then $d(y)=a+b$, it
together with (25), we get
$d(x_1)=d(x_2)=\cdots=d(x_{j-1})=d(y)=a+b$. By applying (2) with
$(\xi, \eta)=(x_1, {\bf x})$,
\begin{displaymath}
a=\frac{S(x_1)-S({\bf x})}{d(x_1)-d({\bf
x})}=\frac{d(u)+d(x_2)+d(x_1)-2-(a+b)}{a+b-1}
\end{displaymath}
\begin{equation}
=\frac{4+d(x_2)+a+b-2-(a+b)}{a+b-1}=\frac{d(x_2)+2}{a+b-1}
\end{equation}

{\bf Subcase 12.1} $j>2$. Then $d(x_2)=a+b$. By applying (26) with
$d(x_2)=a+b$, we get
\begin{displaymath}
a=\frac{d(x_2)+2}{a+b-1}=\frac{a+b+2}{a+b-1}=1+\frac{3}{a+b-1}
\end{displaymath}
it together with Lemma 4.3 and $a, a+b\in N^+$ implies that $a+b=4$,
$a=2$. By applying (2) with $(\xi, \eta)=(u, {\bf x})$,
\begin{displaymath}
a=\frac{S(u)-S({\bf x})}{d(u)-d({\bf
x})}=\frac{d(x_1)+d(y_1)+d(z_1)+d(p_1)-(a+b)}{4-1}
\end{displaymath}
\begin{displaymath}
=\frac{4+d(y_1)+d(z_1)+d(p_1)-4}{3}=\frac{d(y_1)+d(z_1)+d(p_1)}{3}
\end{displaymath}
then $d(y_1)+d(z_1)+d(p_1)=6$, so $d(y_1)=d(z_1)=d(p_1)=2$. By
applying (2) with $(\xi, \eta)=(y_1, {\bf x})$,
\begin{displaymath}
a=\frac{S(y_1)-S({\bf x})}{d(y_1)-d({\bf
x})}=\frac{d(u)+d(y_2)-(a+b)}{2-1}=a+b+d(y_2)-(a+b)=d(y_2)
\end{displaymath}
then $d(y_2)=2$, $l(P_2)\geq3$, by (25), we get
$d(y_3)=\cdots=d(y_{k-1})=d(y_1)=2$, it together with Lemma 2.3(i)
implies that $l(P_2)=3$, $d(y_1)=d(y_2)=2$. In the same way, we get
$l(P_3)=3$, $d(z_1)=d(z_2)=2$, $l(P_4)=3$, $d(p_1)=d(p_2)=2$. Hence,
we get $G=G_{36}$. It is easy to check that $G_{36}$ is 2-walk (2,
2)-linear.

{\bf Subcase 12.2} $j=2$. Then $x_2=v$, $d(x_2)\neq a+b$, it together
with (25) implies that $x_2=v$ and $d(x_2)=d(v)=4$, then $a+b\neq
4$. By applying (26) with $d(x_2)=4$, we get
$a=\frac{2+d(x_2)}{a+b-1}=\frac{6}{a+b-1}$, it together with Lemma
4.3 and $a+b\neq 4$, $a, a+b\in N^+$ implies that $a+b=3$, $a=3$. By
applying (2) with $(\xi, \eta)=(u, {\bf x})$,
\begin{displaymath}
a=\frac{S(u)-S({\bf x})}{d(u)-d({\bf
x})}=\frac{d(x_1)+d(y_1)+d(z_1)+d(p_1)-(a+b)}{4-1}
\end{displaymath}
\begin{displaymath}
=\frac{a+b+d(y_1)+d(z_1)+d(p_1)-(a+b)}{3}=\frac{d(y_1)+d(z_1)+d(p_1)}{3}
\end{displaymath}
then $d(y_1)+d(z_1)+d(p_1)=9$, by the symmetry, we may assume that
$d(y_1)\geq d(z_1)\geq d(p_1)$. This implies that (i) $d(y_1)=4$,
$d(z_1)=a+b=3$, $d(p_1)=2$; or (ii) $d(y_1)=a+b=3$, $d(z_1)=a+b=3$,
$d(p_1)=a+b=3$.

(i) $d(y_1)=4$, $d(z_1)=3$, $d(p_1)=2$. Then $l(P_2)=1$, i.e.,
$k=1$. By applying (2) with $(\xi, \eta)=(z_1, {\bf x})$
\begin{displaymath}
a=\frac{S(z_1)-S({\bf x})}{d(z_1)-d({\bf
x})}=\frac{d(z_0)+d(z_2)+d(z_1)-2-(a+b)}{a+b-1}
\end{displaymath}
\begin{displaymath}
=\frac{4+d(z_2)+3-2-3}{3-1}=\frac{d(z_2)+2}{2}
\end{displaymath}
then $d(z_2)=4$, i.e, $z_2=v$, then $l(P_3)=2$. By applying (2) with
$(\xi, \eta)=(p_1, {\bf x})$
\begin{displaymath}
a=\frac{S(p_1)-S({\bf x})}{d(p_1)-d({\bf
x})}=\frac{d(p_0)+d(p_2)-(a+b)}{2-1}=4+d(p_2)-3=d(p_2)+1
\end{displaymath}
then $d(p_2)=2$, it together with (25) and Lemma 2.3(i) implies that
$l(P_4)=3$, $d(p_1)=d(p_2)=2$. Hence, we get $G=G_{37}$. It is easy
to check that $G_{37}$ is 2-walk (3, 0)-linear.

(ii) $d(x_1)=3$, $d(y_1)=3$, $d(z_1)=3$. By applying (2) with $(\xi,
\eta)=(x_1, {\bf x})$
\begin{displaymath}
a=\frac{S(x_1)-S({\bf x})}{d(x_1)-d({\bf
x})}=\frac{d(u)+d(x_2)+d(x_1)-2-(a+b)}{3-1}
\end{displaymath}
\begin{displaymath}
=\frac{4+d(x_2)+3-2-3}{2}=\frac{d(x_2)+2}{2}
\end{displaymath}
then $d(x_2)=4$, then $x_2=v$. In the same way, we get $d(y_2)=4$,
$y_2=v$; $d(z_2)=4$, $z_2=v$ and $d(p_2)=4$, $p_2=v$. Hence, we get
$G=G_{38}$. It is easy to check that $G_{38}$ is 2-walk (3,
0)-linear.

{\bf Case 13}. $G=T_{15}$, where $x_0=y_0=q_0=u$, $x_j=y_k=f_0=v$,
$z_0=p_0=q_n=w$ and $z_l=p_m=f_e=z$.

{\bf Claim 1}. $d(u)=d(v)$.

By way of contradiction, assume that $d(u)\neq d(v)$, then $l(P_1)$, $l(P_2)\leq 2$, i.e.,
$j,k\geq 2$. By Lemma 4.1,
we get $d(u)\in \{d_{G_0}(u),a+b\}=\{3,a+b\}$, $d(v)\in
\{d_{G_0}(v), a+b\}=\{3,a+b\}$. By the symmetry, we may assume that
$d(u)=a+b$, $d(v)=3$, it together with Lemma 4.3 implies that
$a+b>3$. By the symmetry, we may assume that $l(P_1)\geq l(P_2)$.
$G$ is simple, then $l(P_1)\geq 2$, So $l(P_1)=2$. By applying Lemma 4.4(i) with
$R=P_1$, we get $d(x_1)=d(x_{j-1})$. If $k=1$, then $d(y_1)=d(v)$,
$d(y_{k-1})=d(u)$; If $k=2$, by applying Lemma 4.4(i) with $R=P_2$,
then $d(y_1)=d(y_{k-1})$. So we get
\begin{displaymath}
S(u)-S(v)=d(x_1)+d(y_1)+d(q_1)+d(u)-3-d(x_{j-1})-d(y_{k-1})-d(f_1)
\end{displaymath}
\begin{displaymath}
=
 \left\{ \begin{array}{ll}
d(q_1)-d(f_1) & \mbox{if $k=1$};\\
d(q_1)-d(f_1)+a+b-3 & \mbox{if $k=2$}
\end{array} \right.
\end{displaymath}
then
\begin{equation}
a=\frac{S(u)-S(v)}{d(u)-d(v)}=
\left\{
\begin{array}{ll}
\frac{d(q_1)-d(f_1)}{a+b-3}  &\mbox{if $k=1$}; \\ \\
\frac{a+b-3+d(q_1)-d(f_1)}{a+b-3} &\mbox{if $k=2$}
\end{array} \right.
\end{equation}
If $d(q_1)\leq d(f_1)$, then
\begin{displaymath}
a=\left\{
\begin{array}{ll}
\frac{d(q_1)-d(f_1)}{a+b-3}<0  &\mbox{if $k=1$};\\ \\
1+\frac{d(q_1)-d(f_1)}{a+b-3}<1  &\mbox{if $k=2$}
\end{array} \right.
\end{displaymath}
which contradicts with Lemma 4.3. So $d(q_1)>d(f_1)$. By Lemma 4.1,
we get $d(q_1)\in \{d_{G_0}(q_1),a+b\}\subseteq \{2,3,a+b\}$,
$d(f_1)\in \{d_{G_0}(f_1),a+b\}\subseteq \{2,3,a+b\}$. So we only
need to discuss the following three cases by $d(q_1)>d(f_1)$.

(i) $d(q_1)=3,d(f_1)=2$.

By applying (27), we get
\begin{displaymath}
a= \left\{\begin{array}{ll} \frac{1}{a+b-3} &\mbox{if $k=1$}; \\
\\
1+\frac{1}{a+b-3} &\mbox{if $k=2$}
\end{array} \right.
\end{displaymath}
which together with Lemma 4.3 and $a,a+b\in N^+$ implies that $k=2$,
$a+b=4$, $a=2$. Since $d(q_1)=3$, $d(q_1)\neq a+b$. So,
$d(q_1)=d_{G_0}(q_1)$, $q_1=w$. By applying (2) with $(\xi,
\eta)=(u,{\bf x})$,
\begin{displaymath}
a=\frac{S(u)-S({\bf x})}{d(u)-d({\bf
x})}=\frac{d(x_1)+d(y_1)+d(q_1)+d(u)-3-(a+b)}{a+b-1}
\end{displaymath}
\begin{displaymath}
=\frac{d(x_1)+d(y_1)+3+4-3-4}{3}=\frac{d(x_1)+d(y_1)}{3}
\end{displaymath}
then $d(x_1)+d(y_1)=6$. By Lemma 4.1, we get $d(x_1)\in
\{d_{G_0}(x_1),a+b\}\subseteq\{2,3,4\}$. If $d(x_1)=d(y_1)=3$,
then $x_1=y_1=v$, implying $l(P_1)=l(P_2)=1$; but $G$ is simple, a contradiction. So
$d(x_1)=2$, $d(y_1)=4$ or $d(x_1)=4$,
$d(y_1)=2$. By the symmetry, we only need to discuss $d(x_1)=4$,
$d(y_1)=2$. By applying (2) with $(\xi, \eta)=(x_1, {\bf x})$,
\begin{displaymath}
a=\frac{S(x_1)-S({\bf x})}{d(x_1)-d({\bf
x})}=\frac{d(u)+d(x_2)+d(x_1)-2-(a+b)}{4-1}
\end{displaymath}
\begin{displaymath}
=\frac{4+d(x_2)+4-2-4}{3}=\frac{d(x_2)+2}{3}
\end{displaymath}
then $d(x_2)=4$. Similarly, by applying (2) with $(\xi,
\eta)=(x_2,{\bf x}),(x_3,{\bf x}),\cdots,(x_{j-1},{\bf x})$, we get
$d(x_3)=\cdots=\d(x_{j})=4$, but $x_j=v$ and $d(v)=3$, a contradiction.

(ii) $d(q_1)=a+b$, $d(f_1)=2$.

By applying (27),
\begin{displaymath}
a= \left\{\begin{array}{ll} 1+\frac{1}{a+b-3} &\mbox{if $k=1$}; \\
\\
2+\frac{1}{a+b-3} &\mbox{if $k=2$}
\end{array}\right.
\end{displaymath}
which together with Lemma 4.3 and $a, a+b\in N^+$ implies that
$a+b=4$, and
\begin{displaymath}
a= \left\{\begin{array}{ll} 2 &\mbox{if $k=1$}; \\
\\
3 &\mbox{if $k=2$}
\end{array}\right.
\end{displaymath}

(a) $k=1$, i.e., $l(P_2)=1$.

 Since $k=1$, i.e, $y_1=v$,
$d(y_1)=d(v)=3$. By applying (2) with $(\xi, \eta)=(u, {\bf x})$,
\begin{displaymath}
a=\frac{S(u)-S({\bf x})}{d(u)-d({\bf
x})}=\frac{d(x_1)+d(y_1)+d(p_1)+d(u)-3-(a+b)}{a+b-1}
\end{displaymath}
\begin{displaymath}
=\frac{d(x_1)+3+4+4-3-4}{4-1}=\frac{d(x_1)+4}{3}
\end{displaymath}
then $d(x_2)=2$. By applying (2) with $(\xi, \eta)=(x_1, {\bf x})$
\begin{displaymath}
a=\frac{S(x_1)-S({\bf x})}{d(x_1)-d({\bf
x})}=\frac{d(x_0)+d(x_2)-(a+b)}{2-1}=4+d(x_2)-4=d(x_2)
\end{displaymath}
then $d(x_2)=2$. By applying Lemma 4.4(ii) with $R=P_1$. But the degrees of
end-vertices of $P_1$ are $d(u)=4$ and $d(v)=3$, respectively, a contradicts with Lemma 2.3(ii).
(b)  $k=2$, i.e., $l(P_2)=2$.

If $l(P_5)>1$, i.e., $n>1$, by applying (2) with
$(\xi, \eta)=(q_1, {\bf x})$,
\begin{displaymath}
a=\frac{S(q_1)-S({\bf x})}{d(q_1)-d({\bf
x})}=\frac{d(q_0)+d(q_2)+d(q_1)-2-(a+b)}{a+b-1}
\end{displaymath}
\begin{displaymath}
=\frac{4+d(q_2)+4-2-4}{3}=\frac{d(q_2)+2}{3}
\end{displaymath}
then $d(q_2)=7$, it is impossible.

If $l(P_5)=1$, then $q_1=w$. By applying (2) with $(\xi,
\eta)=(f_1,{\bf x})$,
\begin{displaymath}
a=\frac{S(f_1)-S({\bf x})}{d(f_1)-d({\bf
x})}=\frac{d(v)+d(f_2)-(a+b)}{2-1}=3+d(f_2)-4=d(f_2)-1
\end{displaymath}
then $d(f_2)=4$. If $l(P_6)>2$, i.e., $e>2$, then $d(f_1)=2$,
$d(f_2)=4$, which contradicts with Lemma 4.4. So $l(P_6)=2$. By
applying (2) with $(\xi, \eta)=(v,{\bf x})$
\begin{displaymath}
a=\frac{S(v)-S({\bf x})}{d(v)-d({\bf
x})}=\frac{d(x_{j-1})+d(y_{k-1})+d(f_1)-(a+b)}{3-1}=\frac{d(x_{j-1})+d(y_{k-1})-2}{2}
\end{displaymath}
then $d(x_{j-1})+d(y_{k-1})=8$. So $d(x_{j-1})=4$, $d(y_{k-1})=4$.
By applying (2) with $(\xi, \eta)=(x_{j-1}, {\bf x})$
\begin{displaymath}
a=\frac{S(x_{j-1})-S({\bf x})}{d(x_{j-1})-d({\bf
x})}=\frac{d(v)+d(x_{j-2})+d(x_{j-1})-2-(a+b)}{4-1}
\end{displaymath}
\begin{displaymath}
=\frac{3+d(x_{j-2})+4-2-4}{3}=\frac{d(x_{j-1})+1}{3}
\end{displaymath}
then $d(x_{j-2})=8$, it is impossible.

(iii) $d(q_1)=a+b$, $d(f_1)=3$. Then $f_1=z$.

Since $d(q_1)>d(f_1)$, then $a+b>3$. By applying (27),
\begin{displaymath}
a= \left\{ \begin{array}{ll} 1 &\mbox{if $k=1$}\\ \\
2 &\mbox{if $k>1$}
\end{array}\right.
\end{displaymath}
By Lemma 4.3, we get $a=2$, we only need to discuss $k=2$. By
applying (2) with $(\xi, \eta)=(u, {\bf x})$,
\begin{displaymath}
a=\frac{S(u)-S({\bf x})}{d(u)-d({\bf
x})}=\frac{d(x_1)+d(y_1)+d(q_1)+d(u)-3-(a+b)}{a+b-1}
\end{displaymath}
\begin{displaymath}
=\frac{d(x_1)+d(y_1)+a+b+a+b-3-(a+b)}{a+b-1}=\frac{d(x_1)+d(y_1)+a+b-3}{a+b-1}
\end{displaymath}
then $d(x_1)+d(y_1)=a+b+1$, implying $d(x_1)=2$; Otherwise, $d(x_1)=a+b$ by Lemma 4.1,
and $d(y_1)=1$, a contradiction. By applying (2) with
$(\xi, \eta)=(x_1,{\bf x})$
\begin{displaymath}
a=\frac{S(x_1)-S({\bf x})}{d(x_1)-d({\bf
x})}=\frac{d(x_0)+d(x_2)-(a+b)}{2-1}=d(x_2)
\end{displaymath}
then $d(x_2)=2$. So $l(P_1)>2$, which contradicts with $k=2$.

The proof of Claim 1 is completed.    \hfill $\Box$

By the symmetry, we may assume that $d(u)\geq d(w)$, and $d(x_1)\geq
d(y_1)$. By Lemma 4.1, we get $d(x_1)\in\{d_{G_0}(x_1),a+b\}$. We
discuss the following two cases according to $d(x_1)$.

{\bf Subcase 13.1}. $d(x_1)=a+b$.

By Lemma 4.1, we get $d(y_1)\in \{d_{G_0}(y_1),a+b\}$

(I) $d(y_1)=a+b$.

By applying (2) with $(x_1,{\bf x})$,
\begin{displaymath}
a=\frac{S(x_1)-S({\bf x})}{d(x_1)-d({\bf
x})}=\frac{d(u)+d(x_2)+a+b-2-(a+b)}{a+b-1}=\frac{d(u)+d(x_2)-2}{a+b-1}
\end{displaymath}
then $d(u)+d(x_2)=a(a+b-1)+2\geq 2(a+b-1)+2=2(a+b)$, which implies $d(u)=a+b$,
$d(x_2)=a+b$ and $a=2$ since $d(u)\leq a+b$ and $d(x_2)\leq a+b$ by Lemma 4.1.
By applying (2) with $(\xi, \eta)=(u, {\bf x})$,
\begin{displaymath}
a=\frac{S(u)-S({\bf x})}{d(u)-d({\bf
x})}=\frac{d(x_1)+d(y_1)+d(q_1)+d(u)-3-(a+b)}{a+b-1}
\end{displaymath}
\begin{displaymath}
=\frac{a+b+a+b+d(q_1)+a+b-3-(a+b)}{a+b-1}=\frac{2(a+b)+d(q_1)-3}{a+b-1}
\end{displaymath}
then $d(q_1)=1$, it is impossible.

(II) $d(y_1)=d_{G_0}(y_1)$.

(i) $k=1$, i.e., $y_1=v$. Then $d(v)=d_{G_0}(v)=3$. So
$d(u)=d(v)=3$ by Claim 1.

In the following, we prove $a+b>3$. Otherwise, $a+b\leq 3$. It together with Lemma
4.3 implies that $a+b=3$. By Corollary1(ii), we get
$d(x_0)=d(x_1)=\cdots=d(x_j)=3=a+b$. Then all the vertices on the
cycle $C=P_1\cup P_2$ have the same degree $a+b$, which contradicts
with Lemma 4.5. Hence, $a+b>3$.

If $l(P_1)=2$, then $x_2=v$, and $d(x_2)=d(v)=3$. By applying (2)
with $(\xi, \eta)=(x_1, {\bf x})$
\begin{displaymath}
a=\frac{S(x_1)-S({\bf x})}{d(x_1)-d({\bf
x})}=\frac{d(u)+d(v)+d(x_1)-2-(a+b)}{a+b-1}
\end{displaymath}
\begin{displaymath}
=\frac{3+3+a+b-2-(a+b)}{a+b-1}=\frac{4}{a+b-1}<\frac{4}{3}<2
\end{displaymath}
which contradicts with Lemma 4.3.

If $l(P_1)\geq 3$, by applying with Lemma 4.4(i) with $R=P_1$, we
get $d(x_2)=d(x_3)=\cdots=d(x_{j-1})=d(x_1)=a+b$. By applying (2)
with $(\xi, \eta)=(x_1, {\bf x})$
\begin{displaymath}
a=\frac{S(x_1)-S({\bf x})}{d(x_1)-d({\bf
x})}=\frac{d(u)+d(x_2)+d(x_1)-2-(a+b)}{a+b-1}
\end{displaymath}
\begin{displaymath}
=\frac{3+a+b+a+b-2-(a+b)}{a+b-1}=\frac{a+b+1}{a+b-1}=1+\frac{2}{a+b-1}<\frac{5}{3}<2
\end{displaymath}
which contradicts with Lemma 4.3.

(ii) $k>1$. i.e., $l(P_2)>1$. Then $d(y_1)=d_{G_0}(y_1)=2$.

By applying (2) with $(\xi, \eta)=(u, {\bf x})$,
\begin{displaymath}
a=\frac{S(u)-S({\bf x})}{d(u)-d({\bf
x})}=\frac{d(x_1)+d(y_1)+d(q_1)+d(u)-3-(a+b)}{d(u)-1}
\end{displaymath}
\begin{equation}
=\frac{a+b+2+d(q_1)+d(u)-3-(a+b)}{d(u)-1}=\frac{d(u)+d(q_1)-1}{a+b-1}=1+\frac{d(q_1)}{d(u)-1}
\end{equation}

(a) $d(u)>d_{G_0}(u)=3$, which together with Lemma 4.1 implies that
$d(u)=a+b$. By Lemma 4.1, we get $d(q_1)\in \{d_{G_0}(q_1),a+b\}$.

(a1) $d(q_1)=d_{G_0}(q_1)$.

If $n>1$, then $d(q_1)=d_{G_0}(q_1)=2$. By applying (28) with $d(q_1)=2$, we get
$a=1+\frac{2}{a+b-1}<\frac{5}{3}$. which contradicts with
 Lemma 4.3.

If $n=1$, i.e., then $d(q_1)=d(w)=d_{G_0}(w)=3$. By applying (28) with $d(q_1)=3$, we get
$a=1+\frac{3}{a+b-1}$, which together with Lemma 4.3 and $a, a+b\in
N^+$, implies that $a+b=4$, $a=2$, then $d(x_1)=a+b=4,d(u)=a+b$,
$d(v)=d(u)=a+b=4$. By applying Lemma 4.4(i) with $R=P_1$, we get
$d(x_2)=\cdots=d(x_{j-1})=d(x_1)=a+b=4$.

By applying (2) with $(\xi, \eta)=(y_1, {\bf x})$,
\begin{displaymath}
a=\frac{S(y_1)-S({\bf x})}{d(y_1)-d({\bf
x})}=\frac{d(y_0)+d(y_2)-(a+b)}{2-1}=4+d(y_2)-4=d(y_2)
\end{displaymath}
then $d(y_2)=2$. By applying Lemma 4.4(ii) with $R=P_2$, we get
$l(P_2)=3$, $d(y_1)=d(y_2)=2$. By applying (2) with $(\xi,
\eta)=(u,{\bf x})$,
\begin{displaymath}
a=\frac{S(u)-S({\bf x})}{d(u)-d({\bf
x})}=\frac{d(x_1)+d(y_1)+d(q_1)+d(u)-3-(a+b)}{4-1}
\end{displaymath}
\begin{displaymath}
=\frac{4+2+d(q_1)+4-3-4}{3}=\frac{d(q_1)+3}{3}
\end{displaymath}
then $d(q_1)=3\neq a+b$. So $d(q_1)=d_{G_0}(q_1)$ and $q_1=w$.

In the same way, $f_1=z$ and $d(z)=d_{G_0}(z)=3$. By applying (2) with
$(\xi, \eta)=(w,{\bf x})$,
\begin{displaymath}
a=\frac{S(w)-S({\bf x})}{d(w)-d({\bf
x})}=\frac{d(u)+d(z_1)+d(p_1)-(a+b)}{3-1}
\end{displaymath}
\begin{displaymath}
=\frac{4+d(z_1)+d(p_1)-4}{2}=\frac{d(z_1)+d(p_1)}{2}
\end{displaymath}
then $d(z_1)+d(p_1)=4$, implying $d(z_1)=d(p_1)=2$. By applying (2)
with $(\xi, \eta)=(z_1, {\bf x})$,
\begin{displaymath}
a=\frac{S(z_1)-S({\bf x})}{d(z_1)-d({\bf
x})}=\frac{d(w)+d(z_2)-(a+b)}{2-1}=3+d(z_2)-4=d(z_2)-1
\end{displaymath}
then $d(z_2)=3\neq a+b$. So $d(z_2)=d_{G_0}(z_2),z_2=z$. In the same
way, $p_2=z$.

Hence, we get $G=G_{39}$. It is easy to check that $G_{39}$ is
2-walk (2, 2)-linear.

(a2) $d(q_1)=a+b$. By applying (28), we get
$a=1+\frac{a+b}{a+b-1}=2+\frac{1}{a+b-1}$, implies that $a$ is not an integer, a
contradiction.

(b) $d(u)=d_{G_0}(u)=3$. By applying (28), we get
$a=1+\frac{d(q_1)}{2}$. By Lemma 4.1, we get $d(q_1)\in
\{d_{G_0}(q_1),a+b\}$.

(b1) $d(q_1)=d_{G_0}(q_1)$.

If $n=1$, i.e., $q_1=w$, then $d(q_1)=d(w)=d_{G_0}(w)=3$, implying $a=1+\frac{d(q_1)}{2}=\frac{5}{2}$, a
contradiction.

If $n>1$, $d(q_1)=d_{G_0}(q_1)=2$, implying $a=2$. By applying (2) with $(\xi,
\eta)=(y_1,{\bf x})$
\begin{displaymath}
a=\frac{S(y_1)-S({\bf x})}{d(y_1)-d({\bf
x})}=\frac{d(u)+d(y_2)-(a+b)}{2-1}=3+d(y_2)-(a+b)
\end{displaymath}
then $d(y_2)=a+b-1<a+b$, implying $d(y_2)=d_{G_0}(y_2)\in \{2,3\}$ by  Lemma 4.1.
If $d(y_2)=3$, i.e., $y_2=v$, then
$a+b=4$. So $d(x_1)=a+b=4$. By applying (2) with $(\xi, \eta)=(x_1,
{\bf x})$,
\begin{displaymath}
a=\frac{S(x_1)-S({\bf x})}{d(x_1)-d({\bf
x})}=\frac{d(u)+d(x_2)+d(x_1)-2-(a+b)}{4-1}
\end{displaymath}
\begin{displaymath}
=\frac{3+d(x_2)+a+b-2-(a+b)}{3}=\frac{d(x_2)+1}{3}
\end{displaymath}
then $d(x_2)=5$, it is impossible since $a+b=4<5$. Hence, $d(y_2)=2$ and $a+b=3$.
By applying Lemma 4.4(ii) with $R=P_2$, we get
$l(P_2)=3$, $d(y_1)=d(y_2)=2$. By
applying (2) with $(\xi, \eta)=(x_1, {\bf x})$,
\begin{displaymath}
a=\frac{S(x_1)-S({\bf x})}{d(x_1)-d({\bf
x})}=\frac{d(u)+d(x_2)+d(x_1)-2-(a+b)}{a+b-1}
\end{displaymath}
\begin{displaymath}
=\frac{3+d(x_2)+3-2-3}{3-1}=\frac{d(x_2)+1}{2}
\end{displaymath}
then $d(x_2)=3$. By applying Corollary 1 with $R=P_1$, we get
$d(x_1)=\cdots=d(x_j)=a+b=3$.

By applying (1) with $\xi=u, v$,
respectively, we get $S(u)=3a+b=S(v)$. Since
$S(u)=d(x_1)+d(y_1)+d(q_1)=3+2+2=7$ and
$S(v)=d(x_{j-1})+d(y_2)+d(f_1)=5+d(f_1)$, $d(f_1)=2$.
By applying (2) with $(\xi, \eta)=(f_1, {\bf x})$,
\begin{displaymath}
a=\frac{S(f_1)-S({\bf x})}{d(f_1)-d({\bf
x})}=\frac{d(v)+d(f_2)-(a+b)}{2-1}=3+d(f_2)-3=d(f_2)
\end{displaymath}
then $d(f_2)=2$. By applying Lemma 4.4(ii) with $R=P_6$, we get
$l(P_6)=3$, $d(f_1)=d(f_2)=2$. In the same way, $l(P_5)=3$,
$d(q_1)=d(q_2)=2$.

 By Lemma 4.1, we get
$d(w)\in\{d_{G_0}(w),a+b\}=\{3\}$. By applying (2)
with $(\xi, \eta)=(w, {\bf x})$
\begin{displaymath}
a=\frac{S(w)-S({\bf x})}{d(w)-d({\bf
x})}=\frac{d(q_1)+d(z_1)+d(p_1)-(a+b)}{3-1}
\end{displaymath}
\begin{displaymath}
=\frac{2+d(z_1)+d(p_1)-3}{2}=\frac{d(z_1)+d(p_1)-1}{2}
\end{displaymath}
then $d(z_1)+d(p_1)=5$, implying $d(z_1)=2$ and $d(p_1)=3$ or $d(z_1)=3$ and
$d(p_1)=2$. By the symmetry, we only need to discuss $d(z_1)=2$ and
$d(p_1)=3$. Using the method above on the path $P_2$ to $P_3$, we
get $l(P_3)=3$, $d(z_1)=d(z_2)=2$; Also, using the method above on
the path $P_1$ to $P_4$, we get $d(p_1)=d(p_2)=\cdots=d(p_{m})=3$.

Hence, we get $G=G_{40}$. It is easy to check that $G_{40}$ is
2-walk (2, 1)-linear.

(b2) $d(q_1)=a+b$. Then $a=1+\frac{a+b}{2}$, it together with Lemma
4.1 and $a, a+b\in N^+$, implies that $a+b\geq 4$, and $a\geq 3$.
By applying (2) with $(\xi, \eta)=(y_1, {\bf x})$
\begin{displaymath}
a=\frac{S(y_1)-S({\bf x})}{d(y_1)-d({\bf
x})}=\frac{d(u)+d(y_2)-(a+b)}{2-1}=3+d(y_2)-(a+b)
\end{displaymath}
then $d(y_2)=\frac{3(a+b)}{2}-2=a+b+\frac{a+b}{2}-2\geq a+b\geq 4>d_{G_0}(y_2)$,
it together with Lemma 4.1 implies that $d(y_2)=a+b$, then
$\frac{3(a+b)}{2}-2=a+b$. So $a+b=4$. It together with
Lemma 4.4(i), implies that $y_2=v$ and $d(v)=a+b=4$, then
$d(u)=d(v)=4$, it contradicts with $d(u)=3$.

{\bf Subcase 13.2}. $d(x_1)=d_{G_0}(x_1)$.

(I) $j=1$, i.e., $x_1=v$.

Then $d(v)=d_{G_0}(v)=3$. Since $G$ is simple, $y_1\neq v$. It
together with Lemma 4.3, we get $d(y_1)\in\{2,a+b\}$.

If $d(y_1)=2$, this is a special case in Subcase 13.1(see the case (b) in Subcase13.1(II)).

If $d(y_1)=a+b$, since $d(y_1)\leq d(x_1)=3$ and $a+b\geq 3$, $a+b=3$.
This is a special case in Subcase 13.1(I).

(II) $j>1$. Then $d(x_1)=2$, implying $d(y_1)=2$ since $d(y_1)\leq d(x_1)=2$.

 By Lemma 4.1, we get $d(q_1)\in
\{d_{G_0}(q_1),a+b\}$.

(i) $d(q_1)=d_{G_0}(q_1)$.

(a) $n>1$. Then $d(q_1)=d_{G_0}(q_1)=2$. By applying (2) with $(\xi, \eta)=(u,
{\bf x})$,
\begin{displaymath}
a=\frac{S(u)-S({\bf x})}{d(u)-d({\bf
x})}=\frac{d(x_1)+d(y_1)+d(q_1)+d(u)-3-(a+b)}{d(u)-1}
\end{displaymath}
\begin{displaymath}
=\frac{3+d(x_0)-(a+b)}{d(x_0)-1}=1+\frac{4-(a+b)}{d(x_0)-1}\leq 1+\frac{1}{3-1}<2
\end{displaymath}
which contradicts with Lemma 4.3.

(b) $n=1$, i.e., $q_1=w$. Then $d(q_1)=d(w)=d_{G_0}(w)=3$. By applying (2) with
$(\xi, \eta)=(u, {\bf x})$,
\begin{displaymath}
a=\frac{S(u)-S({\bf x})}{d(u)-d({\bf
x})}=\frac{d(x_1)+d(y_1)+d(q_1)+d(u)-3-(a+b)}{d(u)-1}
\end{displaymath}
\begin{displaymath}
=\frac{7+d(u)-3-(a+b)}{d(u)-1} =1+\frac{5-(a+b)}{d(u)-1}\leq
1+\frac{5-3}{3-1}=2
\end{displaymath}
then $d(u)=3$, $a+b=3$, $a=2$. By applying (2) with $(\xi,
\eta)=(x_1,{\bf x})$,
\begin{displaymath}
a=\frac{S(x_1)-S({\bf x})}{d(x_1)-d({\bf
x})}=\frac{d(x_0)+d(x_2)-(a+b)}{2-1}=3+d(x_2)-3=d(x_2)
\end{displaymath}
then $d(x_2)=2$. By applying Lemma 4.4(ii) with $R=P_1$, we get
$l(P_1)=3$, $d(x_1)=d(x_2)=2$. In the same way, $l(P_2)=3$,
$d(y_1)=d(y_2)=2$. Since $d(u)=3$, then $d(v)=d(u)=3$. By applying
(2) with $(\xi, \eta)=(v, {\bf x})$,
\begin{displaymath}
a=\frac{S(v)-S({\bf x})}{d(v)-d({\bf
x})}=\frac{d(x_2)+d(y_2)+d(f_1)-(a+b)}{3-1}
\end{displaymath}
\begin{displaymath}
=\frac{2+2+d(f_1)-3}{2}=\frac{d(f_1)+1}{2}
\end{displaymath}
then $d(f_1)=3$. By Lemma 4.1, we get $d(f_e)\in
\{d_{G_0}(f_e),a+b\}=\{3\}$. By applying Corollary 1 with $R=P_6$,
we get $d(f_1)=d(f_2)=\cdots=d(f_e)=a+b=3$. By applying
Corollary 1,we get $d(q_1)=d(q_2)=\cdots=d(q_n)=3$ since $d(u)=3$, $d(w)\in
\{d_{G_0}(w),a+b\}=\{3\}$, and $d(q_0)=d(q_1)=d(q_n)=3$. By applying (2)
with $(\xi, \eta)=(w, {\bf x})$,
\begin{displaymath}
a=\frac{S(w)-S({\bf x})}{d(w)-d({\bf
x})}=\frac{d(q_{n-1})+d(z_1)+d(p_1)-(a+b)}{3-1}
\end{displaymath}
\begin{displaymath}
=\frac{3+d(z_1)+d(p_1)-3}{2}=\frac{d(z_1)+d(p_1)}{2}
\end{displaymath}
then $d(p_1)+d(z_1)=4$, so $d(p_1)=2$, $d(z_1)=2$. Using the method
above on the $P_1$ to $P_3$, $P_4$, we get $l(P_1)=3$, $d(z_1)=2$,
$d(z_2)=2$ and $l(P_2)=3$, $d(p_1)=2$, $d(p_2)=2$.

Hence, we get $G=G_{41}$. It is easy to check that $G_{41}$ is
2-walk (2, 1)-linear.

(iii) $d(q_1)=a+b$. By applying (2) with $(\xi, \eta)=(u, {\bf x})$,
\begin{displaymath}
a=\frac{S(u)-S({\bf x})}{d(u)-d({\bf
x})}=\frac{d(x_1)+d(y_1)+d(q_1)+d(u)-3-(a+b)}{d(u)-1}
\end{displaymath}
\begin{displaymath}
=\frac{2+2+a+b-3-(a+b)}{d(u)-1}=\frac{1}{d(u)-1}<2
\end{displaymath}
which contradicts with Lemma 4.3.

Summarizing above, the proof of Theorem 4.1 is completed.

\end{document}